\documentclass[%
 aip,
 amsmath,amssymb,
 reprint,%
]{revtex4-1}
\usepackage[utf8]{inputenc}
\usepackage[T1]{fontenc}
\usepackage{textcomp}
\usepackage{color}
\usepackage{comment}
\usepackage{xcolor}
\usepackage{amsmath}
\usepackage{amsfonts}
\usepackage{bm}
\usepackage{mathptmx}
\usepackage{dcolumn}
\usepackage{siunitx}

\usepackage{graphicx}
\usepackage{float}
\usepackage{booktabs}

\usepackage{listings}

\usepackage{etoolbox}
\usepackage{csquotes}
\usepackage{tabularx}
\usepackage[colorlinks=true, linkcolor=blue]{hyperref}

\usepackage{fontawesome5} 

\usepackage{longtable,booktabs,array}
\newcolumntype{S}{>{\raggedright\arraybackslash}p{0.30\textwidth}} 
\newcolumntype{M}{>{\raggedright\arraybackslash}p{0.70\textwidth}} 
\renewcommand{\arraystretch}{1.06}

\def\unit#1{~ \ensuremath{\mathrm{\,#1}}}
\def\celsius{\ensuremath{\mathrm{^{\circ}C}}}

\DeclareUnicodeCharacter{2019}{'}

\hypersetup{
    colorlinks=true,
    linkcolor=black,
    urlcolor=black,
    citecolor=black
}

\makeatletter
\def\@email#1#2{%
 \endgroup
 \patchcmd{\titleblock@produce}
  {\frontmatter@RRAPformat}
  {\frontmatter@RRAPformat{\produce@RRAP{*#1\href{mailto:#2}{#2}}}\frontmatter@RRAPformat}
  {}{}
}%
\makeatother
\begin{document}

\preprint{AIP/123-QED}

\title[The Hitchhiker's Guide to DDM]{The Hitchhiker's Guide to Differential Dynamic Microscopy}
\author{Enrico Lattuada}
\affiliation{Faculty of Physics, University of Vienna, Boltzmanngasse 5, 1090 Vienna, Austria}

\author{Fabian Krautgasser} 
\affiliation{Faculty of Physics, University of Vienna, Boltzmanngasse 5, 1090 Vienna, Austria}

\author{Maxime Lavaud} 
\affiliation{Faculty of Physics, University of Vienna, Boltzmanngasse 5, 1090 Vienna, Austria}

\author{Fabio Giavazzi}
\affiliation{Dipartimento di Biotecnologie Mediche e Medicina Traslazionale, Universit\`{a} degli Studi di Milano, via F.lli Cervi 93, 20054 Segrate (MI), Italy}

\author{Roberto Cerbino}
\affiliation{Faculty of Physics, University of Vienna, Boltzmanngasse 5, 1090 Vienna, Austria}
\email{roberto.cerbino@univie.ac.at}


\date{\today}

\begin{abstract}
Over nearly two decades, Differential Dynamic Microscopy (DDM) has become a standard technique for extracting dynamic correlation functions from time-lapse microscopy data, with applications spanning classical soft matter systems such as colloidal suspensions, liquid crystals, polymer solutions, gels and glasses, as well as active fluids and biological systems. In its most common implementation, DDM analyzes image sequences acquired with a conventional microscope equipped with a digital camera, yielding time- and wavevector-resolved information analogous to that obtained in multi-angle Dynamic Light Scattering (DLS). With a widening array of applications and a growing, heterogeneous user base, lowering the technical barrier to performing DDM has become a central objective. In this tutorial article, we provide a step-by-step guide to conducting DDM experiments—from planning and acquisition to data analysis—intended as a resource for both new and experienced practitioners. We also introduce the open-source software package \texttt{fastDDM}, designed to efficiently process large image datasets using optimized, parallel algorithms that reduce analysis times by up to four orders of magnitude on typical datasets (e.g., 10,000 frames), thereby enabling high-throughput workflows, reproducibility, and broader adoption across disciplines.
\end{abstract}

\maketitle

\section{\label{sec:intro}Introduction: a guide to the Guide}
The study of dynamics in complex systems is crucial for many scientific fields, ranging from materials science to biology, and includes the multidisciplinary domain of soft matter~\cite{barrat2023soft}. A well-established approach to probing these dynamics is Dynamic Light Scattering (DLS), which has proven invaluable for characterizing the size, shape, and motion of molecules, macromolecules, and a wide array of suspended particles~\cite{berne2000dynamic}. Yet, when confronted with systems that are structurally complex, dynamically heterogeneous, or inherently inhomogeneous, DLS and similar conventional methods can encounter significant interpretive and instrumental limitations~\cite{scheffold2007new}.

Since its introduction by Cerbino and Trappe in 2008~\cite{cerbino2008differential}, Differential Dynamic Microscopy (DDM) has gained prominence as a robust, versatile, and increasingly standardized technique capable of probing dynamical processes across multiple length and time scales in diverse complex fluids and biological systems~\cite{cerbino2017perspective,cerbino2022differential,al2022differential}. DDM integrates key principles of DLS with imaging-based methodologies grounded in optical microscopy. Rather than relying solely on scattered intensity fluctuations, DDM operates directly on time-lapse microscopy images, thereby capitalizing on the inherent strengths of real-space visualization. These include flexible imaging modalities (e.g., bright-field, fluorescence, phase-contrast), the capacity to spatially resolve and target regions of interest within heterogeneous samples, the use of conventional and easily accessible illumination sources, and comparatively straightforward sample preparation protocols suitable for a broad range of systems. Moreover, DDM’s capability to inherently remove static background contributions and its substantial insensitivity to multiple scattering~\cite{giavazzi2014viscoelasticity,eitel2020hitchhiker,nixon2022probing} render it an exceptionally well-suited tool for the quantitative study of sample dynamics, either alongside or in place of traditional DLS.

Thanks to these attributes, DDM has found wide-ranging applications in fields spanning soft matter physics, biophysics, active matter, and microbiology. For example, bright-field DDM can be employed to study small proteins -- either isolated~\cite{guidolin2023protein} or in clusters~\cite{safari2015differential} -- as well as colloids~\cite{cerbino2008differential,giavazzi2016structure,shokeen2017comparison,hitimana2022diffusive,bradley2023sizing}, and motile bacteria~\cite{martinez2012differential,germain2016differential}. Beyond bright-field microscopy, DDM can be extended to a variety of imaging modalities, including wide-field~\cite{he2012diffusive,shokeen2017comparison}, light-sheet~\cite{wulstein2016light}, and confocal~\cite{lu2012characterizing,verwei2022quantifying} techniques. These methods can also be combined, for example by coupling phase-contrast and confocal microscopy to visualize different components within cell tissues and quantify their respective dynamics~\cite{drechsler2017active,giavazzi2018tracking}. This flexibility and cross-compatibility make DDM an increasingly adopted tool not only for the study of complex systems but also for educational use in teaching laboratories and training environments~\cite{ferri2011kinetics,germain2016differential,schatz2023advancing}.

As recently highlighted in Ref.~\onlinecite{zhang2023differential}, two major limitations of conventional DDM workflows are the scarcity of ready-to-use software and the computational demands of analyzing long image sequences, which may contain thousands of frames. In practical terms, these constraints can lead to analysis times extending to several hours, especially in high-throughput or high-frame-rate experiments. This lengthy process represents a significant bottleneck in workflows where rapid feedback or real-time interpretation is required. To address this bottleneck, we have developed a user-friendly, open-source software package, \texttt{fastDDM}, which is presented and discussed in detail in this tutorial~\cite{fastddm}. \texttt{fastDDM} employs optimized, parallelized algorithms~\cite{norouzisadeh2021modern} to significantly streamline the DDM analysis process, thereby reducing the processing time to about one minute or less, depending on the dataset size and hardware configuration. The public release of \texttt{fastDDM}, accompanied by detailed documentation and illustrative examples, is intended to encourage broader adoption and lower the technical barrier for new users. We are committed to maintaining and expanding \texttt{fastDDM} in collaboration with the scientific community, incorporating user feedback and contributions.

Beyond the software, this tutorial provides a comprehensive yet accessible introduction to DDM, covering its physical principles, formal connection to DLS, and wide range of experimental applications. It is intended for a diverse readership -- from experienced researchers looking to expand their analytical toolkit to students and early-career scientists entering the field -- and serves as a practical guide to navigating the broad, evolving, and methodologically rich landscape of DDM.

\subsection*{How to read this paper}
This article is organized so that readers can follow either a \emph{tutorial track} (first-time users) or an \emph{application track} (experienced users). Sections marked with the GitHub logo (\faGithub) are accompanied by \textit{Jupyter notebooks} that enable readers to reproduce and explore the analyses step by step, all available on GitHub \cite{fastddm-tutorials}. All datasets used in the paper are publicly available and linked to the corresponding notebooks.

\textit{Tutorial track.} Sections~\ref{sec:anotherspace}--\ref{sec:particle-sizing} establish the foundations: Section~\ref{sec:anotherspace} introduces image formation and the DDM/light-scattering analogy; Section~\ref{sec:hoodwheel} presents the algorithmic foundations and the optimized implementation through \texttt{fastDDM}; Section~\ref{sec:particle-sizing} provides a worked tutorial on particle sizing in dilute Brownian suspensions, including quantitative “take-home messages” for experimental design. Practical setup advice is offered in Appendix~\ref{app:practical}. Passages marked with an asterisk (*) collect optional or advanced topics and can be skipped on first reading without loss of continuity.

\textit{Application track.} Section~\ref{sec:beyondsizing} collects four case studies: (i) protein solutions, where we extract diffusivity, hysdrodynamics radius and interaction parameters from bright-field data (Section~\ref{subsec:proteins}); (ii) bacterial motion, spanning motile swimmers in dense suspensions studied with confocal DDM, and non-motile anisotropic bacteria with roto–translational Brownian dynamics studied with dark-field DDM (Section~\ref{subsec:bacteria}); (iii) microrheology, converting microscope videos into viscoelastic moduli for a polymer solution (Section~\ref{subsec:ddm-microrheology}); and (iv) time-resolved dynamics of confluent cell monolayers, highlighting ageing and super-diffusive behavior (Section~\ref{subsec:cells-time-dependence}). All examples ship with companion notebooks and datasets.

\section{\label{sec:anotherspace}Understanding DDM: From images to scattering}
The relationship between direct and reciprocal space is a source of enduring fascination and, at times, confusion for those working with imaging and scattering techniques. Reciprocal space representations are deeply rooted in the historical development of crystallography and scattering, while direct space is the natural domain of microscopy. DDM operates at the interface of these two frameworks: it extracts dynamical information by analyzing temporal fluctuations in real-space images, yet quantifies this information in reciprocal space through the structure function\footnote{We refer to $D(q, \Delta t)$ as the structure function, also known in the literature as the image structure function. For brevity and consistency, we use the shorter term throughout this manuscript.}. This duality, which is the key feature of DDM, is also a potential source of conceptual ambiguity.

In this section, we describe the formal and practical relationships between direct and reciprocal space in DDM, with particular attention to how the structure function \( D(\mathbf{q}, \Delta t) \) used in DDM relates to the intermediate scattering function \( f(\mathbf{Q}, \Delta t) \) used in DLS (bold characters represent vectors, in this case scattering vectors). By clarifying how the spatial Fourier components of image differences encode dynamics, we aim to resolve common doubts that arise when interpreting DDM results. We also discuss how different microscopy modalities influence the mapping between real and reciprocal space, and we highlight the limitations and assumptions inherent in applying reciprocal space concepts to microscopy data. 

Our goal is to provide both intuition and formalism, helping readers navigate the apparent paradox of performing reciprocal space analysis on real-space data, a hallmark of DDM and a powerful tool when properly understood.

\subsection{Light scattering probes the reciprocal space}
\label{subsec:LSRP}
\begin{figure}[htbp]
    \centering
    \includegraphics[width=\columnwidth]{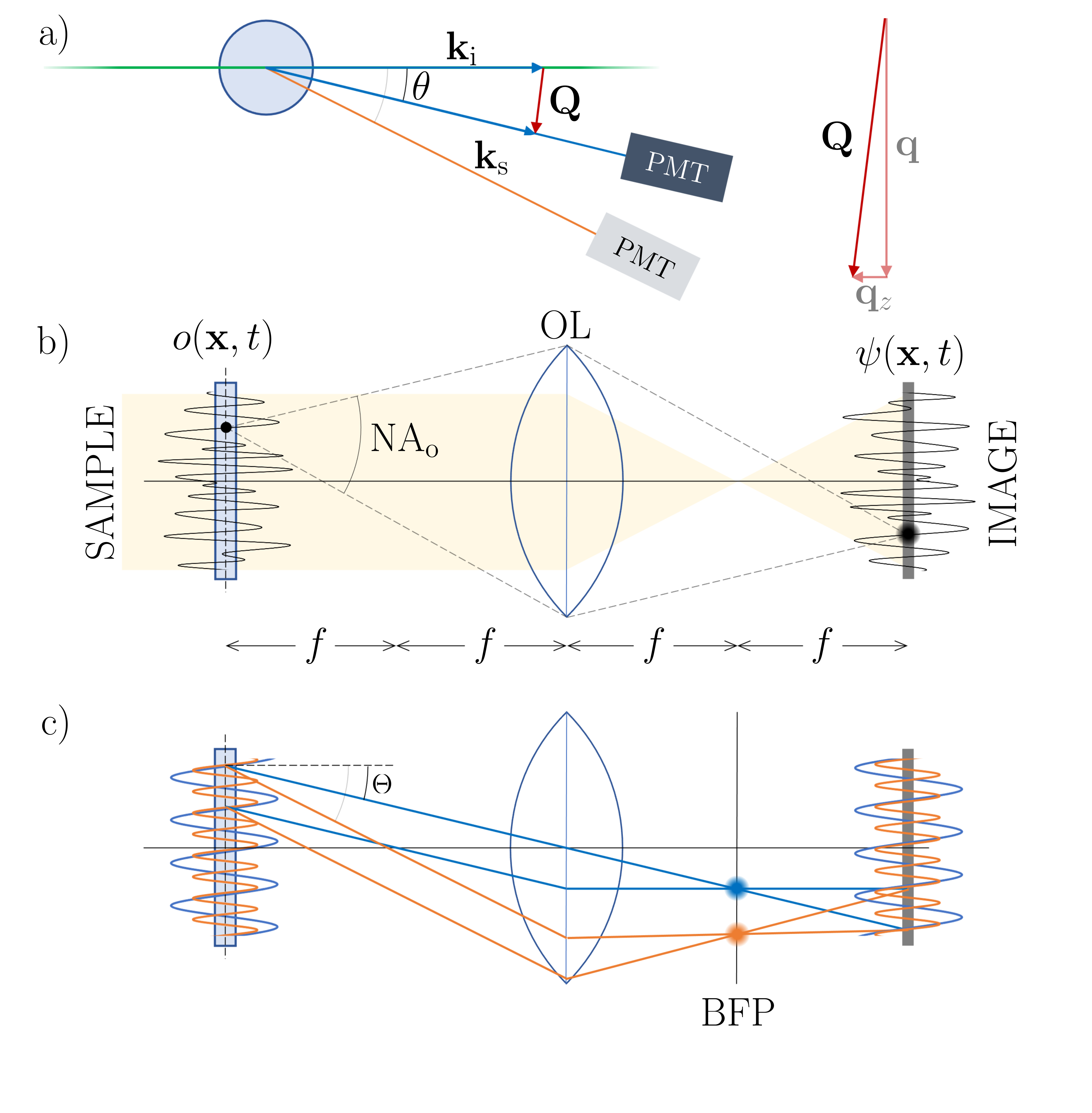}
    \caption{
         \textbf{a)} In a DLS experiment, a monochromatic plane wave with wavevector $\mathbf{k}_\mathrm{i}$ illuminates the sample. Scattered light with wavevector $\mathbf{k}_\mathrm{s}$ is collected at angle $\theta$ in the far field using, e.g., a photomultiplier tube (PMT) or another point detector. This arrangement is effectively equivalent to detecting light in the back focal plane (BFP) of a microscope, making explicit the geometric correspondence between scattering and imaging. The scattering vector $\mathbf{Q}$ can be decomposed using $\mathbf{q}_z$ along the incident beam direction and a transverse one $\mathbf{q}$.
        \textbf{b)} In microscopy, the sample is illuminated by a monochromatic plane wave, and the objective lens (OL) of focal length $f$ and numerical aperture $\mathrm{NA_o}$ collects the scattered light. Geometrical optics maps each scattering direction to a point in the BFP, while each object point is reconstructed in the image plane. In practice, diffraction blurs the image $\psi(\mathbf{x},t)$ over a finite region determined by the point spread function $\kappa(\mathbf{x})$ [Eq.~\eqref{eq:psf}].
        \textbf{c)} According to the Abbe–Fourier theory, an object can be represented as a superposition of sinusoidal gratings, each diffracting light into plane waves at angles $\pm\Theta$ (only $+\Theta$ shown). The lens focuses these into the BFP, and their interference reconstructs the spatial modulations in the image plane. Thus, the image is the sum of sinusoidal contributions (smeared by Eq.~\eqref{eq:convo2D}) that correspond one-to-one with modulations in the sample.
    }
    \label{fig:ddm-dls}
\end{figure}

In light scattering experiments (Fig.~\ref{fig:ddm-dls}a), a monochromatic plane wave with wavevector $\mathbf{k}_\mathrm{i}$ illuminates the sample. Spatial inhomogeneities in the refractive index scatter the light into different directions. A detector placed in the far field at angle $\theta$ relative to the incident beam collects scattered light traveling along the wavevector $\mathbf{k}_\mathrm{s}$. The momentum transferred is described by the \textit{scattering vector} $\mathbf{Q} = \mathbf{k}_\mathrm{s} - \mathbf{k}_\mathrm{i}$. Under elastic conditions, $|\mathbf{k}_\mathrm{s}| = |\mathbf{k}_\mathrm{i}| = n_\mathrm{s} k_0 = 2\pi n_\mathrm{s}/\lambda_0$, where $\lambda_0$ is the wavelength in vacuum, $n_\mathrm{s}$ is the refractive index of the sample medium, and the magnitude of $\mathbf{Q}$ is given by $Q = 2n_\mathrm{s}k_0 \sin(\theta/2)$. Scattering at a prescribed angle $\theta$ occurs only if the sample contains spatial modulations with wavevector $\mathbf{Q}$, corresponding to structures with spatial period $2\pi/|\mathbf{Q}|$. By varying $\theta$, one accesses different $\mathbf{Q}$ values, thus probing a range of length scales within the sample. As illustrated in Fig.~\ref{fig:ddm-dls}a, the scattering vector can be decomposed as $\mathbf{Q} = (\mathbf{q}, \mathbf{q}_z)$, where $\mathbf{q}_z$ lies along the incident beam direction and $\mathbf{q}$ spans the transverse plane. This decomposition becomes essential when comparing light scattering with optical microscopy, which primarily accesses the transverse components.

\begin{figure*}[htbp]
    \centering
    \includegraphics[width=\textwidth]{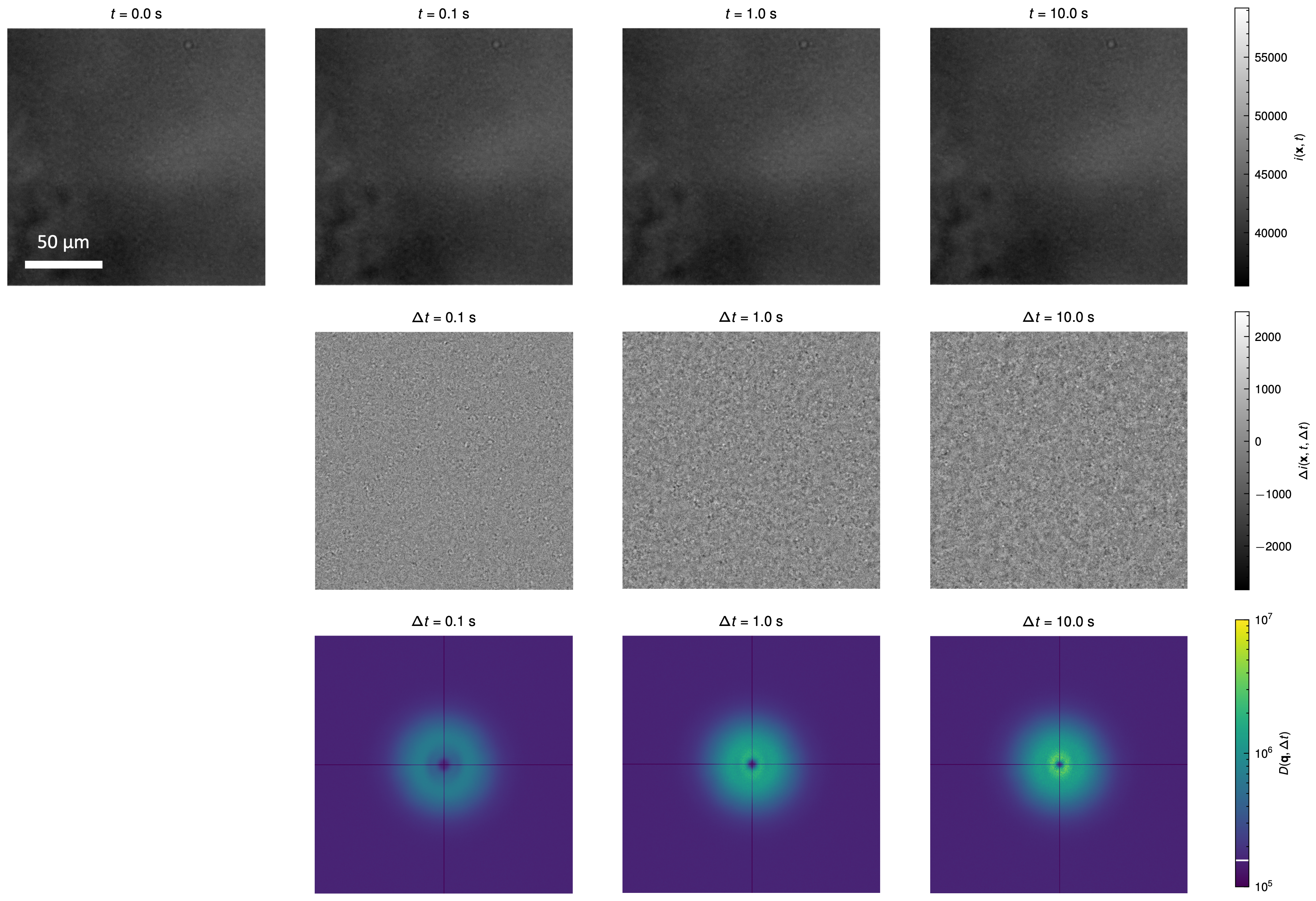}
    \caption{
        \textit{First row}: Bright-field microscopy images of Brownian particles (diameter $240\unit{nm}$) acquired at different times. Each panel spans approximately $166\unit{\mu m}$. The weak signal $\psi$ from the particles is masked by a dominant static contribution $i_{0}$ from dust and other imperfections on optical surfaces and the detector. 
        \textit{Second row}: Difference images $\Delta i(\mathbf{x},t,\Delta t)$ obtained by subtracting a reference image acquired at $t=0$ from subsequent frames at delays $\Delta t=0.1,1.0,10.0$ s. This subtraction removes $i_0$ and highlights the contribution from particle motion. The increasing contrast with $\Delta t$ reflects the Brownian displacements. The average size of the granularity (speckles) visible in the last difference image ($\Delta t=10$ s) gives an estimate of the microscope resolution, as the particle size lies below the resolution limit. 
        \textit{Third row}: Structure functions $D(\mathbf{q},\Delta t)$ computed for $\Delta t=0.1,1.0,10.0$ s, by averaging over 5980, 5800, and 4000 difference images, respectively, characterized by the same $\Delta t$ but different $t$. The contrast increases with $\Delta t$, consistent with enhanced decorrelation due to the particle dynamics. The central black cross masks processing artifacts. The white line on the color bar indicates the estimated noise floor $B$.
    }
    \label{fig:group}
\end{figure*}

To describe light scattering more formally, consider $N_{\mathrm{par}}$ identical particles that occupy positions $\{\mathbf{r}_j(t)\}_{j=1}^{N_{\mathrm{par}}}$ at time $t$. A central quantity in scattering theory is the \emph{intermediate scattering function}
\begin{equation}\label{eq:isfu}
F(\mathbf{Q}, \Delta t) = \frac{1}{N_{\mathrm{par}}} \sum_{j=1}^{N_{\mathrm{par}}} \sum_{k=1}^{N_{\mathrm{par}}} \left\langle \exp\left[ -\mathrm{i} \mathbf{Q} \cdot \left( \mathbf{r}_j(0) - \mathbf{r}_k(\Delta t) \right) \right] \right\rangle~,
\end{equation}
which characterizes the time evolution of spatial correlations at wavevector $\mathbf{Q}$. Here, the brackets $\langle\cdots\rangle$ indicate an ensemble average obtained over many statistically independent equivalent copies of the same system, which for an ergodic system is equivalent to an average over a sufficiently long time. In addition, $\Delta t$ denotes the time lag between two observations of the system, with particle positions evaluated at initial time $t = 0$ and at a later time $t = \Delta t$. At equal times, this reduces to the \emph{static structure factor},
\begin{equation}
S(\mathbf{Q}) = F(\mathbf{Q}, 0)~,
\end{equation}
and the \emph{normalized intermediate scattering function} is defined as
\begin{equation}
f(\mathbf{Q}, \Delta t) = \frac{F(\mathbf{Q}, \Delta t)}{S(\mathbf{Q})}~,
\end{equation}
which decays from $f(\mathbf{Q}, 0) = 1$ to zero as particles lose positional correlations over time. The intermediate scattering function can be separated into \emph{self} and \emph{distinct} parts~\cite{hansen2013theory}
\begin{equation}
\label{eq:selfdist}
f(\mathbf{Q}, \Delta t) = f_\mathrm{s}(\mathbf{Q}, \Delta t) + f_\mathrm{d}(\mathbf{Q}, \Delta t)~,
\end{equation}
where $f_\mathrm{s}$ accounts for correlations of a particle with itself ($j = k$) and $f_\mathrm{d}$ captures correlations between different particles ($j \ne k$). 

In \emph{static light scattering} (SLS), the key observable is the time-averaged intensity
\begin{equation}
\label{eq:SLS}
I_\mathrm{s}(\mathbf{Q}) = \langle I_\mathrm{s}(\mathbf{Q}, t) \rangle_t = N_{\mathrm{par}} P(\mathbf{Q}) S(\mathbf{Q})~,
\end{equation}
where $P(\mathbf{Q})$ is the particle form factor, determined by size and shape.

In \emph{dynamic light scattering} (DLS), the main observable is the normalized intensity autocorrelation function
\begin{equation}
g^{(2)}(\mathbf{Q}, \Delta t) = \frac{\langle I_s(\mathbf{Q}, t) I_s(\mathbf{Q}, t + \Delta t) \rangle_t}{\langle I_s(\mathbf{Q}, t) \rangle_t^2}~.
\end{equation}
Under single-scattering and ideal coherence of light, this relates to $f(\mathbf{Q}, \Delta t)$ via the Siegert relation~\cite{berne2000dynamic}
\begin{equation}
g^{(2)}(\mathbf{Q}, \Delta t) = 1 + |f(\mathbf{Q}, \Delta t)|^2~.
\end{equation}
Experimental imperfections reduce the observed contrast, described by a coherence factor $\beta_{0} \in (0,1]$:
\begin{equation}
g^{(2)}(\mathbf{Q}, \Delta t) = 1 + \beta_{0}\, |f(\mathbf{Q}, \Delta t)|^2~.
\end{equation} 

\textbf{Example: Non-interacting Brownian particles.} For non-interacting Brownian particles with diffusion coefficient $D_0$, the distinct part of the intermediate scattering function vanishes and the self part becomes
\begin{equation}
f_\mathrm{s}(\mathbf{Q}, \Delta t) = \exp(-D_0 Q^2 \Delta t)~,
\end{equation}
indicating exponential relaxation of correlations with rate $\Gamma(\mathbf{Q}) = D_0 Q^2$. This scale-dependent relaxation -- commonly observed across diverse systems, albeit with potentially different $\Gamma(\mathbf{Q})$ -- is a hallmark of scattering-based dynamics: large $Q$ values probe short distances, whereas small $Q$ values access long-wavelength collective modes. 

\vspace{0.5em}
Beyond this elementary case, a broad variety of models for the intermediate scattering function $f(\mathbf{Q},\Delta t)$ are available in the literature, reflecting the diversity of soft and biological matter. The interested reader can find general overviews in Refs.~\onlinecite{berne2000dynamic,lindner_neutrons_2024}. The following examples are only illustrative: in colloidal suspensions, interactions, polydispersity, and hydrodynamic effects are routinely incorporated into $f(\mathbf{Q},\Delta t)$. In macromolecular and polymer solutions, models account for internal modes of chain relaxation and concentration-dependent entanglement effects. In emulsions and foams, coalescence and coarsening dynamics give rise to distinct temporal decays. Bacterial suspensions and, more generally, active fluids exhibit ballistic or super-diffusive contributions to $f(\mathbf{Q},\Delta t)$, requiring models that couple self-propulsion with collective motion. In liquid crystals, orientational order, director fluctuations, and viscoelastic couplings strongly influence the functional form of $f(\mathbf{Q},\Delta t)$. Related developments exist also for gels and glass-forming systems, where stretched or compressed exponential relaxations are widely reported. These examples illustrate that the choice of model for $f(\mathbf{Q},\Delta t)$ depends critically on the system under study and on the relevant length and time scales probed in the experiment.

\subsection{\href{https://github.com/somexlab/fastddm-tutorials/blob/main/Tutorial_0-Introduction/tutorial0.ipynb}{\faGithub} ~ Microscopy probes the direct space}
\label{subsec:microdirect}
Optical microscopy forms two-dimensional (2D) images of three-dimensional (3D) samples in real space. To establish a quantitative framework for image formation, we begin with the simplifying assumption that the sample can be approximated as a 2D object. This approximation is often adequate in practice, particularly for thin specimens or systems dominated by in-plane features. We describe the sample by an \emph{object function} $o(\mathbf{x}, t)$, where $\mathbf{x} = (x, y)$ are spatial coordinates in the object plane and $t$ denotes time. This function encodes the spatial distribution of a relevant physical quantity, such as fluorescence intensity, colloid concentration, or director field orientation in liquid crystals~\cite{giavazzi2014digital}.

Upon illumination (Fig.~\ref{fig:ddm-dls}b), the optical signal $\psi(\mathbf{x}, t)$ collected by the microscope can, under broad experimental conditions~\cite{giavazzi2014digital}, be modeled as the convolution of the object function with the point spread function (PSF) $\kappa(\mathbf{x})$ of the imaging system
\begin{equation}
\label{eq:psf}
\psi(\mathbf{x}, t) = o(\mathbf{x}, t) \ast \kappa(\mathbf{x})~.
\end{equation}
The PSF $\kappa(\mathbf{x})$ describes the response of the imaging system to a point source and depends on both the optical design and illumination conditions~\cite{giavazzi2009scattering}. Due to this convolution, sharp features in the object are blurred, limiting spatial resolution.

The measured intensity $i(\mathbf{x}, t)$ typically contains additional contributions
\begin{equation}
\label{eq:realimages}
i(\mathbf{x}, t) = i_0(\mathbf{x}) + \psi(\mathbf{x}, t) + n(\mathbf{x}, t)~,
\end{equation}
where $i_0(\mathbf{x})$ is a static background arising from imperfections such as dust, lens reflections, or uneven illumination, and $n(\mathbf{x}, t)$ denotes stochastic detection noise.

To extract meaningful information about the sample dynamics, the fluctuating component $\psi$ must be isolated from the dominant static background. In real space, this is often achieved via particle localization or image segmentation. For a system of $N_{\text{par}}$ point-like particles, the object function can be expressed as a sum of Dirac delta functions {\footnote{The generalization to the case where each particle is not described as a point-like object but rather by a spatial distribution $p(\mathbf{x})$ of finite width centered on the particle's centroid can be obtained \textit{via} the formal substitution $\kappa(\mathbf{x})\rightarrow{} \kappa(\mathbf{x})\ast p(\mathbf{x})$, incorporating the effect of particle shape in the optical transfer function. }}
\begin{equation}\label{eq:posbropa}
o(\mathbf{x}, t) = \sum_{j=1}^{N_{\text{par}}} \delta(\mathbf{x} - \mathbf{x}_j(t))~,
\end{equation}
which yields the image signal
\begin{equation}\label{eq:PSI}
\psi(\mathbf{x}, t) = \sum_{j=1}^{N_{\text{par}}} \kappa
(\mathbf{x} - \mathbf{x}_j(t))~.
\end{equation}

When the particle signal exceeds both background and noise, positions $\mathbf{x}_j(t)$ can be extracted using localization algorithms. From the resulting trajectories, one typically computes the 2D van Hove correlation function
\begin{equation}
G(\mathbf{x}, \Delta t) = \frac{1}{N_{\text{par}}} \sum_{j,k} \left\langle \delta(\mathbf{x} - \mathbf{x}_j(t + \Delta t) + \mathbf{x}_k(t)) \right\rangle_t~,
\end{equation}
which gives the probability distribution for displacements $\mathbf{x}$ over a time lag $\Delta t$. This function serves as the direct-space analog of the intermediate scattering function introduced in Sec.~\ref{subsec:LSRP}.

These two descriptions are formally connected via a Fourier transform,~\cite{hansen2013theory}
\begin{equation}
F(\mathbf{q}, \Delta t) = \int G(\mathbf{x}, \Delta t) \, \mathrm{e}^{-\mathrm{i} \mathbf{q} \cdot \mathbf{x}} \, \mathrm{d}\mathbf{x}~,
\end{equation}
highlighting that spatial correlations in reciprocal space and displacement distributions in real space are dual representations of dynamical behavior. This equivalence holds independently of the specific microscopic dynamics.

While localization and tracking grant access to particle-level motion, they require high optical contrast. In many bright-field microscopy experiments, particularly with sub-micron particles or dilute suspensions, the dynamic signal $\psi$ is buried beneath the static background $i_0$. This is illustrated in Fig.~\ref{fig:group}, where the raw images appear nearly identical despite ongoing Brownian motion of polystyrene particles ($240\unit{nm}$ diameter, volume fraction $\phi_0 = 10^{-5}$). The motion becomes discernible only upon computing difference images $\Delta i(\mathbf{x}, t, \Delta t) = i(\mathbf{x}, t + \Delta t) - i(\mathbf{x}, t)$, which highlight time-dependent fluctuations by suppressing the static background.

One might attempt to analyze these temporal pixel-wise fluctuations directly, but this generally fails~\cite{dzakpasu2004dynamic_I,dzakpasu2004dynamic_II}. From a scattering perspective, this failure stems from the broad angular integration performed by each pixel, determined by the objective's numerical aperture $\mathrm{NA_o}$. This integration blends signals from multiple wavevectors and relaxation times, precluding a clean link between pixel intensity and specific dynamic processes. A solution to this limitation is to restrict detection to a narrow angular range, as in Photon Correlation Imaging~\cite{duri_resolving_2009}. A more versatile approach is to abandon the pixel-wise (direct space) view and analyze the data in Fourier space. This leads naturally to DDM, which we introduce in the next section as a synthesis of scattering and imaging methodologies.

\subsection{\href{https://github.com/somexlab/fastddm-tutorials/blob/main/Tutorial_0-Introduction/tutorial0.ipynb}{\faGithub} ~ DDM is microscopy\dots just in another space}
\label{subsec:DDMIntro}

Although based on images acquired in real space, DDM builds on the principles of DLS by analyzing microscope image sequences in reciprocal space. Unlike conventional microscopy, which aims at resolving object details in direct space, DDM focuses on the temporal evolution of spatial Fourier modes, thereby extracting dynamic information without requiring particle resolution or tracking. This makes it particularly effective for systems where the signal is weak, the particles are sub-resolution, or the image is dominated by background and noise. Also, while conventional microscopy aims to reconstruct the spatial structure of an object with maximal fidelity, DDM is unconcerned with resolving individual features. Instead, it targets the fluctuation dynamics encoded in image sequences. This shift in emphasis allows DDM to operate under imaging conditions typically deemed suboptimal, such as slight defocus, halo artifacts, or the presence of unresolved features -- see Fig.~\ref{fig:group}, first row. As shown in Sec.~\ref{sec:particle-sizing}, such conditions may even enhance the sensitivity of DDM.

The DDM workflow begins with a time-lapse image sequence acquired under fixed illumination and imaging conditions. The static background $i_0(\mathbf{x})$ is suppressed by computing difference images
\begin{equation}
\Delta i(\mathbf{x}, t, \Delta t) = i(\mathbf{x}, t + \Delta t) - i(\mathbf{x}, t)~.
\end{equation}
This operation isolates the dynamic signal $\psi$ and eliminates time-invariant contributions. The resulting difference images (Fig.~\ref{fig:group}, second row) are then Fourier transformed:
\begin{equation}
\Delta I(\mathbf{q}, t, \Delta t) = \mathcal{F}[\Delta i(\mathbf{x}, t, \Delta t)]~,
\end{equation}
and their squared modulus is averaged over all starting times $t$:
\begin{equation}
D(\mathbf{q}, \Delta t) = \left\langle |\Delta I(\mathbf{q}, t, \Delta t)|^2 \right\rangle_t~.
\label{eq:ImSF}
\end{equation}
The resulting quantity, known as the structure function, captures the temporal decorrelation of Fourier modes at wavevector $\mathbf{q}$ and lag time $\Delta t$.

Since the measured signal is a convolution $\psi(\mathbf{x}, t) = o(\mathbf{x}, t) \ast \kappa(\mathbf{x})$, its Fourier transform takes the form
\begin{equation}
\Psi(\mathbf{q}, t) = O(\mathbf{q}, t) \cdot K(\mathbf{q})~,
\label{eq:convo2D}
\end{equation}
where $\Psi$, $O$, and $K$ denote the Fourier transforms of $\psi$, $o$, and $\kappa$, respectively. The optical transfer function $K(\mathbf{q})$ acts as a spatial frequency filter.

This formalism aligns with the Abbe-Fourier theory of image formation~\cite{goodman2017introduction}, which models objects as superpositions of sinusoidal gratings. Upon illumination by a plane wave $\mathbf{k}_\mathrm{i}$ along the optical axis, each grating diffracts light into angles $\pm\Theta$ related to $\mathbf{q}$ via $\Theta = \arcsin(q/k_0)$\footnote{In scattering experiments one typically measures light scattered at a scattering angle $\theta$. If we assume that both incident and scattered light travel perpendicularly to the interface between the scattering medium with refractive index $n_\mathrm{s}$ and the external medium with refractive index $n_0$, the scattering angle remains unchanged upon crossing the interface. To calculate the correct scattering vector amplitude one needs to use the relationship $Q=2n_\mathrm{s}k_0\sin{(\theta/2)}$. By contrast, in microscopy experiments the light is usually collected in the external medium after obliquely crossing an interface perpendicular to the optical axis. This implies a refraction of the scattered light, which occurs however without a variation of the transverse scattering vector $\mathbf{q}$. By contrast, the scattering angle changes from $\theta$ to $\Theta=\arcsin{[(n_\mathrm{s}/n_0)\sin(\theta)]}$, which in the paraxial approximation becomes $\Theta\simeq(n_\mathrm{s}/n_0)\theta$ and in turn $q\simeq \Theta k_0=\theta k_0 n_\mathrm{s}/n_0$}.
The objective lens focuses these waves into its back focal plane, and they reinterfere in the image plane, reconstructing the object modulation (Fig.~\ref{fig:ddm-dls}c). The consequence is that image formation encodes a one-to-one correspondence between spatial modulations in the object and the Fourier components of the image. Applying the Fourier transform to each image frame thus isolates specific spatial modes $\mathbf{q}$ and allows one to monitor their temporal evolution. In this way, DDM transforms the microscope into a multi-angle scattering instrument.

The mapping between the DDM wavevector $\mathbf{q}$ and the scattering vector $\mathbf{Q}$ used in traditional light scattering becomes especially simple at small angles. In this regime, the axial component $q_z$ of $\mathbf{Q} = (\mathbf{q}, q_z)$ becomes negligible, so that $\mathbf{Q} \approx \mathbf{q}$ and $Q \approx q = k_0 \Theta$ (Fig.~\ref{fig:ddm-dls}a). This approximation holds exactly in 2D systems and remains valid in 3D samples if they are optically thin or weakly scattering.

To interpret $D(\mathbf{q}, \Delta t)$, it is useful to relate it to the autocorrelation function of the object. Under typical conditions (stationary dynamics and detection noise uncorrelated with the signal) the structure function takes the form~\cite{giavazzi2009scattering,giavazzi2014digital}
\begin{equation}
D(\mathbf{q}, \Delta t) = A(\mathbf{q}) \left[1 - f_\mathrm{R}(\mathbf{q}, \Delta t)\right] + B(\mathbf{q})~.
\label{eq:ImSF2}
\end{equation}
In the above equation, the amplitude $A(\mathbf{q})$ of the fluctuating signal is given by $A(\mathbf{q}) = 2 \left\langle |O(\mathbf{q}, t)|^2 \right\rangle_t |K(\mathbf{q})|^2$. For the specific case of the $N_\mathrm{par}$ point-like particles described by Eq.~\eqref{eq:posbropa}, we obtain the useful expression $A(\mathbf{q}) = 2 N_\mathrm{par}S(\mathbf{q}) |K(\mathbf{q})|^2$, which becomes
\begin{equation}\label{eq:DDMSPT}
    A(\mathbf{q}) = 2 N_\mathrm{par}S(\mathbf{q})P(\mathbf{q})|K(\mathbf{q})|^2
\end{equation}
in the general case where the particles are not point-like and their form factor $P(\mathbf{q})$ needs to be taken into account. The factor $|K(\mathbf{q})|^2$, is sometimes denoted $T(\mathbf{q})$ in the DDM literature\cite{lu2012characterizing}, encodes the instrumental response of the optical system. Eq.~\eqref{eq:DDMSPT} can be considered the DDM equivalent of Eq.~\eqref{eq:SLS}, valid for light scattering experiments. 
The term $B(\mathbf{q}) = 2 \left\langle |N(\mathbf{q}, t)|^2 \right\rangle_t$ - where $N$ denotes the Fourier transform of the noise $n$ -  accounts for the temporally uncorrelated detection noise, and
\begin{equation}
    f_\mathrm{R}(\mathbf{q}, \Delta t) = \operatorname{Re} \left[ \frac{ \langle O(\mathbf{q}, t + \Delta t) O^*(\mathbf{q}, t) \rangle_t }{ \langle |O(\mathbf{q}, t)|^2 \rangle_t } \right] ~,
\label{eq:SNACF}
\end{equation}
is the real part of the normalized object autocorrelation function. 
Thus, $D(\mathbf{q}, \Delta t)$ captures the time evolution of spatial Fourier modes in the sample. 
If the dynamics is non-stationary -- for instance, due to irreversible processes such as phase separation, ageing, or sedimentation, all of which also affect DLS -- the validity of time averaging in Eq.~\eqref{eq:ImSF} becomes questionable. In such cases, DDM can still be applied in a time-resolved fashion, by dividing longer recordings into shorter intervals over which the dynamics can be considered quasi-stationary~\cite{ferri2011kinetics,gao2015microdynamics,giavazzi2016structure,cho2020emergence,martineau2022engineering}.

With these theoretical foundations established, we next consider a paradigmatic case: the Brownian motion of non-interacting particles. This serves as an ideal testbed for interpreting the structure function and understanding how dynamic parameters are extracted from DDM.

\subsection{Quantifying dynamics in DDM: The case of Brownian motion}
\label{subsec:intro-brownian}

To consolidate the theoretical framework developed thus far, we now consider a canonical example: a suspension of very small and non-interacting Brownian particles. This model illustrates the use of DDM to extract both dynamic and static information and provides explicit forms for the structure function under idealized conditions.

We begin by computing the structure function $D(\mathbf{q}, \Delta t)$ from the time-lapse image sequence, as defined in Eq.~\eqref{eq:ImSF}. The relationship between $D(\mathbf{q}, \Delta t)$ and the sample dynamics is encoded in Eq.~\eqref{eq:ImSF2}, where the dynamic term $f_\mathrm{R}(\mathbf{q}, \Delta t)$ reflects the temporal autocorrelation of the object function, and the amplitude $A(\mathbf{q})$ captures its spatial fluctuations.

Consider an idealized system of $N_{\text{par}}$ point-like particles (form factor $P(\mathbf{q}) = 1$) described by Eq.~\eqref{eq:posbropa}, undergoing independent Brownian motion with diffusion coefficient $D_0$. Because the particles are non-interacting and randomly distributed, the static structure factor is unity, $S(\mathbf{q}) = 1$. According to Eq.~\eqref{eq:DDMSPT}, the DDM amplitude simplifies to
\begin{equation}
    A(\mathbf{q}) = 2 N_{\text{par}} |K(\mathbf{q})|^2~,
\end{equation}
where $|K(\mathbf{q})|^2$ reflects the microscope optical transfer function, as defined in Eq.~\eqref{eq:convo2D}. Furthermore, by analogy with the DLS result obtained in 3D, the intermediate scattering function becomes
\begin{equation}\label{eq:ISF}
f_\mathrm{R}(\mathbf{q}, \Delta t) = \exp(-D_0 q^2 \Delta t)~,
\end{equation}
indicating exponential decay with a characteristic relaxation time $\tau(q) = 1 / (D_0 q^2)$. This function is the Fourier transform of the self part of the 2D van Hove function
\begin{equation}\label{eq:VHF}
G(\mathbf{x}, \Delta t) = G_s(\mathbf{x}, \Delta t) = \left( \frac{1}{4 \pi D_0 \Delta t} \right) \exp\left( - \frac{|\mathbf{x}|^2}{4 D_0 \Delta t} \right)~.
\end{equation}

This example demonstrates that DDM grants access to both dynamic quantities (through $f_\mathrm{R}(\mathbf{q}, \Delta t)$) and static contrast (through $A(\mathbf{q})$), even when the particles are below the resolution limit or the signal is dominated by background. The exponential decay in Eq.~\eqref{eq:ISF} provides a direct readout of the diffusion coefficient $D_0$, and deviations from this behavior -- due to interactions, confinement, polydispersity, or non-Brownian dynamics -- can be analyzed using generalized models adapted from DLS theory~\cite{berne2000dynamic,lindner_neutrons_2024}.

Having clarified the DDM signal structure and its interpretation in ideal two-dimensional systems, we now turn to the more general case of three-dimensional samples. This requires accounting for axial contributions to the imaging signal and for the full scattering vector $\mathbf{Q} = (\mathbf{q}, q_z)$, as discussed in the following section. Readers primarily interested in the practical implementation of DDM may proceed directly to Section~\ref{sec:hoodwheel}, while those wishing a deeper theoretical background will find in \ref{subsec:2Dvs3D} and \ref{subsec:3DBF} the necessary generalization to 3D systems.

\subsection{\label{subsec:2Dvs3D}(*) Beyond 2D: Incorporating axial structure and dynamics}

Up to this point, our discussion has assumed a quasi-2D system, either an intrinsically 2D sample or a 3D one whose axial dimension can be neglected. This approximation holds in many practical cases. However, a wide range of soft matter systems -- including colloidal suspensions, biological fluids, and active matter -- exhibit significant structure and dynamics along the optical axis ($z$). It is therefore important to specify the validity range of the quasi-2D approximation and to outline how to generalize DDM to fully 3D samples.

\begin{figure*}[ht]
    \centering
    \includegraphics[width=\textwidth]{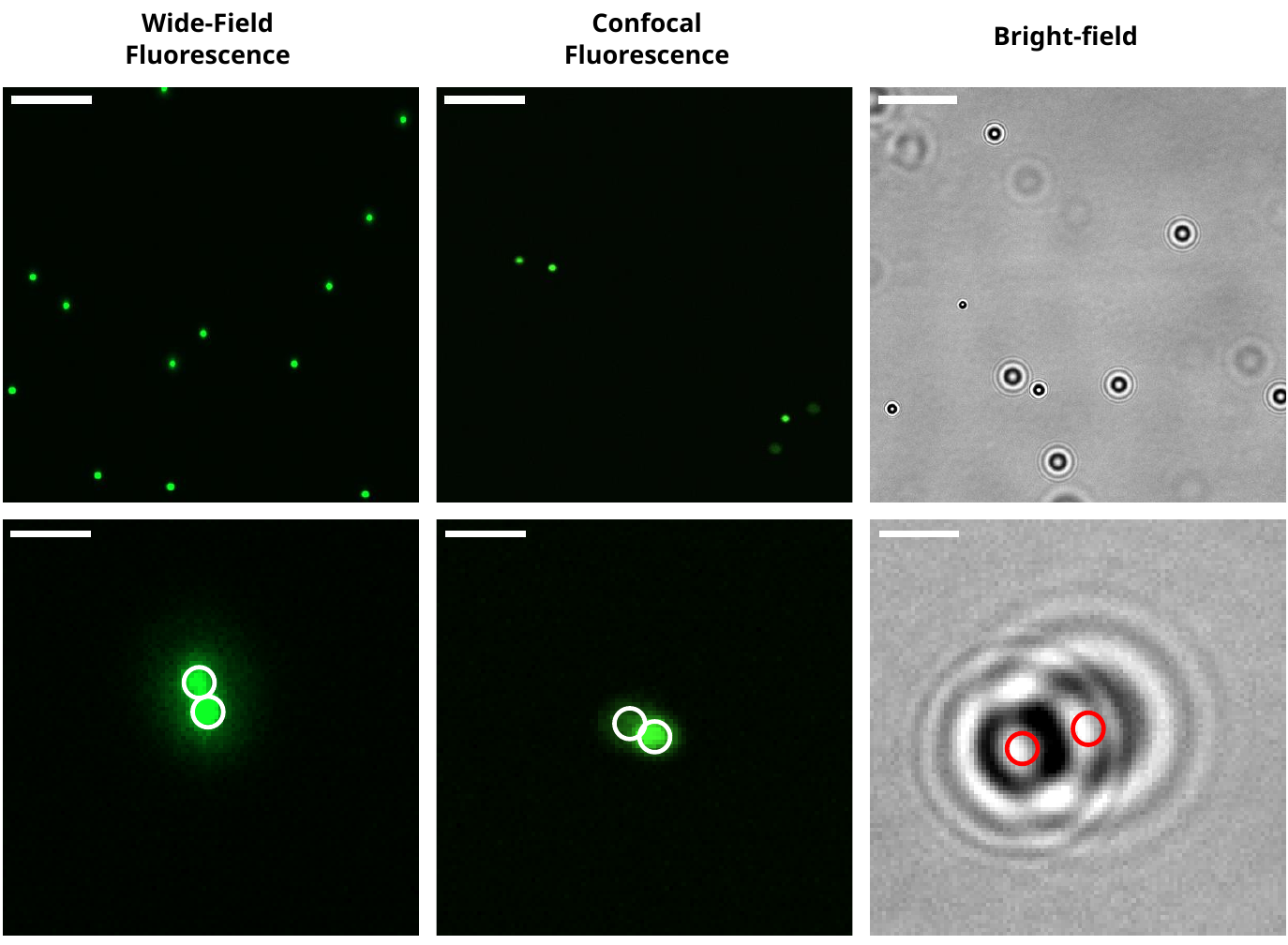}
    \caption{\textit{Top row}: Wide-field fluorescence (left), confocal fluorescence (center), and bright-field (right) microscopy images of fluorescent polystyrene particles suspended in water (nominal diameter $1.9\unit{\mu m}$, volume fraction $\phi_0 \simeq 2 \times 10^{-5}$; Fluoro-Max G0200, Thermo Scientific). These three modalities differ in how they weigh contributions along the optical axis, thus affecting the effective depth from which signal is collected. Scale bars: $32.5\unit{\mu m}$. \textit{Bottom row}: Magnified views of regions containing two closely spaced particles. The visibility, contrast, and apparent size of the particles differ across modalities, illustrating the varying axial and transverse point spread functions. White (fluorescence) and red (bright-field) circles indicate particle positions. These differences reflect the varying axial response of each technique and motivate the formal treatment of axial averaging discussed in this section. Scale bars: $6.2\unit{\mu m}$.}
    \label{fig:z}
\end{figure*}

Extending the DDM framework to 3D requires understanding how the 3D object function $o(\mathbf{x}, z, t)$ contributes to the observed 2D image signal $\psi(\mathbf{x}, t)$. This contribution depends strongly on the imaging modality. For instance, wide-field fluorescence, confocal fluorescence, and bright-field microscopy all differ in their axial resolution and depth weighting, as illustrated in Fig.~\ref{fig:z}.

In general, the image signal is an axial projection of the 3D object convolved with a depth-dependent optical kernel
\begin{equation}
\label{eq:lsi}
\psi(\mathbf{x}, t) = \int \mathrm{d}z \, o(\mathbf{x}, z, t) \ast \kappa(\mathbf{x}, z)~,
\end{equation}
where $\kappa(\mathbf{x}, z)$ is the 3D point spread function (or impulse response), incorporating both transverse and axial resolution. Moving to reciprocal space and denoting the 3D Fourier transforms as $O(\mathbf{Q}, t)$ and $K(\mathbf{Q})$, with $\mathbf{Q} = (\mathbf{q}, q_z)$, yields
\begin{equation}
\label{eq:lsiq}
\Psi(\mathbf{q}, t) = \int \mathrm{d}q_z \, O(\mathbf{Q}, t) K(\mathbf{Q})~.
\end{equation}
This equation shows that the 2D spatial frequency $\mathbf{q}$ in the image receives contributions from a continuum of 3D modes $\mathbf{Q}$, weighted by the transfer function $K(\mathbf{Q})$.

Accordingly, the structure function retains the same functional form as in Eq.~\eqref{eq:ImSF2}
\begin{equation}
\label{eq:ImSF3D}
D(\mathbf{q}, \Delta t) = A(\mathbf{q}) [1 - f_\mathrm{R}(\mathbf{q}, \Delta t)] + B(\mathbf{q})~,
\end{equation}
but the temporal autocorrelation now becomes~\cite{giavazzi2009scattering,giavazzi2014digital}
\begin{equation}
\label{eq:SNACF3}
f_{R}(\mathbf{q},\Delta t) = \operatorname{Re}\left[ \frac{\int \mathrm{d}q_z \, |K(\mathbf{Q})|^2 \langle O(\mathbf{Q}, t+\Delta t) O^{*}(\mathbf{Q}, t) \rangle_t}{\int \mathrm{d}q_z \, |K(\mathbf{Q})|^2 \langle |O(\mathbf{Q}, t)|^2 \rangle_t} \right]~.
\end{equation}
Unlike the 2D case [Eq.~\eqref{eq:SNACF}], the transfer function $K(\mathbf{Q})$ cannot be factored out of the numerator and denominator due to its $q_z$-dependence. As a result, axial dynamics influence the relaxation behavior of each transverse mode $\mathbf{q}$ through a weighted integration over $q_z$.

The magnitude of this influence depends on both the sample (through $O$) and the imaging system (through $K$). A natural question is whether this impact can be neglected. Following~\cite{giavazzi2014digital}, we define the effective axial spread of the optical kernel as
\begin{equation}
\label{eq:deltaq}
\Delta q(\mathbf{q}) \doteq \sqrt{\frac{\int \mathrm{d}q_z \, q_z^2 |K(\mathbf{Q})|^2}{\int \mathrm{d}q_z \, |K(\mathbf{Q})|^2}}~,
\end{equation}
which characterizes the range of axial wavevectors contributing to each transverse mode $\mathbf{q}$. Following Ref.~\onlinecite{giavazzi2014digital}, and writing 
\begin{equation}
    f_\mathrm{R}(\mathbf{Q}, \Delta t) = \operatorname{Re}[f(\mathbf{Q}, \Delta t)]=\operatorname{Re}\frac{\langle O(\mathbf{Q}, t+\Delta t)\, O^{*}(\mathbf{Q}, t) \rangle_t}{\langle|O^(\mathbf{Q}, t)|^2\rangle_t} ~,
\end{equation}
one can derive a practical criterion for assessing when axial effects can be neglected:
\begin{equation}
\label{eq:neglectz}
\left| \frac{1}{f_\mathrm{R}} \left( \frac{\partial f_\mathrm{R}}{\partial q_z} \Delta q + \tfrac{1}{2} \frac{\partial^2 f_\mathrm{R}}{\partial q_z^2} (\Delta q)^2 \right) \right|_{q_z = 0} \ll 1~.
\end{equation}
This expansion, introduced here for the first time, is obtained by Taylor expanding around $q_z=0$ and assumes that higher-order derivatives in $q_z$ can be neglected. When condition~\eqref{eq:neglectz} is satisfied, axial contributions are minor and the sample can be treated, to an excellent approximation, as effectively two-dimensional.

In the following section, we demonstrate how this criterion can be applied in practice by analyzing bright-field DDM data from weakly scattering 3D samples. While our discussion centers on a specific case, the methodology extends naturally to other imaging modalities and dynamical regimes.

\subsection{\label{subsec:3DBF}(*) A worked example: Bright-field DDM of weakly scattering 3D particle suspensions}

Bright-field DDM provides a conceptually transparent setting for illustrating how scattering-like information can be extracted from time-lapse microscopy data. In this section, we work through a detailed theoretical example relevant to dilute suspensions of weakly scattering particles, such as colloids in water imaged under standard bright-field conditions. Building on the framework introduced by Giavazzi et al.~\cite{giavazzi2009scattering}, we summarize the key quantities governing image formation and present a novel explicit expression for the amplitude $A(q)$ of the image structure function (Eq.~\eqref{eq:A-fit}). This expression incorporates both sample-specific and optical parameters and generalizes previous treatments by including the effect of mixed amplitude–phase contrast. While the essential definitions are provided here for completeness, we refer the reader to Ref.~\onlinecite{giavazzi2009scattering} for a detailed derivation and broader discussion.

\vspace{0.5em}
\noindent
We begin by recalling that, under weak scattering conditions, the sample induces only small modulations in the amplitude and phase of the transmitted optical field. The effect of the sample on the incident light is encoded in its complex transmission function
\begin{equation}
o(\mathbf{x}, z, t) = |o(\mathbf{x}, z, t)| \exp[\mathrm{i} \phi(\mathbf{x}, z, t)]~,
\end{equation}
where \( \mathbf{x} \) is the lateral coordinate in the object plane, \( z \) the axial position, and \( t \) time. Physically, \( o(\mathbf{x}, z, t) \) describes the local modification of the complex field amplitude caused by the object — accounting for both attenuation and phase delay. 

\vspace{0.5em}
\noindent
For weakly scattering media, the amplitude and phase modulations are small, and the transmission function can be linearized as
\begin{equation}
o(\mathbf{x}, z, t) \approx 1 + o_\mathrm{A}(\mathbf{x}, z, t) + \mathrm{i} o_\mathrm{P}(\mathbf{x}, z, t)~,
\end{equation}
where \( o_\mathrm{A} \ll 1 \) and \( o_\mathrm{P} \approx \phi \ll 1 \) represent, respectively, the amplitude and phase perturbations induced by the sample.

\vspace{0.5em}
\noindent
For a suspension of particles with refractive index $n = n_R + \mathrm{i} n_I$ and spatial concentration $c(\mathbf{x}, z, t)$, the modulations $o_\mathrm{A}$ and $o_\mathrm{P}$ are linearly related to the local concentration
\begin{align}
o_\mathrm{A} &= a_\mathrm{A} c(\mathbf{x}, z, t)~, \\
o_\mathrm{P} &= a_\mathrm{P} c(\mathbf{x}, z, t)~,
\end{align}
with the prefactors
\[
a_\mathrm{A} = k_0 \frac{\partial n_I}{\partial c}~, \qquad a_\mathrm{P} = -k_0 \frac{\partial n_R}{\partial c}~,
\]
quantifying how changes in concentration modulate the transmitted amplitude and phase.

\vspace{0.5em}
\noindent
Following again Giavazzi et al.~\cite{giavazzi2009scattering}, to describe the optical response of the system, we now introduce two instrumental parameters that control the spatial and spectral coherence of the illumination:

\begin{itemize}
  \item The spatial incoherence parameter $M_\mathrm{co} = \sigma_\mathrm{c} / \sigma_\mathrm{o}$, with $\sigma_\mathrm{c} = \text{NA}_\mathrm{c}/2$ and $\sigma_\mathrm{o} = \text{NA}_\mathrm{o}/2$, quantifies the ratio of the condenser and objective angular apertures.
  \item The spectral width $\Delta \lambda$ of the source determines the longitudinal coherence of the illumination.
\end{itemize}

\vspace{0.5em}
\noindent
These parameters enter the expression for the contrast transfer function $C(q)$, which describes how the visibility of spatial features at wavevector $q$ is modulated by the microscope
\begin{equation}
C(q) = \frac{\exp\left[ -\frac{1}{2} \frac{\left( \frac{q}{q_{\mathrm{ro}}} \right)^2}{1 + \left( \frac{q}{q_{\mathrm{ro}}} \right)^2 \left( \frac{\Delta \lambda}{\lambda_0} \right)^2} \right]}{\sqrt{1 + \left( \frac{q}{q_{\mathrm{ro}}} \right)^2 \left( \frac{\Delta \lambda}{\lambda_0} \right)^2}}~,
\end{equation}
where the cutoff wavevector $q_\mathrm{ro}$ is given by
\begin{equation}
q_{\mathrm{ro}} = k_0 \sigma_\mathrm{o} \sqrt{\frac{1 + 2M_\mathrm{co}^2}{1 + M_\mathrm{co}^2}}~.
\end{equation}

\vspace{0.5em}
\noindent
Next, to account for the 3D nature of the sample, we incorporate the finite optical thickness $\ell$ of the illuminated slab, which enters the axial spread of wavevectors:
\begin{equation}
\label{eq:delta-q}
\Delta q(q) = \sqrt{q^2 \left[ \sigma_\mathrm{c}^2 + \frac{1}{4} \left( \frac{q}{k_0} \right)^2 \left( \frac{\Delta \lambda}{\lambda_0} \right)^2 \right] + \frac{1}{\ell^2}}~.
\end{equation}
Here, $\ell = \ell_{\mathrm{eq}}/\sqrt{2\pi}$ is the effective optical thickness used in the model, obtained by approximating the actual sample — a capillary with rectangular cross-section — as a Gaussian slab of equal integrated thickness. The parameter $\ell_{\mathrm{eq}}$ thus denotes the physical thickness of the illuminated region, while the Gaussian profile is a convenient regularization suggested in Ref.~\onlinecite{giavazzi2009scattering} for analytical calculations.

\vspace{0.5em}
\noindent
Combining these elements, we now write an explicit expression for the amplitude $A(q)$ of the image structure function (as defined in Eq.~\eqref{eq:ImSF2}) under the assumption of mid-plane imaging
\begin{equation}
\label{eq:A-fit}
A(q) = 2 \frac{2 a_\mathrm{P}^2}{\sqrt{\pi}} \frac{C^2(q)}{\Delta q} \left[ (1 + \alpha^2) - (1 - \alpha^2) e^{-(\bar{q}_z / \Delta q)^2} \right]~,
\end{equation}
where $\alpha = a_\mathrm{A} / a_\mathrm{P}$ is a dimensionless parameter quantifying the relative strength of amplitude and phase contributions, and
\begin{equation}
\label{eq:qz}
\bar{q}_z(q) = \frac{q^2}{2 k_0} \left[ 1 - 2M_\mathrm{co}^2 - \frac{1}{\sigma_\mathrm{o}^2} \left( \frac{q}{k_0} \right)^2 \left( \frac{\Delta \lambda}{\lambda_0} \right)^2 \right]~.
\end{equation}

\vspace{0.5em}
\noindent
This expression for $A(q)$ generalizes the form presented in Ref.~\onlinecite{giavazzi2009scattering}, where only the special case $\alpha = 0$ was considered (pure phase contrast). Here, we retain the full $\alpha$-dependence, allowing for mixed amplitude–phase scattering, which becomes relevant for particles with non-negligible absorption or in systems where both $\mathrm{Re}(n)$ and $\mathrm{Im}(n)$ contribute to contrast.

\vspace{0.5em}
\noindent
Notably, the phase parameter $\alpha$ quantifies the relative weight of amplitude and phase scattering, while the associated phase shift
\[
\varphi = \frac{\pi}{2} + \alpha
\]
corresponds to the phase of the light scattered by an individual scatterer in the forward direction, relative to the incident radiation~\cite{vandeHulst1981}. In the Rayleigh limit (small, purely refractive particles), $\varphi \approx \pi/2$. As particle size or refractive index contrast increases, $\varphi$ approaches $\pi$, corresponding to optically larger or partially absorbing particles.

\vspace{0.5em}
\noindent
The implications of this generalization for the interpretation of $A(q)$ are discussed further in Sec.~\ref{subsec:tuning_Aq}.

\vspace{1em}
\noindent
\textbf{Assessing the relevance of axial dynamics in bright field DDM.}
\vspace{0.5em}

To determine whether axial dynamics contribute significantly to the measured image structure function, we use the criterion introduced in Eq.~\eqref{eq:neglectz}, based on the axial wavevector spread $\Delta q(q)$ derived above. In practice, $\Delta q(q)$ for bright-field microscopy exhibits three characteristic regimes:

\begin{itemize}
  \item \textbf{Low wavevectors ($q \ll 1/\ell$):} $\Delta q \approx 1/\ell$, determined by the optical thickness of the sample.
  \item \textbf{Intermediate wavevectors:} $\Delta q \approx q \sigma_\mathrm{c}$, dominated by the condenser numerical aperture.
  \item \textbf{High wavevectors:} $\Delta q \approx (q^2 / 2k_0)(\Delta \lambda / \lambda_0)$, due to spectral broadening.
\end{itemize}

Since most experiments operate in the intermediate-$q$ range~\cite{aime2019probing}, a convenient approximation is:
\begin{equation}
(\Delta q)^2 \approx q^2 \sigma_\mathrm{c}^2 + \ell^{-2}~.
\end{equation}

This result provides a straightforward tool to estimate the relevance of axial dynamics in typical bright-field DDM configurations.

\vspace{1em}
\subsubsection*{Example 1: 3D Brownian motion}

We now apply this framework to a suspension undergoing isotropic 3D Brownian motion with diffusion coefficient $D_0$. The corresponding (real) intermediate scattering function is:
\begin{equation*}
f(\mathbf{Q}, \Delta t) = \exp\left[ -D_0 Q^2 \Delta t \right]~.
\end{equation*}

Substituting this form into the criterion from Eq.~\eqref{eq:neglectz}, one finds that axial dynamics can be neglected if:
\begin{equation*}
D_0 \Delta t (\Delta q)^2 \ll 1~,
\end{equation*}
which defines a characteristic axial relaxation time:
\begin{equation*}
\tau^{(D)}_{\mathrm{ax}}(q) = \frac{1}{D_0 (\Delta q)^2}~.
\end{equation*}

By comparison, the transverse relaxation time is $\tau(q) = (D_0 q^2)^{-1}$. Hence, axial dynamics can be ignored if $\Delta q(q) \ll q$, or equivalently, if:
\[
q \gg q^* = \frac{1}{\ell \sqrt{1 - \sigma_\mathrm{c}^2}}~.
\]

For example, using typical values $\ell = 134~\mu$m and $\sigma_\mathrm{c} = 0.037/2$, we estimate $q^* \approx 0.007~\mu\text{m}^{-1}$, which is below the accessible experimental range. Thus, axial Brownian dynamics can safely be neglected.

\vspace{1em}
\subsubsection*{Example 2: 3D Brownian motion with axial drift}

Consider now the case where Brownian motion is accompanied by a constant axial drift velocity $v_z$ (e.g., due to sedimentation, electrophoresis or thermophoresis). The real part of the intermediate scattering function becomes
\begin{equation*}
f_\mathrm{R}(\mathbf{Q}, \Delta t) = \exp(-D_0 Q^2 \Delta t) \cos(q_z v_z \Delta t)~,
\end{equation*}

and the criterion for neglecting axial effects becomes:
\begin{equation*}
D_0 \Delta t (\Delta q)^2 \left( 1 + \frac{v_z^2 \Delta t}{2 D_0} \right) \ll 1~.
\end{equation*}

This introduces an additional axial time scale due to drift:
\begin{equation*}
\tau^{(v)}_{\mathrm{ax}} = \frac{D_0}{v_z^2}~.
\end{equation*}

To detect $v_z$ via DDM, two conditions must be satisfied:

\begin{align*}
\text{(i)} &\quad \tau^{(v)}_{\mathrm{ax}} \ll \tau(q) \quad \Rightarrow \quad q^2 \ll \frac{v_z^2}{D_0^2}~, \\
\text{(ii)} &\quad \frac{\tau^2(q)}{\tau^{(D)}_{\mathrm{ax}} \tau^{(v)}_{\mathrm{ax}}} \gg 1~.
\end{align*}

The first condition defines a crossover length scale $\ell^* = D_0 / v_z$, beyond which drift dominates over diffusion. The detection window is then bounded by:
\begin{gather*}
q \ll \frac{1}{\ell^*}~, \\
q \ll \frac{\sigma_\mathrm{c}}{\ell^*} \sqrt{ \sqrt{1 + \left( \frac{2 \ell^*}{\ell \sigma_\mathrm{c}^2} \right)^2 } + 1 }~.
\end{gather*}

In practice, this is often too restrictive. For example, for sedimenting silica particles of radius $1~\mu\text{m}$, the sedimentation length is $\ell_s \approx 0.1~\mu\text{m}$. Taking $\sigma_\mathrm{c} \approx 0.1$ and $\ell \approx 100~\mu\text{m}$ yields $q \ll 1.4~\mu\text{m}^{-1}$ — marginally accessible. However, for the $2.1~\mu\text{m}$ PS particles used in Sec.~\ref{sec:particle-sizing}, the condition becomes $q \ll 0.2~\mu\text{m}^{-1}$, below the usable range. This confirms that bright-field DDM is largely insensitive to axial drift under typical conditions.

\vspace{1em}
\subsubsection*{Summary and practical guidance}
The worked examples above demonstrate how axial effects—whether due to Brownian motion or drift—can be systematically evaluated using the axial wavevector spread $\Delta q(q)$ and the associated time scales. Under typical experimental conditions and for common particle sizes, axial dynamics are negligible throughout the accessible $q$-range. This justifies the effective two-dimensional analysis often employed in bright-field DDM and clarifies the conditions under which axial contributions may become detectable or significant. 

Nevertheless, this assumption should not be taken for granted. We therefore recommend using Eq.~\eqref{eq:neglectz} to explicitly verify whether axial dynamics can be safely neglected in a given experimental configuration.

\section{\label{sec:hoodwheel}Under the Hood to On the Road: The Differential Dynamic Algorithm and FastDDM}
\subsection{\label{subsec:underthehood} Under the hood—algorithmic foundations}
At the core of a typical DDM experiment lies the acquisition of a sequence of $N_\mathrm{im} \sim 10^3 \text{--} 10^5$ images, each with lateral dimensions $M_{x,y} \sim 10^2 \text{--} 10^3$ pixels. Images are typically captured at a constant frame rate $\gamma_0 = 1/\Delta t_0$, where $\Delta t_0$ denotes the time interval between consecutive frames. Each image is indexed by an integer $m$, such that the time of acquisition is $t = m \Delta t_0$, and is denoted by $i(\mathbf{x}, m)$.

In the classical implementation~\cite{croccolo2006effect,cerbino2008differential}, the structure function defined in Eq.~\eqref{eq:ImSF} is computed by averaging the power spectra of image differences for all image pairs separated by the same time delay. For a given delay $\Delta t = j \Delta t_0$, with $j$ ranging from 1 to $N_\mathrm{im}-1$, the image difference
\begin{equation*}
    \Delta i (\mathbf{x}, m, j) = i (\mathbf{x}, m + j) - i (\mathbf{x}, m)
\end{equation*}
is evaluated, and its spatial Fourier power spectrum is computed using a Fast Fourier Transform (FFT).

Assuming a stationary or quasi-stationary process, the power spectra can be averaged over different initial times $m$ for each fixed lag $j$, yielding the estimator
\begin{equation}
\label{eq:alg-std-ddm}
    D(\mathbf{q}, j) = \frac{1}{N_\mathrm{im} - j} \sum_{m=0}^{N_\mathrm{im} - j - 1} |\mathcal{F}_{\mathbf{x}} \{ \Delta i(\mathbf{x}, m, j) \} |^2~,
\end{equation}
where $\mathcal{F}_{\mathbf{x}}$ denotes the two-dimensional spatial FFT.

The Fourier transform maps the real-space coordinates $\mathbf{x} = \delta_{\mathrm{px}} (n_x, n_y)$ -- where $\delta_{\mathrm{px}}$ is the physical size of each pixel in the object plane -- into spatial frequency coordinates $\mathbf{q} = (q_x, q_y)$, with
\begin{equation*}
    q_{x} = \frac{2\pi}{M_{x} \delta_{\mathrm{px}}} \{-\lfloor M_{x}/2 \rfloor, \dots, -1, 0, 1, \dots, \lfloor (M_{x}-1)/2 \rfloor \}~,
\end{equation*}
and similarly for $q_y$, where $\lfloor \cdot \rfloor$ is the floor function.

In isotropic systems, an azimuthal average over orientations of $\mathbf{q}$ yields the scalar structure function $d(q, \Delta t)$, with $q = \sqrt{q_x^2 + q_y^2}$. Alternatively, directional information can be preserved by computing sector averages along selected orientations~\cite{pal2020anisotropic}.

The standard algorithm in Eq.~\eqref{eq:alg-std-ddm} is computationally intensive. A significant acceleration is achieved by first computing the Fourier transforms of all images and then evaluating the structure function using
\begin{equation}
\label{eq:alg-ddm}
    D(\mathbf{q}, j) = \frac{1}{N_\mathrm{im} - j} \sum_{m=0}^{N_\mathrm{im} - j - 1} |\Delta I(\mathbf{q}, m, j) |^2~,
\end{equation}
where $I(\mathbf{q}, m) = \mathcal{F}_{\mathbf{x}} \{ i(\mathbf{x}, m) \}$ and $\Delta I(\mathbf{q}, m, j) = I(\mathbf{q}, m+j) - I(\mathbf{q}, m)$. This strategy reduces the number of FFT computations from $O(N_\mathrm{im}^2)$ to $O(N_\mathrm{im})$, but the overall algorithm still scales as $O(N_\mathrm{im}^2)$ due to the summation over all image pairs.

As suggested in Ref.~\onlinecite{norouzisadeh2021modern}, a further and more substantial improvement is obtained by applying the Wiener–Khinchin theorem. Expanding Eq.~\eqref{eq:alg-ddm}, one obtains:
\begin{equation}
\label{eq:alg-ddm-expanded}
\begin{split}
        D(j) = \frac{1}{N_\mathrm{im} - j} \sum_{m=0}^{N_\mathrm{im} - j - 1} &[ |I(m + j)|^2 + |I(m)|^2 \\
        &- 2 \operatorname{Re} \{ I^*(m + j) I(m) \} ]~,
\end{split}
\end{equation}

where the spatial frequency dependence on $\mathbf{q}$ is suppressed for brevity.

The first two terms involve the sum of the last and first $N_\mathrm{im} - j$ power spectra, respectively, and scale linearly with $N_\mathrm{im}$. The last term is the temporal autocorrelation of the complex-valued Fourier-transformed images and introduces the $O(N_\mathrm{im}^2)$ scaling.

By invoking the Wiener–Khinchin theorem~\cite{chatfield2013analysis,goodman2015statistical}, this temporal autocorrelation can be computed as the inverse Fourier transform of the temporal power spectrum:
\begin{equation*}
    \sum_{m=0}^{N_\mathrm{im}-j-1} I^*(m + j) I(m) = \left[\mathcal{F}^{-1}_t \left\{ | \mathcal{F}_t( I(m) ) |^2 \right\} \right] (j)~,
\end{equation*}
where $\mathcal{F}_t$ and $\mathcal{F}^{-1}_t$ are the forward and inverse temporal FFTs.

This approach computes all lag times simultaneously with only two temporal FFT operations per spatial frequency. Furthermore, the procedure in Eq.~\eqref{eq:alg-ddm-expanded} is fully parallelizable~\cite{diffmicro,ladner1980parallel}, enabling substantial performance gains on modern Graphics Processing Units (GPUs).

Overall, the algorithmic complexity is reduced from $O(N_\mathrm{im}^2)$ to $O(N_\mathrm{im} \log N_\mathrm{im})$, transforming a process that might otherwise require hours into one that completes within minutes (or even seconds) for typical datasets.


\subsection{\label{subsec:behindwheel} Behind the wheel—fastDDM implementation and benchmarks}

The algorithmic strategies discussed in the previous section provide a solid foundation for implementing efficient and scalable DDM analysis.
However, turning these principles into practice (especially when dealing with large datasets or limited computational resources) requires robust and accessible software tools.
In recent years, several libraries have been developed to run DDM analysis. Some of these packages focus on particular applications~\cite{ddm-toolkit,muntz_openddm,cddm,arko2019cross}, while others prioritize user-friendly interfaces or integration with Jupyter notebooks to facilitate interactive exploration~\cite{ddm-toolkit,ddmsoft,cddm,arko2019cross,pyddm}.
Several implementations rely on GPU acceleration to dramatically reduce computation times~\cite{diffmicro,muntz_openddm}, following some of the algorithmic optimizations described above.

To provide a practical and high-performance implementation of these methods, we introduce \texttt{fastDDM}, a versatile and open-source Python library developed as a companion tool to this tutorial~\cite{fastddm}.
All analyses presented in the following sections were performed using this software.
\texttt{fastDDM} implements the algorithms described earlier with multiple computational backends: a pure-Python backend for accessibility and readability, and optimized C++ and CUDA modules for fast execution on CPUs and GPUs, respectively. A quantitative benchmark of the speed-up achieved by \texttt{fastDDM} is reported in the \textit{Supplemental Material} (§2).  
There, we provide CPU– and GPU–timing curves (Figs.~S2–S3) together with the Python profiling script (Listing~S1) so that readers can reproduce the performance numbers on their own hardware.
All backends are seamlessly accessible through a consistent Python interface, making the tool suitable for both exploratory data analysis and systematic high-volume processing.

Ultimately, the primary objective of the \texttt{fastDDM} project is to establish and promote a collaborative environment for the development and application of DDM. By making advanced analysis tools openly accessible and extensible, we aim to support a growing community of users and developers in integrating novel methods and adapting them to diverse experimental needs. In this spirit, all the examples and analyses presented in the following sections are based on publicly available datasets that can be explored using \texttt{fastDDM}. This approach serves a dual purpose: it allows readers to develop hands-on experience with real data, and it provides a neutral framework for critically assessing the accessible information (e.g., time and wavevector ranges), free from the interpretive shortcuts or implicit biases that often accompany the analysis of one's own experiments.

\section{\label{sec:particle-sizing} Sizing up DDM: Mastering particle sizing}

Having established the mathematical and computational tools of DDM, we now apply them to a specific, common task: quantifying particle size from image sequences. As in DLS, the ability to extract quantitative information on particle dimensions from the time-dependent decay of dynamic correlation functions lies at the heart of many DDM experiments~\cite{cerbino2008differential,giavazzi2009scattering,bayles2016dark,eitel2020hitchhiker,bradley2023sizing}.

This section offers a detailed guide to particle sizing using DDM, with emphasis on both conceptual understanding and practical implementation.
We present the fundamental principles, describe standard experimental and analytical procedures, and discuss limitations and sources of error.
The methods illustrated here rely on the computational framework introduced above, in particular the open-source \texttt{fastDDM} package, which supports efficient and reproducible data analysis.

To gain familiarity with the whole DDM analysis, we begin by examining open-access experimental datasets by Bradley et al. ~\cite{bradley2023sizing}. Working with such publicly available data replicates the typical experience of a newcomer who has not yet performed a DDM experiment but wishes to analyze existing videos in order to grasp the workflow and identify the key parameters that matter in practice. These datasets allow the full analysis pipeline to be reproduced step by step. 

We then assess how several key experimental parameters influence the accuracy and reliability of sizing results. In particular, we explore how combining multiple acquisitions at different frame rates extends the accessible dynamical range; how image windowing, a standard tool in Fourier analysis, enables access to higher $q$s; how the condenser numerical aperture shapes image contrast and transfer functions; how the choice of objective magnification and numerical aperture set the ultimate resolution and sensitivity of the measurement.

Through these examples, readers will familiarize themselves with both the theoretical insight and the technical skills necessary to perform robust particle sizing with DDM, laying the groundwork for more advanced applications addressed in later sections.

\subsection{\label{sec:particle-sizing-edinburgh} \href{https://github.com/somexlab/fastddm-tutorials/blob/main/Tutorial_1-Particle_sizing/tutorial1.ipynb}{\faGithub} ~ A (mostly harmless) introduction to particle sizing}

Particle sizing is the process of determining the size distribution of particles in a suspension.
Light scattering techniques are commonly used for this purpose because they are fast, non-invasive, and robust~\cite{berne2000dynamic}.
In this introductory section, we focus on measuring the size of relatively monodisperse suspensions, where the particles have a narrow size range around a mean value.
Similar to DLS, our aim is to measure the diffusion coefficient of the suspended particles and calculate their size from it.

To guide readers through the DDM-based particle sizing workflow and provide hands-on experience, we refer to the example notebooks \texttt{Tutorial0} and \texttt{Tutorial1} included in the \texttt{fastDDM} repository. These open-source tutorials offer a step-by-step implementation of the methodology discussed below, allowing readers to replicate the results and build familiarity with the analysis pipeline before attempting their own experiments.

To demonstrate the process of particle sizing using DDM, we use the open-source microscopy image sequences~\cite{edinburgh} of monodisperse particle suspensions accompanying the article by Bradley et al. ~\cite{bradley2023sizing}.
We briefly describe the sample preparation and image acquisition methods. All samples were prepared by dispersing polystyrene (PS) nanoparticles of different diameters ($60\unit{nm}$, $120\unit{nm}$, $240\unit{nm}$, $500\unit{nm}$, $1.1\unit{\mu m}$, and $2.1\unit{\mu m}$, Thermo Scientific) in Milli-Q water to a volume fraction $\phi_0 = 10^{-5}$.
The samples were then loaded into $0.4 \times 4 \times 50\unit{mm}$ glass capillaries (Vitrocom Inc.), which were subsequently sealed to prevent evaporation.
Bright-field images were acquired using a Nikon Ti-E inverted microscope equipped with a Orca Flash 4.0 (Hamamatsu) fast digital CMOS camera.
Each video ($N_\mathrm{im} = 6000$ frames with $M_x = M_y = 512$ pixels) was acquired at a frame rate $\gamma_0 = 200\unit{fps}$ using a 20$\times$, $\mathrm{NA_o}=0.5$ objective, yielding an effective pixel size $\delta_{\mathrm{px}} = 0.325\unit{\mu m}$.

To obtain the particle hydrodynamic radius $R_\mathrm{h}$, we proceed as follows:
\begin{enumerate}
    \item We first compute the structure function $D(\mathbf{q}, \Delta t)$ from the image sequence.
    \item Assuming isotropic dynamics for Brownian motion, we then compute its azimuthal average $d(q, \Delta t)$.

As anticipated in Sec.~\ref{subsec:intro-brownian}, for dilute suspensions of non-interacting, monodisperse particles undergoing Brownian diffusion with a diffusion coefficient $D_0$, the intermediate scattering function takes the form $f_{R}(q, \Delta t) = \exp{\left[ -\Gamma(q) \Delta t \right]}$, where the relaxation rate is given by $\Gamma(q) = D_0 q^2$.
For each $q$, we fit 
\begin{equation}
\label{eq:dqt-model}
    d(q, \Delta t) = A(q) [1 - f_{R}(q, \Delta t)] + B(q)
\end{equation}
to the data using the model for $f_{R}(q, \Delta t)$ to obtain the relaxation rate $\Gamma(q)$.

\item A final fit of $\Gamma(q)$ with a quadratic model $D_0 q^2$ yields the diffusion coefficient, from which the hydrodynamic radius is obtained via the Stokes-Einstein equation:
\begin{equation}
\label{eq:stokes-einstein}
    R_\mathrm{h} = \frac{k_{\mathrm{B}} T}{6 \pi \eta D_0}~,
\end{equation}
where $k_{\mathrm{B}} = 1.38 \times 10^{-23}\unit{J/K}$ is the Boltzmann constant, $T$ the absolute temperature, and $\eta$ the solvent viscosity.
\end{enumerate}
\begin{figure}
    \centering
    \includegraphics[width=\columnwidth]{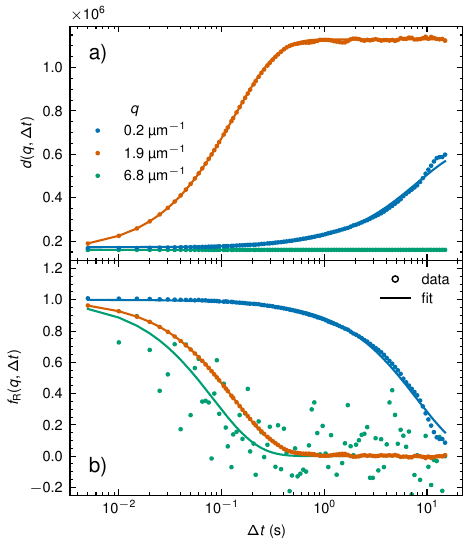}
    \caption{a) Structure function $d(q, \Delta t)$ (symbols) as a function of the delay time $\Delta t$ for three different values of $q$ (see legend), obtained from the analysis of the $240\unit{nm}$ PS particles sample of Ref.~\onlinecite{bradley2023sizing}. The solid lines show the corresponding fits to Eq.~\eqref{eq:dqt-model}, as discussed in the text. b) Intermediate scattering functions $f_{R}(q, \Delta t)$ (symbols) and best fits (solid lines) derived using the same parameters as in panel a).}
    \label{fig:part-siz-dqt-fit}
\end{figure}
To illustrate the analysis workflow in detail, we focus on the $240\unit{nm}$ particle sample. The corresponding image sequence and its structure function $D(\mathbf{q}, \Delta t)$ are shown in Fig.~\ref{fig:group}. As previously noted, the contrast in $D(\mathbf{q}, \Delta t)$ increases with $\Delta t$. This effect is more evident in the azimuthally averaged structure function $d(q, \Delta t)$ shown in Fig.~\ref{fig:part-siz-dqt-fit}a.

Already at this stage, some key features of DDM analysis emerge. At large wavevectors (green curve), the structure function is dominated by detection noise and lacks signal from the sample. At intermediate wavevectors (orange curve), the structure function increases with $\Delta t$ and reaches a well-defined plateau. This regime typically provides the most reliable data. At low wavevectors (blue curve), the signal often fails to fully develop due to the long characteristic diffusion times. 

By fitting Eq.~\eqref{eq:dqt-model} to $d(q, \Delta t)$ for each $q$, we extract the amplitude $A(q)$, noise term $B(q)$, and relaxation rate $\Gamma(q)$. Before describing the fitting procedure, we outline some strategies to obtain good initial parameter estimates. Several methods -- briefly discussed in the Supplementary material -- have been proposed to estimate $B(q)$, particularly in the context of microrheology~\cite{kurzthaler2018probing, bayles2017probe, escobedo2018microliter, cerbino2017dark, giavazzi2018tracking, gu2021uncertainty}. In particle sizing, where precise quantification of $B(q)$ is less critical, we find that a quadratic fit of the first 3–5 data points in $\Delta t$ yields a reasonable estimate of $B(q)$. Once $B(q)$ is known, $A(q)$ can be estimated using the long-time plateau: $d(q, \Delta t \to \infty) = A(q) + B(q)$. Following the definition of the structure function~\eqref{eq:ImSF}, the plateau corresponds to the background-subtracted time-averaged image power spectrum: $\langle |I(\mathbf{q}, t) - \langle I(\mathbf{q}, t) \rangle_t|^2 \rangle_t$.

Given estimates of $A(q)$ and $B(q)$, we can reconstruct the intermediate scattering function via $f_{R}(q, \Delta t) = 1 - [d(q, \Delta t) - B(q)] / A(q)$, and extract a rough estimate of $\Gamma(q)$ by finding the time $\Delta t$ such that $f_{R}(q, \Delta t) = 1/\mathrm{e}$. We recommend selecting a reference wavevector (e.g., $q=1.9\unit{\mu m^{-1}}$ in Fig.~\ref{fig:part-siz-dqt-fit}a), where the structure function is well-developed, to obtain initial parameters. The fit can then be propagated incrementally to neighboring $q$ values using previous fit results as initial guesses. Since $A(q)$, $B(q)$, and $\Gamma(q)$ are typically smooth functions of $q$, this approach is well justified and compatible with regularization schemes.


In practice, the quality and stability of the structure function fit depend on the choice of delay times $\Delta t$ included in the fitting, the weighting of data points in the objective function, and whether the data are uniformly distributed on a linear or logarithmic scale. In a standard least-squares fitting algorithm, the objective function is defined as the sum of the squared residuals between the model and the data, each term optionally scaled by a weight that reflects the expected uncertainty of the corresponding data point. These strategies can significantly improve the reliability of the extracted parameters, especially in the presence of experimental noise or non-ideal sampling conditions. Here we use the \texttt{lmfit}~\cite{newville_lmfit_2025} Python library to fit data uniformly distributed on a linear scale, with weights set to $1/\sigma_{d}(q, \Delta t)$, where $\sigma_{d}(q, \Delta t)$ is the standard deviation of the azimuthal average of the structure function (more about this in Section~\ref{subsec:error}). Users should be aware that other fitting libraries may adopt different conventions, and should carefully verify the definitions before passing weights. \footnote{In \texttt{lmfit}, the residuals are internally normalized by the weights. Providing weights $w_i$ is equivalent to minimizing $\sum_i (w_i \,[y_i - f(x_i)])^2$. Thus choosing $w_i = 1/\sigma_i$ corresponds to the conventional definition of a weighted least-squares fit.}

\begin{figure}
    \centering
    \includegraphics[width=\columnwidth]{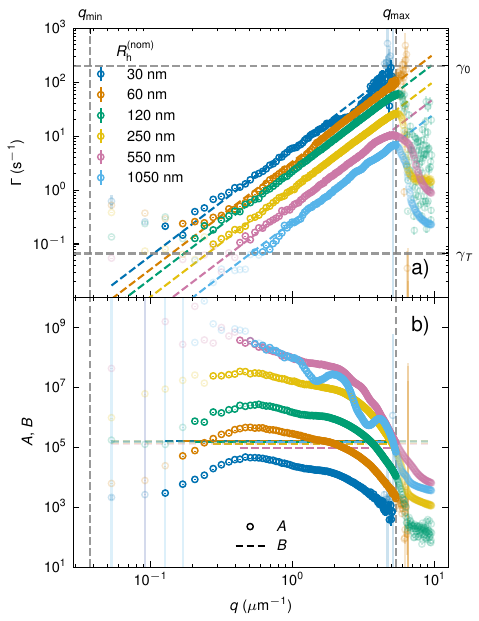}
   \caption{
 a) Relaxation rate $\Gamma(q)$ (open symbols) as a function of wavevector $q$ for different particle sizes, as indicated in the legend. Colored dashed lines are weighted fits of the form $\Gamma(q) = D_0 q^2$, used to extract the diffusion coefficient $D_0$ and the corresponding hydrodynamic radius $R_\mathrm{h}$. The gray dashed horizontal lines mark the accessible relaxation rate range ($\gamma_T \leq \Gamma \leq \gamma_0$), while the gray vertical lines indicate the accessible wavevector window ($q_{\min} \leq q \leq q_{\max}$) for this dataset. Transparent data shows discareded data outside the reliable $q$ range by using the intersection of $D_0 q^2$ with $\gamma_T$ as a lower limit cutoff)
b) Static amplitude $A(q)$ (symbols) and noise floor $B(q)$ (dashed lines) obtained from the fits in panel a. Color coding matches panel a.
}
    \label{fig:part-siz-ab}
\end{figure}
Figure~\ref{fig:part-siz-dqt-fit}b shows the intermediate scattering functions obtained from the data and fits of panel a. The same fitting procedure can be applied to the full dataset. The extracted parameters are shown in Fig.~\ref{fig:part-siz-ab}. We now examine the behavior of the relaxation rate $\Gamma(q)$ (Fig.~\ref{fig:part-siz-ab}a). In the intermediate $q$ range, $\Gamma(q)$ follows the expected Brownian scaling $\Gamma(q) = D_0 q^2$. Deviations occur at both extremes: at large $q$, signal-to-noise limitations dominate; at low $q$, the dynamics evolve too slowly to be captured within the finite acquisition window. A representative collection of intermediate scattering functions -- covering several $q$ values for each particle size --, together with the corresponding single-$q$ fits, is provided in Fig.~S4 of the \textit{Supplemental Material} (§4.1).

These limitations can be rationalized by considering the acquisition parameters. The maximum resolvable relaxation rate $\gamma_0$ is the inverse of the delay between consecutive frames $1 / \Delta t_0$. The minimum observable rate $\gamma_T$ is set by the inverse of the total acquisition time, $1 / (N_\mathrm{im} \Delta t_0)$. Wavevector access is governed by both optical and acquisition parameters. The smallest accessible wavevector is $q_{\text{min}} = 2\pi / (M \delta_{\text{px}})$, where $M$ is the number of pixels per image side and $\delta_{\text{px}}$ is the pixel size. At these low $q$, the number of Fourier-space pixels contributing to the azimuthal average is small, reducing statistical accuracy. Furthermore, optical transfer functions (e.g., in bright-field microscopy, see Fig.~\ref{fig:part-siz-ab}b) may attenuate low-$q$ signal~\cite{giavazzi2009scattering,giavazzi2014digital}.

The upper wavevector limit is set by $\min(q_{\mathrm{px}}, q_{\mathrm{NA}})$, where $q_{\mathrm{px}} = \pi / \delta_{\mathrm{px}}$ is the Nyquist limit determined by the pixel size, and $q_{\mathrm{NA}} = 2\pi \mathrm{NA_o} / \lambda_0$ arises from the optical resolution limit imposed by the microscope objective, keeping into account that the objectve numerical aperture is given by $NA_\mathrm{o}=n_0\sin{\Theta_{max}}=n_0q_{max}/k_0$,. Assuming an illumination wavelength $\lambda_{0} = 550\unit{nm}$, we obtain $q_{\mathrm{NA}} \approx 5.7\unit{\mu m^{-1}}$, which is smaller than $q_{\mathrm{px}} \simeq 10\unit{\mu m^{-1}}$ and therefore sets the effective upper bound in the present dataset. As a side note, $q_{\mathrm{max}}$ may also be limited by the particle form factor, which introduces a high-$q$ cut-off when particle scattering becomes negligible.

These bounds are indicated by gray dashed lines in Fig.~\ref{fig:part-siz-ab} and guide the selection of reliable data. Notably, while low-$q$ data may fall within the formal access window, their interpretation may still be compromised if the structure function plateau is not reached (as seen for the blue curve in Fig.~\ref{fig:part-siz-dqt-fit}a).

Within the valid $q$ range, we realize a first fit of the $\Gamma(q)$ to estimate diffusion coefficients $D_0$, obtained from a weighted fit to $\Gamma(q) = D_0 q^2$. These coefficients can be further used to discard data (shown as transparent in Fig.~\ref{fig:part-siz-ab} a)) outside the reliable $q$ range by using the intersection of $D_0 q^2$ with $\gamma_T$ as a cutoff), realizing a second fit using the most reliable data we measure $D_0$ of Table~\ref{tab:part-sizing}. Using the Stokes–Einstein relation~\eqref{eq:stokes-einstein}, the hydrodynamic radii $R_\mathrm{h}$ are then calculated, assuming $T=25\celsius$ and $\eta = 0.89\unit{mPa \, s}$. Table~\ref{tab:part-sizing} summarizes the measured $D_0$ and $R_\mathrm{h}$, and compares the $D_0$ values with those reported in Ref.~\onlinecite{bradley2023sizing}, noting that a different method for estimating $D_0$ and the associated uncertainty was employed in that study.

\begin{table}
\centering
\begin{tabularx}{\columnwidth}{l c c c}
\toprule
\shortstack{Manufacturer\\radius $R_{\mathrm{nom}}$ (nm)\\\cite{bradley2023sizing}} &
\shortstack{Diffusion coeff.\\$D_0$ ($\mu$m$^2$/s)\\\cite{bradley2023sizing}} &
\shortstack{Diffusion coeff.\\$D_0$ ($\mu$m$^2$/s)\\This work} &
\shortstack{Hydr. radius\\$R_\mathrm{h}$ (nm)\\This work} \\
\midrule
30   & $7.3479\pm0.0267$   & $5.96 \pm0.08$   & $41.1 \pm0.5$  \\
60   & $3.5716\pm0.0033$   & $3.35 \pm0.03$   & $73.3 \pm0.6$  \\
120  & $2.2214\pm0.0009$   & $2.215 \pm0.006$ & $110.8 \pm0.3$ \\
250  & $1.0372\pm0.0003$   & $1.037 \pm0.002$ & $236.9 \pm0.5$ \\
550  & $0.4943\pm0.0002$   & $0.497 \pm0.002$ & $493.5 \pm1.5$ \\
1050 & $0.2614\pm0.0002$   & $0.256 \pm0.001$ & $957 \pm4.5$   \\
\bottomrule
\end{tabularx}
\caption{Measured diffusion coefficients $D_0$ and hydrodynamic radii $R_\mathrm{h}$ for the investigated particle suspensions with manufacturer radii $R_{\mathrm{nom}}$.}
\label{tab:part-sizing}
\end{table}

\paragraph*{Effect of polydispersity.}
In practical samples, a
distribution of particle sizes can is typically accompanied by deviations from a pure single-exponential relaxation that is typical for monodisperse samples. In this case, the intermediate scattering function
is effectively a weighted sum of exponentials corresponding to the different
diffusion coefficients~\cite{berne2000dynamic}. For weak polydispersity, a cumulant expansion provides a useful diagnostic tool~\cite{mailer2015particle}: the first cumulant yields the mean diffusivity, while the second cumulant quantifies the variance of the distribution. More generally, one may fit the data using a simple parametric distribution (e.g. Schultz or log-normal), though care
must be taken to avoid overfitting when the number of free parameters increases. The reader interested in these aspect, as well in multimodal fitting of DDM data is referred to Ref.~\onlinecite{bradley2023sizing}.

\subsection{\href{https://github.com/somexlab/fastddm-tutorials/blob/main/Tutorial_1-Particle_sizing/tutorial1.ipynb}{\faGithub} ~ Tracking-free determination of the particles mean squared displacement}

One interesting application of DDM relies on the inversion of the intermediate scattering function to extract the mean squared displacement (MSD) of Brownian particles, a quantity typically accessed in real space via particle-tracking experiments. For thermally driven motion of non-interacting particles in a homogeneous medium, particle displacements follow a Gaussian distribution~\cite{berne2000dynamic}. Under this assumption, the intermediate scattering function can be written as

\begin{equation}
\label{eq:isf-msd}
f_{R}(q, \Delta t) = \exp\left[ -\frac{q^2}{4} \langle \Delta r^2(\Delta t) \rangle \right]~,
\end{equation}

where $\langle \Delta r^2(\Delta t) \rangle$ is the 2D MSD. Inversion of Eq.~\eqref{eq:isf-msd} provides

\begin{equation}
\label{eq:logf-msd}
\langle \Delta r^2(\Delta t) \rangle=-\frac{4}{q^2}\log (f_{R}(q, \Delta t)) ~,
\end{equation}

which can be used to extract $\langle \Delta r^2(\Delta t) \rangle$ from $f_{R}(q, \Delta t)$ without need of tracking the particles in real space. While the inversion in Eq.~\eqref{eq:logf-msd} appears straightforward, several procedures have been devised to enhance the robustness of the extracted MSDs~\cite{edera2017differential,giavazzi2018tracking,brizioli2022reciprocal}.

Here, we use the procedure outlined in Ref.~\onlinecite{brizioli2022reciprocal}, to analyze the data from Ref.~\onlinecite{bradley2023sizing} and extract the MSD from DDM analysis. Following Ref.~\onlinecite{brizioli2022reciprocal}, for each particle size we consider only wavevectors in the range ($q_\mathrm{min}, q_\mathrm{max}$) and time delays for which $f_{R}(q, \Delta t)$ has relaxed for less than $25\%$. If this interval contains at least 15 points we fit a function of the form $a + (\langle \Delta r^2(\Delta t) \rangle /4) q^2$ to $- \log (f_{R}(q, \Delta t))$, obtaining $a$ and $\langle \Delta r^2(\Delta t) \rangle$ from the best fitting curve where $a$ accounts for possible inaccuracies in estimating $B$. We note that for some of these particles the MSD would be essentially inaccessible with real-space tracking methods, since their size falls well below the resolution limit of optical microscopy.

\begin{figure}
    \centering
    \includegraphics[width=\columnwidth]{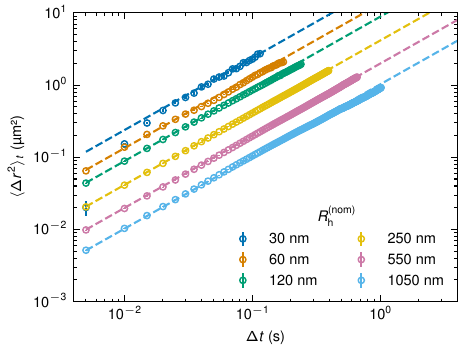}
    \caption{Mean squared displacement (MSD) $\langle \Delta r^2(\Delta t) \rangle$ as a function of lagtime $\Delta t$ for different particles size. Dashed line corresponds to  $\langle \Delta r^2(\Delta t) \rangle = 4D_0\Delta t$, using the value $D_0$ from DDM analysis (see also Table \ref{tab:part-sizing}).}
    \label{fig:msd}
\end{figure}

The result of this procedure is shown in Fig.~\ref{fig:msd} and found in very good agreement with the values of $D_0$ obtained with DDM in section \ref{sec:particle-sizing-edinburgh}, with a systematic deviation from the expected short-time behavior only for the the smallest ($30$ nm) particles. We attribute this reduced accuracy to the difficulty in properly selecting a suitable q-range for the MSD calculation. Representative $\log (f_{R}(q, \Delta t)$ scattering functions -- covering several $q$ values for each particle size --, together with the corresponding single-$q$ fits, is provided in Fig.~S5 of the \textit{Supplemental Material} (§4.2).

\subsection{\label{subsec:error} \href{https://github.com/somexlab/fastddm-tutorials/blob/main/Tutorial_0-Introduction/tutorial0.ipynb}{\faGithub} ~ Uncertainty of the structure function}

\begin{figure}
    \centering
    \includegraphics[width=\columnwidth]{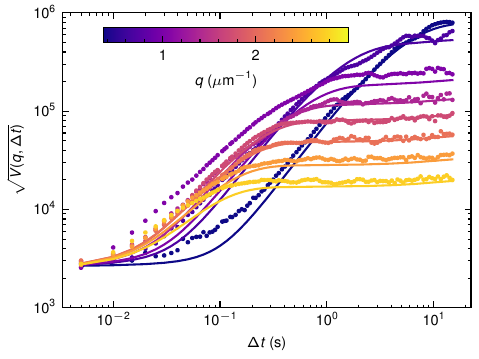}
    \caption{Standard deviation $\sqrt{V(q,\Delta t)}$ of the structure function for the $R_{\mathrm{nom}}=120\unit{nm}$ sample of Ref.~\onlinecite{edinburgh} estimated from the azimuthal average (symbols) and analytically (lines) for different wavevectors, equally spaced in the range $q \in [0.36, 3]\unit{\mu m^{-1}}$. Colors indicate the corresponding $q$ values (see color bar).}
    \label{fig:error}
\end{figure}
As briefly mentioned in the previous subsection, the experimentally determined structure function is a statistical estimator subject to time- and wavevector-dependent intrinsic fluctuations, whose quantification is important for proper data treatment. 

According to Eq. \ref {eq:alg-ddm}, each point $D(\mathbf{q},\Delta t)$ of the structure function is calculated as the mean of $(N_\mathrm{im}-j)$ distinct values $\{|\Delta I(\mathbf{q}, m, j) |^2\}_m$, each one corresponding to same time interval $\Delta t=j\Delta t_0$ and a different initial time point $t=m\Delta t_0$, while the azimuthally averaged structure function $D(q,\Delta t)$ is obtained as the mean of $N_\mathrm{q}\simeq \pi q/q_{min}$ distinct instances $\{D(\mathbf{q},\Delta t)\}_{|\mathbf{q}|=q}$.
To properly estimate the uncertainties associated with these estimators is important to keep in mind that they are not obtained as the mean of statistically independent variables.

Under the general hypothesis that the underlying process is Markovian, two values $|\Delta I(\mathbf{q}, m_1, j) |^2$ and
$|\Delta I(\mathbf{q}, m_2, j) |^2$ are always uncorrelated if $|m_1-m_2|>j$, i.e. if the two intervals $[t_1,t_1+\Delta t]$ and $[t_2,t_2+\Delta t]$ are not overlapping. However, if the process has a characteristic correlation time $\tau(\mathbf{q})$ shorter than $\Delta t$, $|\Delta I(\mathbf{q}, m_1, j) |^2$ and
$|\Delta I(\mathbf{q}, m_2, j) |^2$ can be still considered uncorrelated as long as $|t_1-t_2|>\tau(\mathbf{q})$, i.e. if time elapsed between $t_1$ and $t_2$ is larger than the correlation time.
Based on this argument, a conservative estimate of the number $N_{\Delta t}$ of \textit{statistically independent} variables in the set $\{|\Delta I(\mathbf{q}, m, j) |^2\}_m$ can be thus obtained as 
$N_{\Delta t}\simeq\max(N_\mathrm{im}/j,N_\mathrm{im}\Delta t_0/\tau(\mathbf{q}))$.
Consequently, the standard error associated with $D(\mathbf{q},\Delta t)$ can be estimated as $\sigma_\mathrm{D}(\mathbf{q},\Delta t)\simeq\sqrt{W(\mathbf{q},\Delta t)/N_{\Delta t}}$, where $W(\mathbf{q},\Delta t)$ is the estimated variance of the set $\{|\Delta I(\mathbf{q}, m, j) |^2\}_m$.

This expression can be used to estimate the uncertainty associated with $D(\mathbf{q},\Delta t)$ for each given 2D wavevector $\mathbf{q}$ and delay time $\Delta t$.
Since it requires prior knowledge of the correlation time $\tau(\mathbf{q})$, its calculation may require an iterative approach. For example, one can perform an unweighted fit of $D(\mathbf{q},\Delta t)$ to obtain a first rough estimate for $\tau(\mathbf{q})$. Based on this value, $\sigma_\mathrm{D}(\mathbf{q},\Delta t)$ can be estimated. The fitting procedure can then be repeated using $\sigma_\mathrm{D}(\mathbf{q},\Delta t)$ as a weight, to obtain a more refined estimate of $\tau(\mathbf{q})$ and of the other parameters of interest.

If the azimuthally averaged structure function is considered instead, its standard error can be more directly estimated as $\sigma_\mathrm{d}(q,\Delta t)=\sqrt{V(q,\Delta t)/N_\mathrm{q}}$, where $V(q,\Delta t)$ is the estimated variance of the set $\{D(\mathbf{q}, \Delta t)\}_{|\mathbf{q}|=q}$.
This is a consequence of the fact that, for an isotropic system with a large number of particles, $D(\mathbf{q_1},\Delta t)$ and $D(\mathbf{q_2},\Delta t)$ are statistically independent whenever $\mathbf{q}_1\neq \mathbf{q}_2$.

It can be worth noticing that if more than one image sequence of the same sample is available, estimates obtained from distinct sequences must be considered to be statistically independent.

Besides being estimated experimentally, the expectation value of the intrinsic variance $V_0(\mathbf{q},\Delta t)=\langle D(\mathbf{q},\Delta t)^2 \rangle- \langle D(\mathbf{q},\Delta t) \rangle^2$  of the experimentally determined structure function given in Eq. \ref{eq:alg-std-ddm} can also be calculated analytically, at least under suitable assumptions on the nature of the underlying process. A novel expression valid for a sequence of images of uncorrelated Brownian particles is reported in \textit{Supplemental Material}, §2.  
To evaluate the accuracy of this analytical prediction, we again use the $R_{\mathrm{nom}} = 120\unit{nm}$ dataset from Ref.~\onlinecite{bradley2023sizing}. Figure~\ref{fig:error} shows the standard deviation $\sqrt{V(q,\Delta t)}$ computed directly from the azimuthal average (symbols) alongside the theoretical predictions $\sqrt{V_0(q,\Delta t)}$ (solid lines) for representative $q$-values. The comparison shows good agreement across the accessible delay times, validating the analytical model within its regime of applicability.

\subsection{\label{subsec:join}\href{https://github.com/somexlab/fastddm-tutorials/blob/main/Tutorial_2-Merging/tutorial2.ipynb}{\faGithub} ~ Joining fast and slow acquisitions}

In the previous sections, we discussed key considerations for analyzing \emph{existing} image sequences. We now shift focus to strategies for optimizing DDM acquisitions, particularly in scenarios where experimental planning is still possible.

As detailed in Sec.~\ref{sec:particle-sizing-edinburgh}, a fundamental limitation in DDM analysis arises from the finite duration of the image acquisition. A short experiment limits the number of wavevectors for which the structure function $d(q, \Delta t)$ reaches a well-defined plateau, effectively introducing a minimum reliable decay rate $\gamma_T = 1/(N_\mathrm{im} \Delta t_0)$ (cf. Fig.~\ref{fig:part-siz-ab}).

The simplest remedy is to extend the acquisition duration. However, this quickly becomes impractical due to:
(i) the large file sizes generated\footnote{For instance, a 6000-frame sequence at 512$\times$512 resolution and 16-bit depth already exceeds $3\unit{GB}$. A 100-fold increase would yield $300\unit{GB}$.} and
(ii) the correspondingly long analysis time. A more tractable strategy is to acquire \emph{two} image sequences using different frame rates $\gamma_0$, such that their delay-time ranges partially overlap. A convenient criterion is to choose frame rates that differ by a factor $\sqrt{N_\mathrm{im}}$, where $N_\mathrm{im}$ is the number of frames per sequence. This ensures overlap between the upper half of the fast sequence and the lower half of the slow one, when viewed on a logarithmic $\Delta t$ scale. For example, acquiring 10,000 frames at $100\unit{fps}$ and $1\unit{fps}$, respectively, satisfies this condition.

To demonstrate this method, we acquired two bright-field image sequences of a dilute aqueous suspension of spherical PS nanoparticles with nominal diameter $252\unit{nm}$ (Microparticles GmbH). The sample was gently vortexed, diluted to a volume fraction $\phi_0=10^{-3}$ in density-matched ($\rho_{\text{PS}} = 1.05\unit{g/cm^3}$) glycerol-water solution ($c=21.5\%$ glycerol, filtered at $0.2\unit{\mu m}$), and used to fill a $0.3 \times 3 \times 20\unit{mm}$ rectangular glass capillary (VitroCom Inc.). The capillary was finally sealed on both sides with hematocrit sealing compound to prevent evaporation and fixed on a microscope glass slide using epoxy glue.

Imaging was performed in bright-field on a Nikon Eclipse Ti2 inverted microscope equipped with a Prime BSI Express CMOS camera (Teledyne Photometrics). K\"ohler illumination was achieved using an LED light source and an ELWD condenser ($\mathrm{NA_c}=0.3$). A $20\times$, $\mathrm{NA_o} = 0.45$ objective yielded an effective pixel size $\delta_{\text{px}} = 0.325\unit{\mu m}$. Two sequences of 10,000 frames (256$\times$256 pixels) were acquired at $108\unit{fps}$ and $1\unit{fps}$, respectively. Critically, the exposure time was kept constant at $9\unit{ms}$ in both acquisitions to match the noise and static contrast levels. Experiments were conducted at $T=21\celsius$, assuming a solvent viscosity $\eta = 1.78\unit{mPa \, s}$~\cite{volk2018density}.

\begin{figure}
    \centering
    \includegraphics[width=\columnwidth]{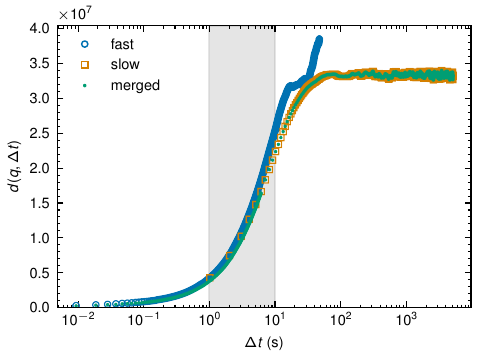}
    \caption{Structure function $d(q, \Delta t)$ at $q = 0.34\unit{\mu m^{-1}}$ obtained from fast ($108\unit{fps}$) and slow ($1\unit{fps}$) acquisitions on the PS $252\unit{nm}$ sample. The merged curve (green) results from aligning the fast acquisition onto the slow via least squares in the overlap region (shaded area).}
    \label{fig:dqt-fast-slow-melt}
\end{figure}
Figure~\ref{fig:dqt-fast-slow-melt} shows $d(q, \Delta t)$ at a representative wavevector from both acquisitions. We merge the two structure functions by computing a multiplicative correction factor from a least squares fit of the first few ($\sim$10) $\Delta t$ values of the slow-acquisition curve, applied to the fast-acquisition data. 
In \texttt{fastDDM}, this merging procedure is performed automatically, including the interpolation and correction steps. Detailed implementation notes are provided in the online documentation. 
We note that this merging operation is valid under the assumption that the contrast terms $A(q)$ and $B(q)$ are similar across acquisitions, a condition met here by equal illumination and exposure settings. The resulting composite structure function (green dots in Fig.~\ref{fig:dqt-fast-slow-melt}) exhibits a smooth and stable behavior for all times.

For particle sizing, the improvement in results may be modest. As shown in Fig.~\ref{fig:cfr-fast-melt-win}a, the diffusion coefficients extracted from the fast and merged datasets are $D_0^{(\mathrm{fast})} = (0.962 \pm 0.003)\unit{\mu m^2 / s}$ and $D_0^{(\mathrm{merged})} = (0.959 \pm 0.002)\unit{\mu m^2 / s}$, both consistent with the reference value $D_0 = 0.957\unit{\mu m^2 / s}$. Nevertheless, this approach is particularly beneficial when one seeks to access smaller wavevectors (lower $q$), where long delay times are essential, or when acquisition resources (e.g., number of frames) are constrained. In such cases, joining acquisitions at different frame rates can significantly expand the accessible dynamic range while preserving analysis quality.

\subsection{\label{subsec:windowing} \href{https://github.com/somexlab/fastddm-tutorials/blob/main/Tutorial_3-Image_windowing/tutorial3.ipynb}{\faGithub}\; The effect of image windowing}

\begin{figure}
    \centering
\includegraphics[width=\columnwidth]{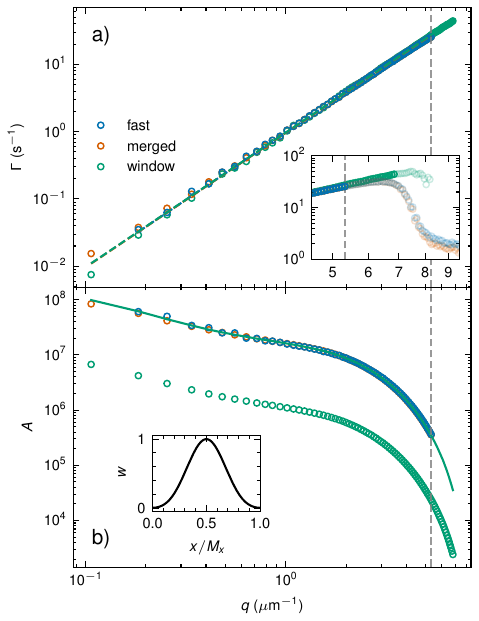}
    \caption{a) Relaxation rate $\Gamma(q)$ extracted from $d(q, \Delta t)$ computed with and without windowing on the PS $252\unit{nm}$ sample. The vertical dashed line indicates the optical cutoff $q_{\text{max}}$ set by the objective’s numerical aperture. The inset zooms in on the high-$q$ region. Transparent markers correspond to excluded data. b) Amplitude $A(q)$ of the structure function with and without windowing. The solid line represents a rescaling of the amplitude obtained with windowing by the mean value of the squared windowing function. Inset shows the Blackman-Harris window function $w(x)$ as a function of the normalized distance $x/M_x$ (see Eq.~\eqref{eq:Blackman}). }
    \label{fig:cfr-fast-melt-win}
\end{figure}

At high wavevectors, the relaxation rate $\Gamma(q)$ extracted from DDM data exhibits an anomalous decrease, deviating from the expected $D_0 q^2$ behavior. Figure~\ref{fig:cfr-fast-melt-win}a shows this artifact for the PS $252\unit{nm}$ sample (see inset). The distortion originates from hard image boundaries: when particles (or their blurred images if the particles cannot be optically resolved) cross the finite field of view, the lack of periodic boundary conditions in the Fourier transform introduces spurious temporal fluctuations. These artifacts primarily contaminate high-$q$ modes, where the genuine dynamical signal is weaker. For Brownian motion, the associated timescale can be estimated as $\tau_\mathrm{B} \simeq R_{\text{app}}^2 / D_0$, where $R_{\text{app}}$ is the larger of the particle diameter and the point spread function (PSF) width.

To suppress these artifacts, each frame $i(\mathbf{x},t)$ can be multiplied by a spatial windowing function $w(\mathbf{x})$ before Fourier transformation, as originally shown in Ref.~\onlinecite{giavazzi2017image}. The function should be non-negative, symmetric, and vanish at the image edges, thereby reducing the contribution from particles crossing the boundaries. This approach mirrors methods in signal processing designed to minimize edge effects in time-domain data~\cite{priemer1991introductory, harris1978use}. A commonly used choice is the two-dimensional Blackman–Harris window, separable along $x$ and $y$, and defined as $w(\mathbf{x}) = w(x) w(y)$, with~\cite{harris1978use}
\begin{equation}
    w(x) = \sum_{j=0}^3 (-1)^j a_j \cos{\left( \frac{2 \pi j x}{M_x} \right)}~,
    \label{eq:Blackman}
\end{equation}
where $0 \le x < M_x$ and
\begin{align*}
    a_0 &= 0.3635819 \,, & a_1 &= 0.4891775 \\
    a_2 &= 0.1365995 \,, & a_3 &= 0.0106411
\end{align*}

Representative intermediate scattering functions with and without windowing, spanning several $q$ values, are reported in Fig.~S6 of the \textit{Supplemental Material} (§4.3). These confirm that windowing eliminates high-$q$ artifacts without distorting the underlying dynamics. The inset of Fig.~\ref{fig:cfr-fast-melt-win}a shows that windowing restores the quadratic scaling $\Gamma(q) = D_0 q^2$ up to $q \simeq 6.9\unit{\mu m^{-1}}$, even beyond the nominal $q_{\text{max}}$ imposed by the optical resolution. The resulting diffusion coefficient is $D_0^{(\mathrm{win})} = (0.969 \pm 0.002)\unit{\mu m^2/s}$, in close agreement with the reference value $D_0 = 0.957\unit{\mu m^2/s}$.

A potential disadvantage of windowing is that it decreases the amplitude $A(q)$ of the structure function, as shown in Fig.~\ref{fig:cfr-fast-melt-win}b. However, unless the decrease lowers the signal to unmeasurable values, it is possible to compensate for this decrease by rescaling the depressed $A(q)$ with the spatial average of the squared windowing function. The corrected $A(q)$ is illustrated in Fig.~\ref{fig:cfr-fast-melt-win}b as a continuous line, in excellent agreement with the $A(q)$ of the unwindowed data.

In summary, image windowing provides a simple software-based enhancement that restores the expected high-$q$ dynamics and extends the usable $q$ range without any hardware modifications.

\subsection{\label{subsec:objective-mag} \href{https://github.com/somexlab/fastddm-tutorials/blob/main/Tutorial_4-Objective_magnification/tutorial4.ipynb}{\faGithub} ~ Effect of the objective lens magnification}

\begin{figure}
    \centering
    \includegraphics[width=\columnwidth]{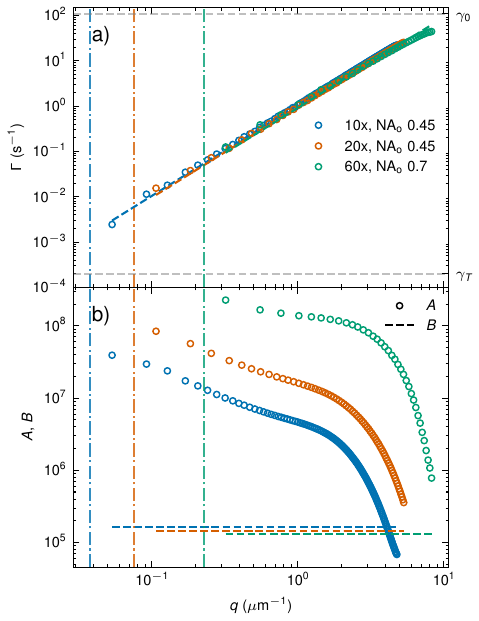}
    \caption{a) Relaxation rate $\Gamma(q)$ (symbols) obtained from $d(q, \Delta t)$ for PS $252\unit{nm}$ particles measured with different objective magnifications (see legend). Dashed lines indicate the low-$q$ cutoff $q_L = 2\pi / L$ corresponding to the different magnifications (color-coded). The gray dashed lines mark the accessible relaxation rate range ($\gamma_T \leq \Gamma \leq \gamma_0$). b) Corresponding amplitude $A(q)$ (symbols) and noise floor $B(q)$ (dashed lines).}
    \label{fig:obj-gamma}
\end{figure}
In Sec.~\ref{subsec:join}, we discussed how extending the acquisition duration improves access to slow dynamics, effectively lowering the minimum measurable relaxation rate $\Gamma$. Here, we shift our focus to an alternative, direct strategy for accessing lower wavevectors $q$: by increasing the field of view $L$ of the imaging system, the lowest accessible wavevector $q_\mathrm{L} = 2\pi / L$ can be decreased. This is particularly relevant when dealing with fast dynamics (e.g., small particles) or when constrained by low-frame-rate cameras.

A simple way to increase $L$ would be to increase the number of pixels in the field of view, but this approach comes at the cost of much larger data volumes and longer analysis times. A more efficient solution is to use a lower magnification objective, thereby increasing the effective pixel size $\delta_{\text{px}}$ and thus enlarging the physical field of view.

To demonstrate this effect, we repeated the experiment of Sec.~\ref{subsec:join} using three different objectives: $10\times$, $20\times$, and $60\times$. For each magnification, we acquired two image sequences (10000 frames, $256\times 256$ pixels) at $108\unit{fps}$ and $1\unit{fps}$ respectively, keeping the exposure time fixed at $9\unit{ms}$. The lamp intensity was adjusted to ensure the same average image brightness across all videos. The final structure functions were obtained by merging the fast and slow acquisitions using the procedure described earlier in Sec.~\ref{subsec:join}. Representative intermediate scattering functions and corresponding fits for different objectives are provided in the \textit{Supplemental Material} (Fig.~S7).

In Fig.~\ref{fig:obj-gamma}a, we show the measured relaxation rates $\Gamma(q)$ as a function of $q$.
As expected, the low-$q$ cutoff $q_L$ shifts to smaller values with decreasing objective magnification, consistent with the relation $q_\mathrm{L} = 2\pi m_\mathrm{o}/ (M \delta^*_{\text{px}})$, where $m_o$ is the objective magnification, $M$ the image matrix size, and $\delta^*_{\text{px}}$ the native pixel size of the sensor. The ability to shift $q_\mathrm{L}$ downward expands the range of accessible dynamics, especially for slowly relaxing systems.
However, a change in magnification also affects the signal amplitude $A(q)$. While the noise floor $B(q)$ remains approximately constant across different magnifications, the signal amplitude increases with increasing $m_\mathrm{o}$, improving the overall signal-to-noise ratio (Fig.~\ref{fig:obj-gamma}b).

\begin{figure}
    \centering
    \includegraphics[width=\columnwidth]{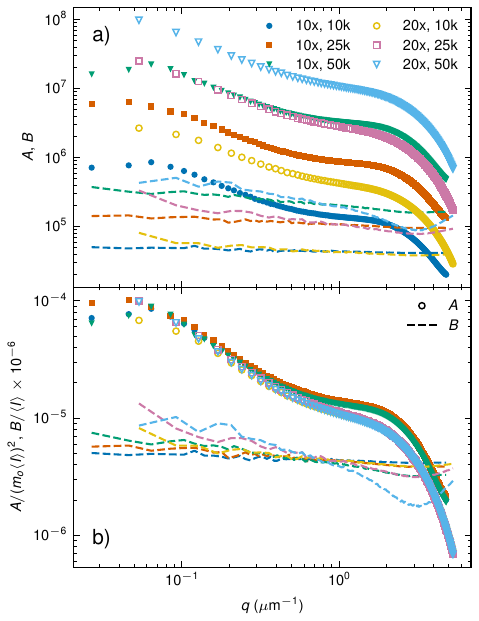}
    \caption{(a) Amplitude $A(q)$ (symbols) and noise floor $B(q)$ (dashed lines) for PS $252\unit{nm}$ particles imaged at $10\times$ (solid symbols) and $20\times$ (open symbols) objective magnification, under identical mean image intensity. (b) Same data after scaling according to the relations discussed in the main text.}
    \label{fig:obj-intensity}
\end{figure}
To further probe this dependence, we performed a systematic series of measurements with an ORCA-Flash4.0 V3 (Hamamatsu) camera and $10\times$ and $20\times$ objectives. The illumination (and/or exposure) was adjusted to set the mean sensor signal to 10k, 25k, and 50k camera counts. Figure~\ref{fig:obj-intensity}a shows the resulting $A(q)$ and $B(q)$; see also Fig.~S8 in the \textit{Supplemental Material} for intermediate scattering functions acquired at different objectives and intensities. The nearly $q$-independent offset $B(q)$ increases linearly with the mean intensity $\langle i\rangle$, as expected when photon shot noise dominates over read noise: in the notation of Eq.~\eqref{eq:realimages}, where the image is written as signal plus a noise term $n$ (with Fourier transform $N$), one has
$B(q)\propto \langle |N|^{2}\rangle \propto \langle i\rangle$ for the CMOS sensor. In contrast, the signal amplitude scales quadratically with both illumination and magnification, $A(q)\ \propto\ \langle i\rangle^{2}\, m_{\mathrm{o}}^{2}$. The $\langle i\rangle^{2}$ dependence follows from the heterodyne nature of bright-field image formation \cite{giavazzi2009scattering}. The $m_{\mathrm{o}}^{2}$ factor arises because $A(q)$ represents the integral of the spectral density over one Fourier-space “pixel” of area $(\Delta q)^{2}$; for a fixed sensor pixel pitch and image matrix size, the $q$-space sampling step $\Delta q\equiv q_{L}$ scales as $q_{L}\propto m_{\mathrm{o}}$, hence $(\Delta q)^{2}\propto m_{\mathrm{o}}^{2}$.

These scalings yield a practical guideline for DDM: increasing the mean intensity (without saturating the detector) and using higher-magnification objectives both boost the measurable signal. In particular, the amplitude-to-offset ratio improves as
\begin{equation}\label{eq:sigtono}
\frac{A(q)}{B(q)}\ \propto\ \langle i\rangle\, m_{\mathrm{o}}^{2}  
\end{equation}
which makes the benefits of higher illumination and magnification explicit for bright-field DDM.

\subsection{\label{subsec:objective-na} \href{https://github.com/somexlab/fastddm-tutorials/blob/main/Tutorial_5-Objective_na/tutorial5.ipynb}{\faGithub} ~ Effect of the objective numerical aperture}

\begin{figure}
    \centering
    \includegraphics[width=\columnwidth]{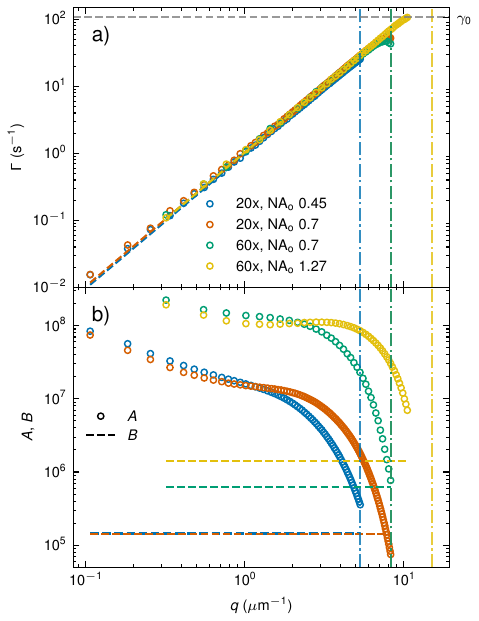}
    \caption{(a) Relaxation rate $\Gamma(q)$ obtained from $d(q, \Delta t)$ for PS $252\unit{nm}$ particles imaged with objectives of different numerical apertures (see legend). Dashed vertical lines indicate the theoretical upper limit $q_{\mathrm{NA}} = 2 \pi \mathrm{NA_o} / \lambda_o$ with $\lambda_o = 550\unit{nm}$. The dashed horizontal line denotes the limit imposed by the frame rate $\gamma_0$. (b) Corresponding signal amplitude $A(q)$ (symbols) and noise floor $B(q)$ (dashed lines). Color coding as in panel (a).}
    \label{fig:obj-na-gamma-a}
\end{figure}
As discussed in Sec.~\ref{subsec:windowing}, image windowing can be employed to suppress artifacts arising from boundary effects, thereby improving the reliability of the structure function at large wavevectors $q$. However, this method does not overcome the fundamental optical limit set by the microscope objective's numerical aperture: $q_{\mathrm{NA}} = 2\pi \mathrm{NA_o} / \lambda_o$. This fundamental equation determines the maximum wavevector that can be resolved in an optical system with illumination wavelength $\lambda_o$ and objective numerical aperture $\mathrm{NA_o}$

To extend the measurable $q$ range beyond this boundary, one must increase the numerical aperture of the imaging objective. Of course, this approach is only meaningful if the Nyquist limit $q_{\mathrm{px}} = \pi / \delta_{\mathrm{px}}$ exceeds the optical cutoff, ensuring that spatial frequencies are adequately sampled.

To illustrate this effect, we acquired bright-field microscopy image sequences of the PS $252\unit{nm}$ sample from Sec.~\ref{subsec:join}, using the same imaging setup and conditions. We tested four objectives:
\begin{itemize}
    \item A low numerical aperture objective (20$\times$, $\mathrm{NA_o} = 0.45$);
    \item A high numerical aperture, water-immersion objective (60$\times$, $\mathrm{NA_o} = 1.27$);
    \item Two objectives with the same numerical aperture ($\mathrm{NA_o} = 0.70$) but different magnifications (20$\times$ and 60$\times$).
\end{itemize}

Each acquisition consisted of two image sequences (10000 frames at 108 and 1 fps, respectively), except for the 60$\times$, $\mathrm{NA_o} = 1.27$ water-immersion objective, where the acquisition was limited to 5000 frames due to evaporation of the immersion medium. All videos were recorded at $256 \times 256$ pixels and $9\unit{ms}$ exposure time. The illumination intensity was adjusted to maintain a consistent average brightness across all acquisitions. Structure functions were merged using the procedure described in Sec.~\ref{subsec:join}. Representative intermediate scattering functions for all optical configurations are shown in Fig.~S9 of the \textit{Supplemental Material}.

Figure~\ref{fig:obj-na-gamma-a}a shows the resulting relaxation rates $\Gamma(q)$ plotted against $q$ for each objective. As expected, increasing $\mathrm{NA_o}$ shifts the upper bound $q_{\mathrm{NA}}$ proportionally, thereby extending the range of accessible wavevectors. For the objective with the highest numerical aperture, the dominant constraint on $\Gamma(q)$ becomes the frame rate $\gamma_0$ of the camera, rather than optical resolution.

The corresponding amplitude $A(q)$ and noise floor $B(q)$ are shown in Fig.~\ref{fig:obj-na-gamma-a}b. As observed previously in Sec.~\ref{subsec:objective-mag}, the noise level $B(q)$ remains approximately constant across different objectives, governed primarily by the average image intensity. In contrast, the amplitude $A(q)$ benefits from both higher magnification and increased $\mathrm{NA_o}$, which preserve the signal over a broader $q$ range before amplitude decay sets in.

Together, these observations confirm that high-NA objectives are crucial for extending DDM measurements to large $q$ values -- especially when characterizing rapidly diffusing or small-scale structures.

\subsection{\label{subsec:condenser-na} \href{https://github.com/somexlab/fastddm-tutorials/blob/main/Tutorial_6-Condenser_na/tutorial6.ipynb}{\faGithub} ~ Effect of the condenser numerical aperture}

In this section, we extend our investigation to a final instrumental parameter: the numerical aperture of the condenser, denoted $\mathrm{NA_c}$. Whereas previous sections primarily addressed strategies to optimize measurements through either post-processing or objective selection, we now examine how varying $\mathrm{NA_c}$ can enhance the signal amplitude in DDM experiments.
This provides a practical and often overlooked method for improving signal quality, with implications that will be revisited in Sec.~\ref{subsec:proteins}.

To evaluate the influence of $\mathrm{NA_c}$, we acquired bright-field image sequences of an aqueous suspension of PS nanoparticles (diameter $252\unit{nm}$) prepared at a volume fraction $\phi_0 = 10^{-4}$, following the same protocol described in Sec.~\ref{subsec:join}.

\begin{figure}
    \centering
    \includegraphics[width=\columnwidth]{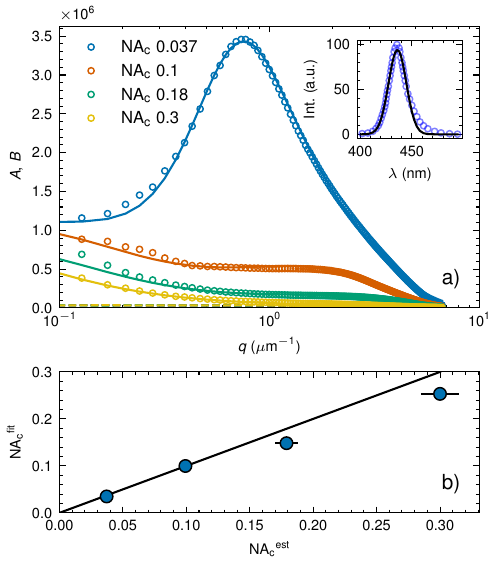}
    \caption{a) Amplitude $A(q)$ (symbols) and noise term $B(q)$ (dashed lines) obtained from fits of $d(q, \Delta t)$ for the PS $252\unit{nm}$ sample ($\phi_0 = 10^{-4}$) under varying condenser numerical apertures $\mathrm{NA_c}$ (legend). Continuous lines show fits using Eq.~\eqref{eq:A-fit}. The inset displays the lamp emission spectrum (Gaussian fit). b) Comparison between fitted values $\mathrm{NA_c}^{\mathrm{fit}}$ from Eq.~\eqref{eq:A-fit} (derived from data in Fig.~\ref{fig:cond-AB}) and estimated values $\mathrm{NA_c}^{\mathrm{est}}$ obtained by measuring the aperture diaphragm size. The continous line indicates the ideal condition $\mathrm{NA_c}^{\mathrm{fit}} = \mathrm{NA_c}^{\mathrm{est}}$.}
    \label{fig:cond-AB}
\end{figure}

Microscopy was performed on a Nikon Eclipse Ti2 inverted microscope equipped with an ORCA-Flash4.0 V3 CMOS camera (Hamamatsu; pixel size $\delta^*_{\text{px}} = 6.5\unit{\mu m}$). The sample was uniformly illuminated using a blue LED lamp (LIDA, Lumencor), emitting a quasi-Gaussian spectrum centered at $\lambda_0 \simeq 436\unit{nm}$ with bandwidth $\Delta \lambda \simeq 9\unit{nm}$ (see inset of Fig.~\ref{fig:cond-AB}).
K\"ohler illumination was established using an ELWD condenser ($\mathrm{NA_c} = 0.3$).
By adjusting the aperture diaphragm, we selected four values of $\mathrm{NA_c}$ (approx. 0.037, 0.1, 0.18, and 0.3).
For each setting, two image sequences of $10000$ frames ($512 ~\times ~512$ pixels) were acquired at $111\unit{fps}$ and $1\unit{fps}$, respectively, using a $20\times$ objective with $\mathrm{NA_o} = 0.7$, yielding an effective pixel size $\delta_{\text{px}} = 0.325\unit{\mu m}$. All other acquisition parameters -- including exposure time ($9\unit{ms}$) and average image intensity -- were kept constant across experiments by adjusting the LED power accordingly.

Figure~\ref{fig:cond-AB}a) displays the amplitude $A(q)$ and the corresponding noise term $B(q)$ for each value of $\mathrm{NA_c}$.
A clear trend emerges: reducing $\mathrm{NA_c}$, and thereby increasing the spatial coherence of illumination, enhances the measured signal amplitude.
At higher $\mathrm{NA_c}$, $A(q)$ decays monotonically with $q$.
Upon lowering $\mathrm{NA_c}$, the amplitude increases significantly and may exhibit a non-monotonic behavior, including a pronounced maximum at intermediate $q$, followed by a decay at large $q$ due to limitations imposed by the objective’s numerical aperture.
All curves appear to converge toward a common asymptotic value at low $q$.

To quantitatively interpret the data, we performed a global fit of $A(q)$ using Eq.~\eqref{eq:A-fit}.
We fixed the objective numerical aperture $\mathrm{NA_o} = 0.7$ and the illumination parameters ($\lambda_0 = 436\unit{nm}$ and $\Delta\lambda = 8.6\unit{nm}$) based on a Gaussian fit of the LED spectrum (see inset of Fig. \ref{fig:cond-AB}a)).
We further constrained the fits by assuming a common capillary depth $\ell$ and a shared contrast parameter $\alpha$, which depends on the optical properties of the particles and the medium.

The fitted condenser numerical apertures $\mathrm{NA_c}^{\mathrm{fit}}$ are plotted against the independently estimated values $\mathrm{NA_c}^{\mathrm{est}}$ in Fig.~\ref{fig:cond-AB}b).
The agreement is satisfactory, especially at low $\mathrm{NA_c}$.
The deviations at higher values are likely attributable to breakdowns in the small-angle approximation.

The global fit also yields an equivalent
flat-top capillary thickness~\cite{giavazzi2009scattering} of $\ell_{\mathrm{eq}} = \sqrt{2\pi}, \ell = 336 \pm 3,\unit{\mu m}$, in good agreement with the nominal depth of $300 \pm 30,\unit{\mu m}$. In addition, the fit returns a phase delay per particle of $\varphi \simeq \pi/2+\alpha =1.838 \pm 0.003,\unit{rad}$, closely matching the Mie-theory estimate of $\varphi = 1.81 \pm 0.02,\unit{rad}$, calculated from the particle radius, illumination wavelength $\lambda_0$, and the refractive indices of polystyrene ($n_{\mathrm{PS}} = 1.586$) and solvent ($n_\mathrm{s} = 1.363$), including their respective uncertainties. This result suggests that DDM experiments with partially coherent light can, in principle, be used to infer the optical thickness of particles, extending the approach introduced in Ref.~\onlinecite{potenza_how_2010} for coherent laser illumination.

\begin{figure*}
    \centering
    \includegraphics[width=\textwidth]{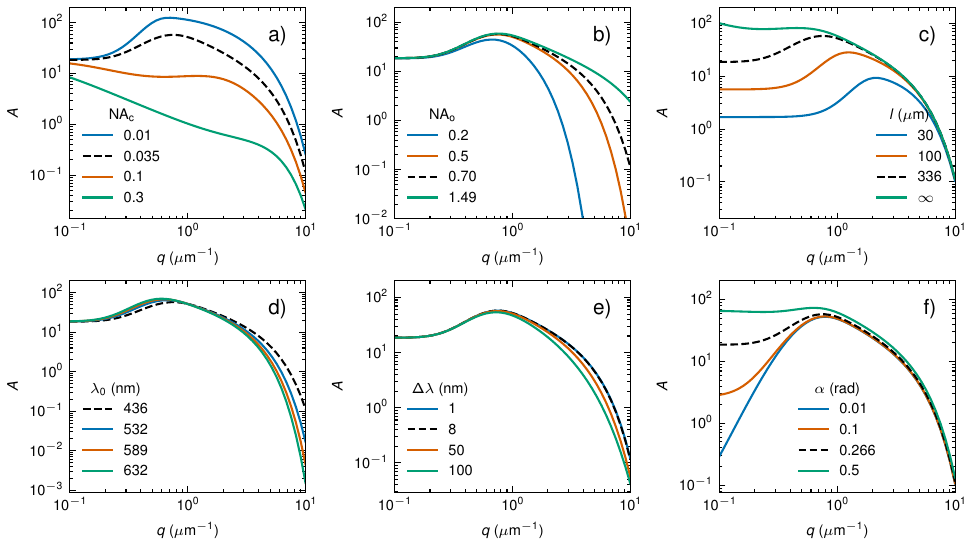}
    \caption{Dependence of the model in Eq.~\eqref{eq:A-fit} on key experimental parameters: condenser numerical aperture (a), objective numerical aperture (b), capillary thickness (c), illumination wavelength (d), spectral bandwidth (e), and phase delay $\varphi$ (f). In each case, all other parameters are held constant to match the best-fit values for $\mathrm{NA_c} = 0.037$ in Fig.~\ref{fig:cond-AB}a). The dashed line serves as a reference curve.}
    \label{fig:A-vs-params}
\end{figure*}

\subsection{\label{subsec:tuning_Aq} \href{https://github.com/somexlab/fastddm-tutorials/blob/main/Tutorial_6-Condenser_na/tutorial6.ipynb}{\faGithub} ~Tuning the amplitude $A(q)$}
Building upon the insights from the previous section, we now explore in more detail how the functional form of Eq.~\eqref{eq:A-fit} depends on key experimental parameters. To assess parameter sensitivity, Fig.~\ref{fig:A-vs-params} displays the model predictions for $A(q)$ as individual experimental parameters are varied while holding the others constant (set to values fitted for $\mathrm{NA_c} = 0.037$).
Panels (a)–(f) show how condenser and objective apertures, capillary depth, wavelength, bandwidth, and phase delay each influence the shape and magnitude of $A(q)$.
\begin{itemize}
    \setlength\itemsep{0.3em}
    
    \item \textbf{Condenser numerical aperture (Panel a):}  
    As already discussed in Sec.~\ref{subsec:condenser-na}, decreasing $\mathrm{NA_c}$ increases the spatial coherence of the illumination and enhances the signal amplitude. 
    Notably, a local maximum in $A(q)$ emerges at high $q$ and progressively shifts toward lower $q$ as $\mathrm{NA_c}$ decreases, becoming more pronounced.

    \item \textbf{Objective numerical aperture (Panel b):}  
    Increasing $\mathrm{NA_o}$ allows access to higher $q$ values by expanding the upper bound $q_{\mathrm{NA}} = 2\pi \mathrm{NA_o} / \lambda_{o}$.
    The amplitude at high $q$ is also improved, as the collection of higher-angle scattered light becomes more efficient.

    \item \textbf{Capillary depth (Panel c):}  
    Increasing the depth $\ell$ of the sample enhances the amplitude across all $q$ values, owing to the larger number of particles contributing to the signal.
    In addition, the increased optical path length introduces averaging effects, and the position of the local maximum in $A(q)$ shifts toward lower $q$.

    \item \textbf{Illumination wavelength (Panel d):}  
    A decrease in the central wavelength $\lambda_0$ causes the peak of $A(q)$ to shift to higher $q$, allowing access to faster dynamics and finer structural details.
    The change is moderate but may be useful in systems where the optical setup limits access to large $q$ values.

    \item \textbf{Illumination bandwidth (Panel e):}  
    Increasing the spectral bandwidth $\Delta \lambda$ broadens the illumination spectrum and reduces the spatial coherence of the source, resulting in a slight contraction of the observable $q$ range and a modest suppression of the peak amplitude.

    \item \textbf{Particle-induced phase delay (Panel f):}  
    The contrast parameter $\alpha$ (proportional to the phase delay $\varphi$ induced by a particle) governs the efficiency of image modulation by the sample. 
    Larger $\alpha$ values result in stronger signals at low $q$, improving detection of long-wavelength fluctuations or collective dynamics.

\end{itemize}

These dependencies illustrate the flexibility of DDM and the importance of tailoring the experimental configuration to the specific range of spatial and temporal scales of interest. In particular, the condenser numerical aperture and capillary depth offer convenient means for signal tuning, while optical choices such as wavelength and objective properties determine the resolution limits of the method. We remind the reader that changing these experimental parameters at constant intensity, leaves the background $B(q)$ substantially unaltered, whereas it changes $A(q)$, thus impacting directly on the signal-to-noise ratio $A/B$ in DDM experiments.

\subsection{\label{subsec:surprise} DDM vs DLS}

\begin{figure}
    \centering
    \includegraphics[width=\columnwidth]{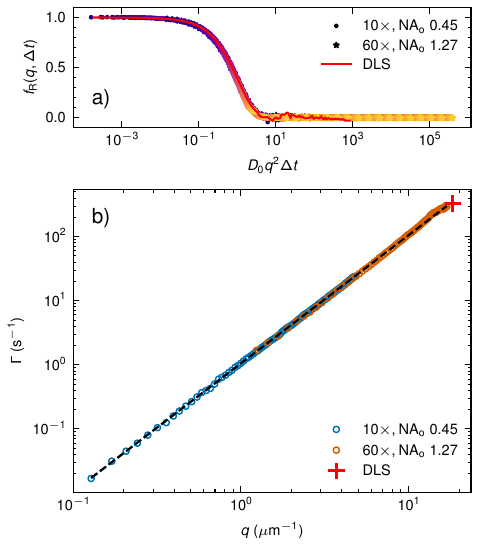}
    \caption{a) Intermediate scattering function $f_{R}(q, \Delta t)$ for the PS $252\unit{nm}$ sample. The data points correspond to different wavevectors $q$, obtained from DDM measurements using two objectives with different magnifications and numerical apertures (see legend). All curves are plotted as a function of the rescaled delay time $D_{0} q^2 \Delta t$, using the diffusion coefficient $D_{0}$ extracted from the relaxation rates $\Gamma(q)$ shown in panel b. The solid line represents the average intermediate scattering function measured from six independent acquisitions using a commercial DLS instrument, scaled with the same $D_{0}$ from DDM. b) Relaxation rate $\Gamma(q)$ vs $q$ from DDM (symbols). The plus symbol denotes the mean relaxation rate obtained from DLS.}
    \label{fig:windowing-join-obj}
\end{figure}

We conclude this section by demonstrating how DDM compares to DLS, particularly in terms of accessible wavevector range and data consistency.

To achieve the widest possible $q$ coverage, we carefully selected two sets of acquisitions on the PS $252\unit{nm}$ sample.
To access low-$q$ values, we revisited the measurements performed with a $10\times$, $\mathrm{NA_o} = 0.45$ objective, as discussed in Sec.~\ref{subsec:objective-mag}.
To extend toward higher $q$ values, we employed the dataset acquired using the $60\times$, $\mathrm{NA_o} = 1.27$ water-immersion objective, introduced in Sec.~\ref{subsec:objective-na}.
To mitigate boundary artifacts and optimize high-$q$ accuracy, all images were preprocessed using windowing (see Sec.~\ref{subsec:windowing}).

Figure~\ref{fig:windowing-join-obj}a presents the intermediate scattering functions $f_{R}(q, \Delta t)$ computed from DDM for a broad range of $q$ values.
All curves collapse onto a single master curve when plotted against the dimensionless time variable $D_{0} q^2 \Delta t$, where $D_{0}$ is the diffusion coefficient obtained from the global fit of the relaxation rates $\Gamma(q)$ shown in Fig.~\ref{fig:windowing-join-obj}b. For comparison, we superimpose the average field correlation function $g_1(\Delta t)$ measured from six independent acquisitions using a commercial DLS instrument.
These experiments were performed on the same particle-solvent system, prepared at a slightly lower volume fraction ($\phi_0 = 2 \times 10^{-4}$) and held at $T = 24^{\circ}\mathrm{C}$.
When rescaled using the diffusion coefficient obtained from DDM, the DLS data shows excellent agreement with the DDM results, both in the intermediate scattering function (panel a) and in the corresponding relaxation rate (panel b).

In summary, by judiciously selecting imaging parameters, we have extended the measurable DDM dynamic range in $q$ by more than two decades. The close match between DDM and DLS demonstrates the robustness of DDM and its potential as a complementary or even alternative tool for dynamic characterization across a wide range of length scales.

\section{Beyond particle sizing: worked examples of advanced applications of DDM\label{sec:beyondsizing}}
The preceding sections establish the fundamentals of DDM and its efficient computation (\emph{cf.} Section~III) and provide a tutorial path for particle sizing in dilute Brownian suspensions. In this section, we present a curated set of \emph{worked examples} that extend DDM beyond colloidal particle sizing and illustrate how the method is applied, stress-tested, and interpreted in more complex scenarios. These examples are chosen to highlight distinct methodological ingredients rather than to be exhaustive surveys of any given field. For each subsection we provide an accompanying Jupyter notebook and make the underlying dataset publicly available. The notebooks reproduce the reported figures and numerical results from the released data and specify all steps of the analysis with minimal code commenting.

These worked examples are deliberately concise and application-agnostic in their claims: they serve as \emph{illustrative} guides for applying DDM in regimes that depart from dilute Brownian diffusion, while the notebooks and data should hopefully enable readers to adapt the workflows to their own systems.

\subsection{\label{subsec:proteins} \href{https://github.com/somexlab/fastddm-tutorials/blob/main/Tutorial_7-Intermolecular_interactions/tutorial7.ipynb}{\faGithub} DDM of (semi)dilute protein solutions: Measuring size and intermolecular interactions}

\textit{Aim and goal.} Guidolin et al.~\cite{guidolin2023protein} demonstrated that DDM can extract hydrodynamic radii and interaction parameters for various proteins, objects far smaller than the colloids typically analyzed so far. Building on their dataset of \emph{Bovine Serum Albumin} (BSA) solutions, the aim here is to show, using \texttt{fastDDM}, how even a simple bright-field microscope can operate as a quantitative scattering instrument for dilute and semi-dilute protein solutions. The goals are twofold: (i) to determine the single-particle (dilute-limit) diffusion coefficient $D_0$ and the corresponding hydrodynamic radius $R_\mathrm{h}$, and (ii) to quantify intermolecular interactions via the concentration dependence of the collective diffusion coefficient $D(c)$ and the static amplitude $\langle A \rangle$, yielding the diffusion interaction parameter $k_D$ and the second virial coefficient $B_2$.

\textit{New concepts.} At finite concentrations, interparticle interactions affect the diffusive dynamics
\begin{equation}
\label{eq:gamma_diff}
\Gamma(q) = D(c)\,q^2~,
\end{equation}
where $D(c)$ is the concentration-dependent diffusion coefficient. To first order in $c$,
\begin{equation}
\label{eq:coll_diff_proteins}
D(c) = D_0 \left(1 + k_D c \right),
\end{equation}
with $k_D$ the diffusion interaction parameter~\cite{berne2000dynamic}. 

Interactions also affect the scattering intensity, which in turn change the image-structure-function amplitude $A(q)$, which scales with the scattering intensity and imaging transfer function, $A(q)\sim I(q)\,T(q)$. For proteins much smaller than the optical wavelength, the Debye–Zimm relation~\cite{zimm1948scattering} connects concentration and structure factor,
\begin{equation}
\label{eq:debye-zimm}
\frac{K^* c}{\Delta R} = \frac{1}{M_\mathrm{w}} + 2 B_2 c~,
\end{equation}
leading to the working model for the concentration-averaged amplitude
\begin{equation}
\label{eq:debye-zimm-amplitude}
\langle A \rangle = \frac{k\, M_\mathrm{w}\, c}{1 + c/c_0}, \qquad c_0 = \left(2 B_2 M_\mathrm{w} \right)^{-1},
\end{equation}
where $k$ absorbs instrumental factors and $\langle A \rangle$ is obtained by averaging $A(q)$ in a suitable wavevector range where it is substantially $q$-independent. Equations~\eqref{eq:coll_diff_proteins} and \eqref{eq:debye-zimm-amplitude} thus provide complementary dynamic ($k_D$) and static ($B_2$) measures of interactions. Since the interactions affecting the dynamics are both the direct ones (encoded by $B_2$) and the hydrodynamic ones, the two interaction parameters, $k_D$ and $B_2$, are related to each other. The interested reader should consult the relevant literature, such as for instance Refs.~\onlinecite{harding1985concentration,meechai1999translational,saluja2010diffusion,winzor2021quantifying}.

\textit{Results.} Using image sequences of BSA solutions, structure functions were computed for each video, azimuthally averaged, and replicates at fixed concentration were averaged, to improve statistical accuracy given the very small signal from the protein samples.

\begin{figure}
    \centering
    \includegraphics[width=\columnwidth]{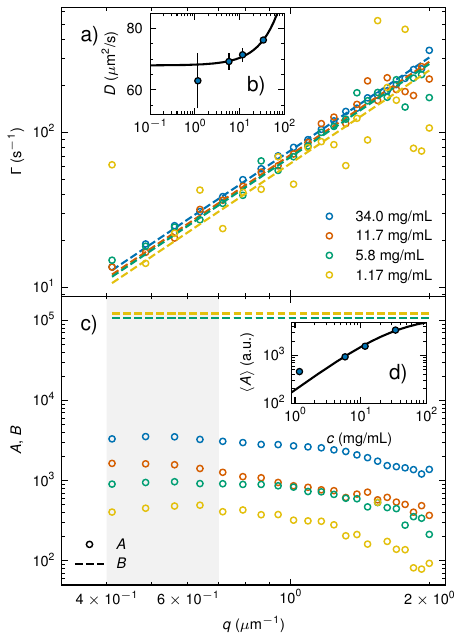}
    \caption{a) Relaxation rate $\Gamma(q)$ at different protein concentrations $c$, as indicated. Dashed lines are quadratic fits $\Gamma = Dq^2$. 
    b) Diffusion coefficients $D$ obtained from the fits in panel a). The continuous line is a fit of Eq.~\eqref{eq:coll_diff_proteins}.
    c) Amplitude $A(q)$ (symbols) and noise $B(q)$ (dashed lines) for all concentrations. The gray shaded area indicates the $q$-range used to compute $\langle A \rangle$. 
    d) Average amplitude $\langle A \rangle$ vs $c$. The continuous line is a fit of Eq.~\eqref{eq:debye-zimm-amplitude}.}
    \label{fig:bsa_gamma_ab}
\end{figure}

Single-exponential decays were fitted to obtain the relaxation rates $\Gamma(q)$ and amplitudes $A(q)$; diffusion coefficients $D$ were then extracted via Eq.~\eqref{eq:gamma_diff}. The resulting data are shown in Fig.~\ref{fig:bsa_gamma_ab}. The relaxation rate $\Gamma(q)$ exhibits a quadratic dependence on $q$ at all concentrations $c$ (Fig.~\ref{fig:bsa_gamma_ab}a), consistent with diffusive dynamics. The diffusion coefficient $D(c)$ extracted from these fits increases with $c$ (Fig.~\ref{fig:bsa_gamma_ab}b), in agreement with Eq.~\eqref{eq:coll_diff_proteins}. A linear regression yields $D_0 = (68 \pm 1)\unit{\mu m^2/s}$ and $k_D = (3.6 \pm 0.5)\unit{mL/g}$, indicating weak net repulsion and agreeing with Ref.~\onlinecite{guidolin2023protein} and prior BSA measurements under comparable conditions~\cite{larsen2021probing}. A positive $k_D$ indicates net repulsive interactions, whereas a negative $k_D$ corresponds to net attractive interactions~\cite{harding1985concentration,connolly2012weak,wei2023improved}. From $D_0$ a hydrodynamic radius $R_\mathrm{h} = (3.18 \pm 0.05)\unit{nm}$ is inferred.

The static channel likewise reports on interactions. Figure~\ref{fig:bsa_gamma_ab}c shows the amplitude $A(q)$ and the (orders of magnite larger) noise $B(q)$; the latter is nearly flat and concentration-independent under fixed imaging conditions (i.e. with the same transfer function $T(q)$), whereas $A(q)$ grows with $c$. This growth is quantified via the average amplitude $\langle A \rangle$ over $0.4 < q < 0.7\unit{\mu m^{-1}}$ (gray band) and compared to Eq.~\eqref{eq:debye-zimm-amplitude}. Fitting Eq.~\eqref{eq:debye-zimm-amplitude} to the data with $M_\mathrm{w} = 66.6\unit{kDa}$ yields $B_2 = (2.2 \pm 1.4)\times 10^{-4}\unit{mol\, mL / g^2}$ (Fig.~\ref{fig:bsa_gamma_ab}d), consistent with Ref.~\onlinecite{guidolin2023protein}. 

The practical detection limit for BSA in this configuration is $\sim 1\unit{mg/mL}$ (not yet at the level of state-of-the-art DLS sensitivity) and is primarily constrained by sensor noise and the number of usable frames, suggesting clear pathways to improved limits with longer acquisitions and lower-noise detectors.

The approach is readily extendable to other systems, including monoclonal antibody solutions, whose hydrodynamic sizes and weak, short-ranged interactions fall within the regime accessible to bright-field DDM given suitable signal-to-noise ratio and frame statistics. 

\textit{Methods.} Microscopy image sequences of aqueous BSA (Sigma Aldrich \#A7638) solutions reported in Ref.~\onlinecite{guidolin2023protein} were analyzed using \texttt{fastDDM}. Samples spanned concentrations $c$ from $1.17$ to $34\unit{mg/mL}$, prepared by serial dilution in phosphate-buffered saline (PBS, ROTH \#1058.1). Concentrations were independently verified with a commercial UV spectrophotometer (ThermoFisher). Solutions were loaded into glass capillaries with a $3 \times 0.3\unit{mm}$ rectangular cross-section (Vitrocom Inc.) and sealed at both ends with epoxy to prevent evaporation. Imaging was performed on a Nikon Eclipse Ti-U inverted microscope in bright-field, equipped with a Hamamatsu Orca Flash 4.0 v2 CMOS camera. The temperature was maintained at $T=(22 \pm 2)\celsius$. Each video comprised 5000 frames at $500\unit{fps}$ with $2\times 2$ binning, yielding $1024\times 128$ pixels and an effective pixel size $\delta_{\mathrm{px}}=0.65\unit{\mu m}$ using a $m_\mathrm{o}=20\times$, $\mathrm{NA_o}=0.5$ objective. For each sample, four videos were acquired at two positions in the mid-plane of the capillary. Exposure time ($1.99\unit{ms}$) and lamp intensity were held fixed across acquisitions to ensure comparable noise levels.

\subsection{\label{subsec:bacteria}  Bacterial motion: From active swimmers to Brownian rotators}
Microorganisms, such as cells and bacteria, are far more complex than passive colloids. Their dynamics often involve active motion~\cite{bray2000cell}. Bacterial dynamics span from active, run–and–tumble swimming (e.g.\ \emph{E.~coli}) to purely Brownian motion when propulsion is suppressed. Dynamic light scattering techniques are, in principle, suitable for characterizing such motion~\cite{nossal1971use}. However, probing the typical swimming length scale of about  $10\unit{\mu m}$ requires access to very small scattering angles~\cite{boon1974light} ($\theta \approx 4^{\circ}$). In this context, DDM offers a unique advantage~\cite{martinez2012differential,germain2016differential}, especially when combined with confocal microscopy, as it enables access to small wavevectors even in dense, optically opaque samples~\cite{lu2012characterizing}. In general, all these distinct motility mechanisms leave distinct fingerprints in the intermediate scattering function \(f_{R}(q,\Delta t)\), the interplay between them depending, among many other things, on the concentration of the bacterial suspensions.

In dilute suspensions imaged in brightfield or phase contrast, DDM provides a high-throughput route to bulk motility parameters by fitting the intermediate scattering function \(f_{R}(q,\Delta t)\) inferred from intensity fluctuations. Seminal studies~\cite{martinez2012differential} used a two-population (swimming and dead bacteria) model for \(f_{R}(q,\Delta t)\) to extract the population swimming-speed distribution \(P(v)\), the motile-cell fraction, and the diffusivity of non-motile cells, with accuracy comparable to single-cell tracking but averaged over \(\sim 10^4\) cells and acquired within minutes.

Building on this dilute-suspension baseline, which can be considered the equivalent of particle sizing for bacterial suspensions, we now examine two complementary regimes that stress different aspects of DDM analysis. First, we consider \emph{confocal} DDM of moderately dense bacterial suspensions to access small $q$ in optically crowded samples. Second, we turn to \emph{non-motile anisotropic} bacteria, where roto–translational Brownian motion approximately factorizes into a $q$-independent rotational relaxation and a $q^{2}$ translational decay, enabling separate estimation of the rotational ($D_{r}$) and translational ($D_{t}$) diffusivities. For both use cases, accompanying notebooks and released datasets reproduce all figures and parameter estimates.

\subsubsection{\label{subsubsec:dense bacteria} \href{https://github.com/somexlab/fastddm-tutorials/blob/main/Tutorial_8-Bacterial_motion/tutorial8a.ipynb}{\faGithub} Motile bacteria in dense suspensions: Ballistic motion and velocity distributions}

\textit{Aim and goals.} Even though DDM is way more insensitive of DLS to multiple scattering~\cite{giavazzi2014viscoelasticity,eitel2020hitchhiker,nixon2022probing}, the study of dense, optically opaque samples (in particular bacteria) remains challenging. A powerful strategy to overcome these challenges is to exploit the optical sectioning capability of confocal microscopy, which restricts detection to the focal plane and thereby reduces background and multiple scattering. This approach, named Confocal DDM (ConDDM) was originally demonstrated in Ref.~\onlinecite{lu2012characterizing} for dense suspensions of hard-spheres and swimming bacteria. Here we showcase the use of ConDDM with dense \emph{bacterial} suspensions of swimming \emph{Bacillus subtilis}, a flagellated bacterium. Confocal image sequences were obtained from Ref.~\onlinecite{lu2012characterizing} (data unpublished). The suspension was macroscopically opaque and consisted of motile \emph{B. subtilis}. We show how to use \texttt{fastDDM} to quantify: (i) bulk ballistic swimming via $\Gamma(q)\propto q$ and the mean speed $v_0$, (ii) the population speed distribution \(P(v)\), and (iii) near-surface deviations from ballistic scaling.

\textit{New concepts.} Active, approximately ballistic swimming of dense bacteria at scale $q^{-1}$ is captured by a compressed-exponential ISF,
\begin{equation}
\label{eq:compressed-exp-isf}
f_{R}(q,\Delta t)=\exp\!\left[-\big(\Gamma(q)\Delta t\big)^{\beta}\right], \qquad \beta>1,
\end{equation}
with bulk ballistic scaling \(\Gamma(q)=v_0\,q\). One can also exploit the general relationship~\cite{berne2000dynamic} between the intermediate scattering function and the velocity distribution $P(v)$
\begin{equation}
\label{eq:isf-velocity-distribution}
    f(\mathbf{q}, \Delta t) = \langle \exp{\left[ \mathrm{i}\mathbf{q} \cdot \mathbf{v} \Delta t \right]} \rangle = \int \mathrm{d}^3 \mathbf{v} \, P(\mathbf{v}) \exp(-\mathrm{i} \mathbf{q} \cdot \mathbf{v} \Delta t)~.
\end{equation}
When the process is isotropic, Eq.~\eqref{eq:isf-velocity-distribution} simplifies to $f_{R}(q, \Delta t) = \int_0^{\infty} \mathrm{d}v \, P(v) J_0(q \Delta t v)$, where $J_0$ is the zeroth-order Bessel function.
Therefore, the velocity distribution can be obtained by applying a Fourier sine transformation to the intermediate scattering function~\cite{cipelletti2003universal}:
\begin{equation}
\label{eq:velocity-distribution}
    P(v) = \frac{2 v}{\pi} \int_0^{\infty} \mathrm{d}x \, x f(x) \sin(x v)~,
\end{equation}
where $x = q \Delta t$.
Eqs.~\ref{eq:compressed-exp-isf}~and~\ref{eq:velocity-distribution} enable estimation of \(v_0\) and \(P(v)\) from DDM data.

\begin{figure}
    \centering
    \includegraphics[width=\columnwidth]{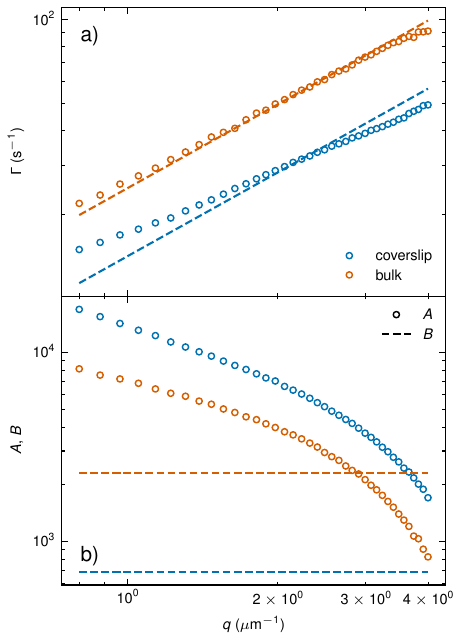}
    \caption{a) Relaxation rate $\Gamma(q)$ obtained from fits of $d(q, \Delta t)$ for swimming bacteria near the coverslip and in the bulk. Dashed line: linear fit $\Gamma(q) = v_0 q$ for bulk data. b) Corresponding static amplitude $A(q)$ (symbols) and noise floor $B(q)$ (dashed lines). Color code as in panel a).}
    \label{fig:bacteria_gamma_AB}
\end{figure}

\begin{figure}
    \centering
    \includegraphics[width=\columnwidth]{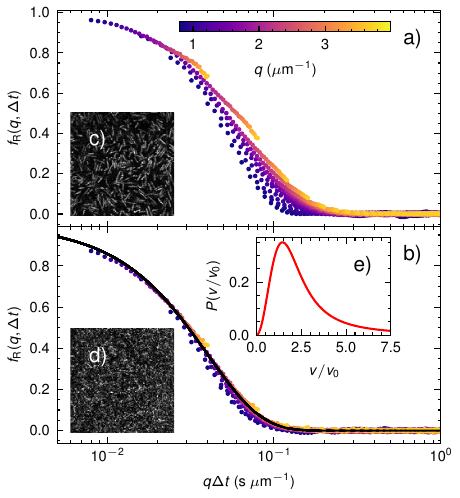}
    \caption{$f_{R}(q, \Delta t)$ as a function of $q\Delta t$ for swimming bacteria near the coverslip (a) and in the bulk (b). The bulk data collapse onto a master curve (black line) described by $\exp[-(q v_0 \Delta t)^{1.35}]$. Panels c) and d) show representative images near the coverslip and in the bulk, respectively. Panel e): velocity distribution $P(v)$ extracted from the bulk data via Eq.~\eqref{eq:velocity-distribution}.}
    \label{fig:bacteria_g1_scaling}
\end{figure}

\textit{Results.} The intermediate scattering function was computed from confocal image sequences and fitted to Eq.~\eqref{eq:compressed-exp-isf} with \(\beta=1.35\). Figure~\ref{fig:bacteria_gamma_AB}a reports \(\Gamma(q)\) near the coverslip and in the bulk; Fig.~\ref{fig:bacteria_gamma_AB}b shows the corresponding \(A(q)\) and \(B(q)\). In the bulk, \(\Gamma(q)\) grows linearly with \(q\), consistent with ballistic motion; a linear fit yields \(v_0 = (24.9 \pm 0.2)\,\unit{\mu m/s}\). Near the coverslip, \(\Gamma(q)\) deviates from linearity, indicating non-ballistic, heterogeneous dynamics. The scaling analysis of \(f_{R}(q,\Delta t)\) vs \(q\Delta t\) (Fig.~\ref{fig:bacteria_g1_scaling}) shows collapse in the bulk, well described by a compressed exponential with \(\beta=1.35\), while no collapse is observed near the surface. The contrast reflects geometric constraints: near the coverslip, trajectories are quasi-2D and surface-aligned (panel c), whereas in the bulk the motion explores 3D orientations (panel d). Using Eq.~\eqref{eq:velocity-distribution} with the bulk \(f\), the velocity distribution \(P(v)\) in Fig.~\ref{fig:bacteria_g1_scaling}e is obtained.

\textit{Methods.} Confocal image sequences were obtained from Ref.~\onlinecite{lu2012characterizing} (data unpublished). The suspension was macroscopically opaque and consisted of motile, flagellated \emph{B.~subtilis}. Two sequences, each comprising 5000 frames (128$\times$128 pixels), were recorded at $100\unit{fps}$ using a point-scanning confocal microscope equipped with a 63$\times$ oil-immersion objective (Leica), yielding a pixel size $\delta_{\mathrm{px}} = 0.591\unit{\mu m}$. Recordings were performed at two depths: at the coverslip and $12\unit{\mu m}$ into the bulk. The sample temperature was maintained at $T=33\celsius$. Following Sec.~\ref{sec:particle-sizing}, image-structure functions were computed, azimuthally averaged, and fitted to Eq.~\eqref{eq:compressed-exp-isf} to extract \(\Gamma(q)\), \(\beta\), and \(A(q)\); bulk \(\Gamma(q)\) was regressed against \(q\) to estimate \(v_0\).

\paragraph*{Note.}
In bright-field DDM, intensity fluctuations arise from variations in transmitted light produced by refractive-index (or thickness) inhomogeneities; the recorded signal integrates contributions along the sample depth set by the objective depth of field. In wide-field \emph{fluorescence} DDM, contrast originates from incoherent emission of labeled structures; the signal is approximately proportional to the local fluorophore density but likewise integrated over the axial extent of the excitation/detection path. The different contrast mechanisms imply different transfer functions $T(q)$ and noise statistics. In fluorescence DDM, photobleaching induces a slow, monotonic decay of the average intensity that can act as an additional, spurious relaxation channel in the ISF, effectively \emph{speeding up} apparent dynamics. Mitigations include frame-wise normalization, restricting fits to time windows with negligible bleaching, and verifying robustness against excitation-power changes and control samples. Analyses should explicitly check and correct for bleaching-driven drifts before interpreting relaxation rates. 

In addition, the confocal detection scheme used in ConDDM introduces optical sectioning with finite slice thickness $\delta$, restricting detection to a thin axial slab. Objects stochastically entering and leaving this slab generate \emph{axial number fluctuations}, which add an extra relaxation pathway that \emph{accelerates} the apparent dynamics at small $q$ relative to true motion.

This constitutes a paradigmatic case in which the axial motion cannot be neglected, and the fully three-dimensional character of the relaxation processes must be taken into account. A quantitative treatment and diagnostics for this low-$q$ effect are provided in Ref.~\onlinecite{lu2012characterizing} and should be consulted when sectioning effects are expected to be significant. 

All datasets analyzed here are restricted to a \emph{high-$q$} regime, $q \gg q_{\delta}$ with $q_{\delta}\equiv 2\pi/\delta$, where confocal-sectioning–induced axial number fluctuations have negligible impact on the measured dynamics. In addition, bleaching was found to be negligible during the acquisition time.

\subsubsection{  \href{https://github.com/somexlab/fastddm-tutorials/blob/main/Tutorial_8-Bacterial_motion/tutorial8b.ipynb}{\faGithub} Non-motile bacteria: Roto–translational dynamics of anisotropic particles\label{subsubsec:non-motile bacteria}}

\textit{Aim and goals.} Quantify translational and rotational Brownian dynamics of anisotropic particles (non-motile bacteria) by dark-field DDM, with goals: (i) estimate \(D_t\) from the \(q^2\) translational mode and (ii) estimate \(D_r\) from the $q$-independent rotational contribution, verifying the expected two-mode structure of the ISF.

\textit{New concepts.} For anisotropic particles exhibiting roto–translational diffusion, the ISF can be modeled as
\begin{equation}
\label{eq:isf-bacteria-rotation}
f_{R}(q,\Delta t)=a\,e^{-\Gamma_1(q)\Delta t} + (1-a)\,e^{-\Gamma_2(q)\Delta t},
\end{equation}
with \(0\le a \le 1\), \(\Gamma_2(q)=D_t q^2\) (pure translation) and \(\Gamma_1(q)=6D_r + D_t q^2\) (rotation plus translation). The mixing coefficient $a$ can be $q$-dependent, reflecting the relative weight of the two modes. The two-mode form in Eq.~\eqref{eq:isf-bacteria-rotation} is observable only if the single-particle image changes with orientation, i.e.\ the recorded intensity couples to the particle’s instantaneous director \(\hat{\mathbf u}(t)\). If the contrast is orientation-insensitive, the rotational contribution does not appear in the ISF and only the translational mode \(\Gamma_2(q)=D_t q^2\) is observed. Orientation-dependent contrast can be achieved, for instance, by (i) dark-field detection (dark field DDM or DF-DDM~\cite{cerbino2017dark}); (ii) polarization optics that isolates the depolarized component (polarized DDM or P-DDM~\cite{giavazzi2016simultaneous,kamal2024dynamics}); or (iii) orientation-sensitive image pre-processing that converts orientation changes into measurable intensity features (squared-gradient DDM or SG-DDM~\cite{giavazzi2021probing}).
SG-DDM works well when the anisotropic particles are not too small compared to the microscope resolution. P-DDM requires sufficient optical anisotropy and depolarization signal. DF-DDM is often a robust alternative under weak depolarization or for small scattering objects.

\begin{figure}
    \centering
    \includegraphics[width=\columnwidth]{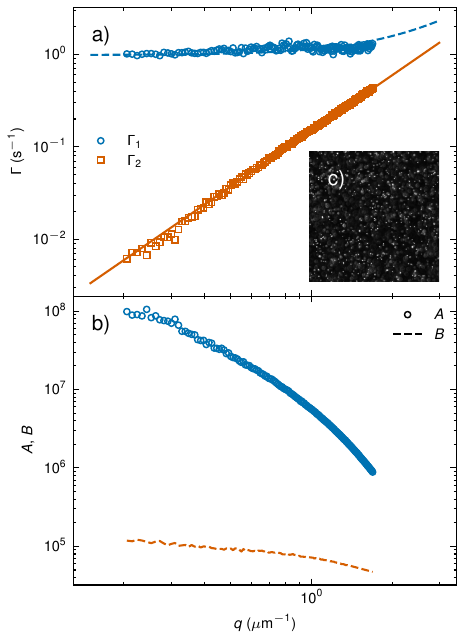}
    \caption{a) Relaxation rates vs $q$ for non-motile bacteria, obtained by dark-field DDM. Dashed and solid lines are fits to $\Gamma_1(q) = 6D_r + D_t q^2$ and $\Gamma_2(q) = D_t q^2$, respectively. \emph{Note the near $q$-independent plateau of $\Gamma_1$ at low $q$}, reflecting the dominant rotational term $6D_r$. b) Corresponding $A(q)$ (symbols) and $B(q)$ (dashed line). c) Dark-field image of non-motile bacteria.}
    \label{fig:df_gamma_ab}
\end{figure}

\textit{Results.} As expected for a non-active system of anisotropic particles, the intermediate scattering function departs from a single-exponential form and exhibits two relaxation modes corresponding to roto–translational diffusion. Relaxation rates extracted from dark-field DDM are shown in Fig.~\ref{fig:df_gamma_ab}a, with \(A(q)\) and \(B(q)\) in Fig.~\ref{fig:df_gamma_ab}b; a representative image is in Fig.~\ref{fig:df_gamma_ab}c. The second mode \(\Gamma_2(q)\) follows a quadratic law, yielding \(D_t=(0.149 \pm 0.001)\unit{\mu m^2 / s}\). At low wavevectors, \(\Gamma_1(q)\) is nearly $q$-independent (dominated by the rotational term \(6D_r\)) and crosses over toward the \(D_t q^2\) contribution at higher \(q\). Fitting \(\Gamma_1(q)=6D_r + D_t q^2\) with \(D_t\) fixed gives \(D_r=(0.165 \pm 0.002)\unit{s^{-1}}\). Both estimates agree with Ref.~\onlinecite{cerbino2017dark}.

\textit{Methods.} Dark-field microscopy image sequences of non-motile, non-flagellated \emph{E.~coli} DH5$\alpha$ were the same used for Ref.~\onlinecite{cerbino2017dark}. The sample was a suspension in PBS; to avoid sedimentation, Percoll\textsuperscript{\textregistered} was added to match the solvent density to that of the bacteria. Percoll\textsuperscript{\textregistered} consists of very small (diameter $\sim 15$--$30\unit{nm}$) colloidal silica particles, routinely used for density-gradient centrifugation\footnote{Due to their small size, the intensity of the light scattered by Percoll\textsuperscript{\textregistered} is negligible compared to that scattered by the bacteria, as checked in Ref.~\onlinecite{cerbino2017dark}.}; their non-toxicity, owing to polyvinylpyrrolidone coating, makes them suitable for biological materials~\cite{pertoft2000fractionation}. To avoid depletion interactions due to Percoll\textsuperscript{\textregistered}, the concentration of bacteria was kept very low: $\sim 10^5\unit{bacteria / mL}$, corresponding to a volume fraction $\phi_0 \simeq 2\times 10^{-7}$. The sample was confined into a $0.3 \times 1 \times 20\unit{mm}$ rectangular glass capillary, carefully sealed on both ends onto a microscope slide with petroleum jelly to prevent evaporation. The solvent viscosity at the experimental temperature, measured with a capillary viscometer, was $\eta = (1.87 \pm 0.02) \times 10^{-3}\unit{Pa \, s}$. Dark-field images were acquired using a Nikon Eclipse Ti-E microscope equipped with an Orca Flash 4.0 v2 (Hamamatsu) fast CMOS camera at $T=24\celsius$. Approximately $10^4$ frames (512$\times$512 pixels after $2\times 2$ binning) were recorded at $20\unit{fps}$ using a $10\times$ objective, $\mathrm{NA}=0.15$, yielding an effective pixel size $\delta_{\mathrm{px}} = 1.29\unit{\mu m}$. Illumination used a condenser stage ($\mathrm{NA_c}=0.4$) coupled with a PH3 phase-contrast ring mask. Following Sec.~\ref{sec:particle-sizing}, structure functions were computed and fitted with Eq.~\eqref{eq:isf-bacteria-rotation} (via the model for $d(q,\Delta t)$ in Eq.~\eqref{eq:dqt-model}) to extract $\Gamma_{1,2}(q)$, $A(q)$, and $B(q)$; \(D_t\) and \(D_r\) were determined from the $q$-dependence of $\Gamma_{2}$ and $\Gamma_{1}$, respectively.

\subsection{\label{subsec:ddm-microrheology} \href{https://github.com/somexlab/fastddm-tutorials/blob/main/Tutorial_9-Microrheology/tutorial9.ipynb}{\faGithub} From Brownian motion to rheological moduli: DDM as a tool for microrheology}
\textit{Premise.} Many soft materials such as gels, creams, biofluids, and polymer solutions are neither purely viscous (Newtonian) liquids nor perfectly elastic (Hookean) solids. Instead, they exhibit a combination of energy storage and dissipation when subjected to deformation. These materials are termed \emph{viscoelastic}, and their rheological response typically depends on both the amplitude and timescale of the imposed perturbation. Conventional rheology measures a material stress response to a controlled deformation, or vice versa. In the linear response regime — where deformations remain small enough to avoid nonlinear effects — the material's behavior is characterized by the complex modulus
\begin{equation*}
G^\star(\omega) = G'(\omega) + \mathrm{i} G''(\omega)~,
\end{equation*}
where $G'(\omega)$ and $G''(\omega)$ are the storage and loss moduli. For example, in an oscillatory strain experiment with $\gamma(t) = \gamma_{\text{max}} \sin(\omega t)$, the resulting stress signal is
\begin{equation*}
\sigma(t) = G' \gamma_{\text{max}} \sin(\omega t) + G'' \gamma_{\text{max}} \cos(\omega t)~.
\end{equation*}
Loosely speaking, if for a given frequency region $G' \gg G''$, the material behaves primarily as a solid; conversely, if $G'' \gg G'$, the behavior is liquid-like.

Despite their versatility, standard rheometers have several limitations. They require relatively large sample volumes (typically milliliters), probe only bulk-average properties, and are constrained to limited frequency ranges due to torque sensitivity (low $\omega$) and inertial effects (high $\omega$).

\textit{Aim and goals.} Here, we use data from Ref.~\onlinecite{escobedo2018microliter} to determine the viscoelastic response of a polymer solution. We aim at establishing DDM microrheology as a minimally invasive route to \emph{linear} viscoelastic moduli over broad time/frequency windows using microliters of sample seeded with tracer particles. \footnote{Crucially, the signal must originate from tracer particles; therefore the medium must be sufficiently transparent (for the chosen imaging modality) so that tracer-induced intensity fluctuations dominate. Alternatively, one could use fluorescent tracers to study samples whose intrinsic scattering is not negligible}. Specifically: (i) obtain $G'(\omega)$ and $G''(\omega)$ from the tracer mean-square displacement (MSD) inferred from DDM; (ii) validate against conventional oscillatory rheometry and DLS microrheology.

\textit{New concepts.} Microrheology infers mechanical properties from the stochastic motion of embedded tracers. For tracers that probe the continuum response (radius $R$ large compared to the microstructure; no slip; linear response), the generalized Stokes–Einstein relation (GSER) links the Laplace-transformed MSD to the complex modulus~\cite{mason1995optical}:
\begin{equation}
\label{eq:generalized-SE}
G^{\star} (\omega) = \left. \frac{2 k_{\mathrm{B}} T}{3 \pi R s \,\langle \Delta \tilde{r}^2 (s) \rangle} \right\rvert_{s=\mathrm{i}\omega} ,
\end{equation}
where $\langle \Delta \tilde{r}^2(s) \rangle$ is the Laplace transform of the 2D MSD $\langle \Delta r^2(\Delta t) \rangle$, and $s$ is the Laplace frequency. A very popular microrheology approach (particle-tracking microrheology) quantifies the tracer MSD by following the motion of the particles in time during microscopy experiments. Alternatively, one could use DDM to analyze microscopy videos of the tracer motions and extract the MSD from Eq.~\eqref{eq:logf-msd}\footnote{A similar procedure is used in DLS and DWS microrheology, where the relevant equations need to account that these techniques probe the 3D MSD}, without need for the tracers to be individually resolved. DDM microrheology has been implemented with various microscopy modalities and tracer systems~\cite{bayles2017probe, edera2017differential, escobedo2018microliter}.

\begin{figure}
    \centering
    \includegraphics[width=\columnwidth]{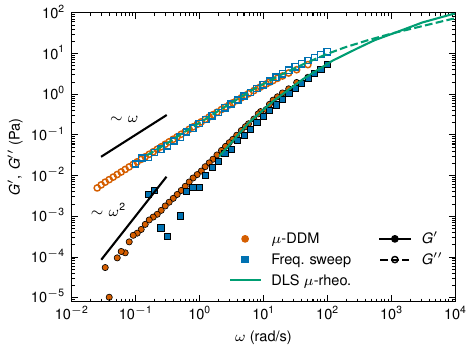}
    \caption{Storage modulus $ G'(\omega)$ and loss modulus $ G''(\omega)$ of a 2.1\% PEO solution measured by DDM microrheology ($\mu$-DDM), standard oscillatory rheology, and DLS microrheology, as indicated in the legend.}
    \label{fig:microrheology-G}
\end{figure}

\textit{Results.} DDM microrheology was applied to a viscoelastic aqueous PEO solution (details below). Following the standard workflow, the azimuthally averaged $D(q,\Delta t)$ and in turn the intermediate scattering function $f_{R}(q,\Delta t)$ were computed; the MSD was obtained from Eq.~\eqref{eq:logf-msd}, and $G'(\omega), G''(\omega)$ were derived using Eq.~\eqref{eq:generalized-SE}. Figure~\ref{fig:microrheology-G} shows the storage modulus $G'(\omega)$ and the loss modulus $G''(\omega)$ obtained by DDM microrheology $\mu$-DDM) and compares them with results from conventional rheometry (oscillatory frequency sweep) and DLS microrheology. The DDM data reveal a characteristic crossover between the elastic and viscous regimes: at low frequencies, $G'' > G'$, indicating that the system flows like a liquid; at higher frequencies, $G'$ increases and eventually surpasses $G''$, suggesting that the medium responds more elastically on short timescales.

All three methods provide quantitatively consistent results across the accessible frequency range, validating the reliability of DDM microrheology even at low frequencies, where rheometers are typically limited by torque sensitivity. Notably, DDM microrheology extends the accessible frequency range downwards compared to both DLS and mechanical rheometry, offering a significant advantage when probing slowly relaxing systems.

\textit{Methods.} The sample was an aqueous solution of poly(ethylene oxide) (PEO, $M_\mathrm{w}=900\unit{kDa}$, Sigma-Aldrich) at mass fraction $c=2.1\%$, prepared by gentle stirring at $T=40\celsius$ to fully dissolve the polymer~\cite{escobedo2018microliter}. Polystyrene tracer particles (Invitrogen) of diameter $330\unit{nm}$ were added to a final volume fraction $\phi_0=7.5\times10^{-4}$. The dispersion was loaded into a rectangular glass capillary with internal cross-section $0.2 \times 2 \unit{mm}$ (VitroCom). Bright-field image sequences were acquired on a Nikon Eclipse Ti-E microscope with a Mako-U130 (Allied Vision Technologies) CMOS camera at $T=20\celsius$. Videos comprised $1.25 \times 10^5$ frames, $256 \times 256$ pixels, acquired at $100\unit{fps}$ using a $20\times$, $\mathrm{NA}=0.5$ objective, yielding $\delta_{\mathrm{px}}=0.24\unit{\mu m}$. Analysis followed Secs.~\ref{sec:particle-sizing} and~\ref{sec:particle-sizing-edinburgh} to compute the azimuthally averaged structure function $d(q, \Delta t)$. From this, we extracted the MSD via Eq.~\eqref{eq:logf-msd} and then determined the viscoelastic moduli using Eq.~\eqref{eq:generalized-SE}.

While the method appears straightforward, there are at least two issues that deserve attention:
\begin{itemize}
    \item Once $\langle \Delta r^2(\Delta t) \rangle$ is obtained, direct use of instead of Eq.~\eqref{eq:generalized-SE} can prove challenging due to the numerically ill-conditioned nature of the direct Laplace inversion~\cite{evans2009direct, marple1987digital, vyas2016passive}. One popular strategy, which is the one used also in this work, is to use an approximate procedure based on \emph{local power-law fits} of the MSD~\cite{mason2000estimating}. While less accurate than other approaches, this procedure is usually easier to implement for first time microrheology users.
    \item Obtaining $f_{R}(q,\Delta t)$ from $D(q,\Delta t)$ is highly sensitive to the accurate determination of the amplitude $A(q)$ and baseline $B(q)$ of the latter. If a model for the MSD is known, it is advisable to fit the data leaving $A(q)$ and $B(q)$ as free parameters. Otherwise, incorrect estimates of these quantities will propagate into errors in $G^\star(\omega)$. Several strategies have been proposed to mitigate this problem. One approach relies on an iterative refinement of $A(q)$ and $B(q)$ to minimize the dispersion in the derived MSD across different $q$ values~\cite{edera2017differential}. More advanced uncertainty-aware estimators have also been proposed~\cite{gu2021uncertainty, lennon2023data}. In the present case, the large number of frames enables a reliable direct estimation of $A(q)$ and $B(q)$. We followed the procedure outlined in Sec.~\ref{sec:particle-sizing-edinburgh}, focusing on the wavevector range $q \in [1.78, 4.74]\,\unit{\mu m^{-1}}$.
\end{itemize}  

Together, these results demonstrate the utility of DDM microrheology as a minimally invasive, high-throughput tool for probing \emph{linear} viscoelastic properties over a wide dynamic range, using only microliters of sample and without requiring mechanical actuation or particle tracking.

\subsection{\label{subsec:cells-time-dependence} \href{https://github.com/somexlab/fastddm-tutorials/blob/main/Tutorial_10-Cell_monolayer/tutorial10.ipynb}{\faGithub} Cellular rearrangements in confluent monolayers}

\textit{Premise.} Biological tissues composed of confluent, densely packed cells continuously undergo division, apoptosis, and shape fluctuations that shape their collective behavior~\cite{park2015unjamming,devany2021cell,atia2018geometric}. Tracking individual cells in such non-stationary, crowded environments is challenging even \emph{in vitro}. DDM offers a robust alternative to particle tracking and particle image velocimetry (PIV), extracting quantitative structural and dynamical information at the single-cell level from intensity fluctuations~\cite{giavazzi2018tracking,fajner2021hecw}.

\textit{Aim and goals.} Quantify the ageing (time-evolving) dynamics of confluent epithelial monolayers by (i) computing time-resolved mean-square displacements (MSDs) from DDM, (ii) identifying the slowdown associated with crowding/jamming, and (iii) fitting a long-time power-law MSD to extract the scaling exponent $\beta$ and an effective diffusivity $D_{\mathrm{eff}}$, with careful control of intercellular spatial correlations via the selection of wavevectors ensuring that the condition $S(q)=1$ is met.

\textit{New concepts.} 
\begin{itemize}
    \item Classical DDM analysis assumes statonarity of the dynamics. In non-stationary systems, we can analyze short time chunks for which the stationarity assumption holds, and concatenate chunk-wise analyses to obtain the time evolution of dynamical parameters (“ageing”).  
    \item The intermediate scattering function decomposes into self and distinct parts (see Eq.~\eqref{eq:selfdist}). To focus on single-cell (self) motion and minimize density-correlation effects, analysis at wavevectors where the static structure factor satisfies $S(q)\!\approx\!1$ is advantageous. Here, $S(q)$ is estimated from segmented cell positions.  
    \item Long-time MSDs are modeled as
\begin{equation}
\label{eq:msd-cells}
\langle \Delta r^2 (\Delta t) \rangle = \Delta r^2_0 + D_{\text{eff}} \,\Delta t^{\beta},
\end{equation}
where $\Delta r_0^2$ accounts for short-time localization/measurement offsets; $\beta>1$ indicates superdiffusion.
\end{itemize}

\begin{figure}
    \centering
    \includegraphics[width=\columnwidth]{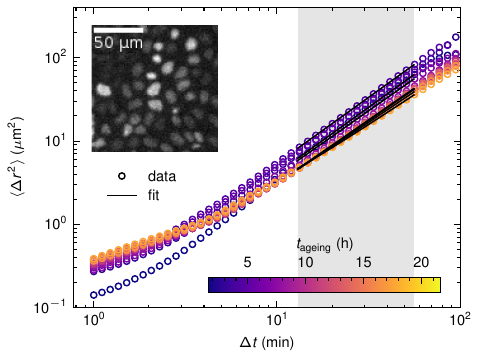}
    \caption{Mean square displacement (symbols) extracted via DDM at different ageing times $t_{\text{ageing}}$ (color-coded). Continuous lines are fits of Eq.~\eqref{eq:msd-cells} over the shaded $\Delta t$ interval. Inset: Fluorescence image (100\,$\times$\,100 pixels) of MDCK cell nuclei.}
    \label{fig:cells-msd}
\end{figure}

\begin{figure}
    \centering
    \includegraphics[width=\columnwidth]{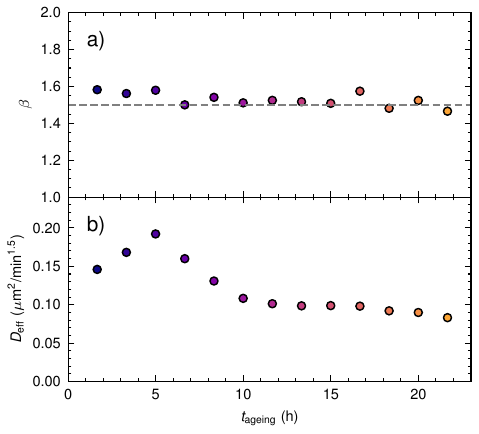}
    \caption{Temporal evolution of (a) the scaling exponent $\beta$ and (b) the prefactor $D_{\text{eff}}$ in Eq.~\eqref{eq:msd-cells}. Color codes correspond to Fig.~\ref{fig:cells-msd}.}
    \label{fig:cells-beta-deff}
\end{figure}

\textit{Results.} Wide-field fluorescence datasets of confluent MDCK monolayers (details below) were analyzed in overlapping, partially shifted windows (200 frames per chunk; shift 100 frames) to resolve ageing. For each chunk and field of view (FOV), $D(q,\Delta t)$ was computed and azimuthally averaged; $S(q)$ was obtained from segmented cell nuclei and used to select wavevectors with $S(q)\!=\!1$ so as to suppress distinct contributions. The MSD was then computed via Eq.~\eqref{eq:logf-msd}.  
Figure~\ref{fig:cells-msd} reports MSDs at increasing ageing times $t_{\mathrm{ageing}}$: the curves do not collapse and progressively slow down, consistent with increasing crowding and an approach to jamming/glassy behavior~\cite{angelini2011glass}. Following Ref.~\onlinecite{giavazzi2018tracking}, an extrapolation to $q\!\to\!0$ was obtained from linear fits in the early-$\Delta t$ regime.  
Long-time MSDs (fit window $\Delta t\in[13,56]\,$min) were fitted with Eq.~\eqref{eq:msd-cells}. The fitted exponent remains $\beta$ remains very close to $1.5$ over the entire experiment (Fig.~\ref{fig:cells-beta-deff}a), indicating persistent superdiffusion. To retain consistent physical units for $D_{\mathrm{eff}}$, fits were repeated with $\beta=1.5$ fixed. The corresponding $D_{\mathrm{eff}}$ decreases by approximately a factor of two over 24\,h (Fig.~\ref{fig:cells-beta-deff}b), quantifying the ageing slowdown, starting approximately $5$ h after the beginning of the image acquisition.  
DDM gives practical advantages over tracking in this regime: computational cost scales with image size rather than cell count; user input is minimal; and reproducibility is enhanced.

\textit{Methods.} Wide-field fluorescence microscopy image sequences were taken from Ref.~\onlinecite{giavazzi2018tracking}. MDCK cells were cultured in 6-well plates ($1.5\times 10^6$ cells/well) in complete medium at $T=37^{\circ}\mathrm{C}$ and 5\% CO$_2$, forming uniform confluent monolayers. Imaging used an Olympus IX81 microscope with an Orca-AG (Hamamatsu) CCD. Six randomly selected FOVs were recorded (to minimize spatial bias). For each FOV, $\sim$1400 frames ($672\times 512$ pixels, $2\times 2$ binning) were acquired every $60\,$s over 24\,h with a $10\times$ objective (effective pixel size $\delta_{\mathrm{px}}=1.29\,\mu\mathrm{m}$).  
To assess ageing, each video was split into partially overlapping chunks of 200 frames, offset by 100 frames. For each chunk, the azimuthally averaged structure function was computed and then averaged across FOVs. Cell nuclei were segmented using \texttt{stardist}~\cite{schmidt2018}; static correlations were quantified with \texttt{freud}~\cite{freud2020} to select wavevectors with $S(q)=1$. The MSD was derived from Eq.~\eqref{eq:logf-msd}, with $A(q)\!+\!B(q)$ estimated from the image power spectrum and $B(q)$ taken from the high-$q$ limit. Long-time fits used $\Delta t\in[13,56]\,$min as in Ref.~\onlinecite{giavazzi2018tracking}; $\beta$ and $D_{\mathrm{eff}}$ were obtained by nonlinear least squares, with a second pass at fixed $\beta=1.5$ to report $D_{\mathrm{eff}}$ in consistent units.

\section{Conclusions and Outlook}
\label{sec:conclusions}

We are now at the end of this trip. In this tutorial, we have traced DDM from its theoretical foundations to its state-of-the-art applications, and we have highlighted the open-source \texttt{fastDDM} library as a practical, high-performance tool for data analysis. To conclude, we distill the journey into three parts: the core lessons we hope readers will retain, a few topics we could only touch upon but that merit further exploration, and possible next steps for the community as DDM continues to evolve.
\\
\paragraph*{Landmarks along the way}  
This part gathers the key stops of our trip: the main methodological and conceptual lessons established throughout the paper. 
\begin{itemize}
    \item \textbf{From optics to dynamics.}  
          We used optics to unify microscopy and scattering through the intermediate scattering function, treating both quasi-2D and fully 3D cases and providing practical criteria for when the full 3D formalism is essential.

    \item \textbf{End-to-end workflow.}  
          A reproducible recipe covering experiment design, multi-rate acquisition, windowing, amplitude/noise extraction, and uncertainty quantification, allows newcomers to obtain reliable multiscale dynamics on the first attempt.

    \item \textbf{Computation made easy: \texttt{fastDDM}.}  
          Fast Wiener–Khinchin algorithms and GPU acceleration shrink analysis from hours to seconds, all packaged in an open Python/C++/CUDA library with example notebooks, tests, and reference datasets.

    \item \textbf{Broad validation and community value.}  
          Within one framework, we reproduced particle sizing down to protein sizes, motile and rotational bacterial dynamics in challenging conditions, polymer-solution microrheology over five frequency decades, and ageing in epithelial monolayers. Because DDM needs only a standard microscope and consumer-grade camera, these protocols and code support frontier research, inter-lab benchmarks, and classroom teaching alike.
\end{itemize}

\paragraph*{Detours worth noting}
Not every path could be explored in detail here. This part highlights themes that remain important for DDM but were left aside in this guide.  
\begin{itemize}
    \item \textbf{Interactions and collective dynamics.} With the notable exception of proteins, this guide has focused mainly on single-particle dynamics that could in principle be equally described by mean squared displacements. Yet, DDM has been applied to study direct and hydrodynamic interactions in dense colloids~\cite{lu2012characterizing}, as well as collective relaxation modes in colloidal gels~\cite{cho2020emergence}, dense~\cite{lu2012characterizing} and glassy~\cite{pastore2022multiscale} suspensions, and liquid crystals~\cite{giavazzi2014viscoelasticity,poy2023hidden,sebastian2024distinctive}, where the probed relaxation is intrinsically cooperative. Such cases deserve further dedicated discussion, which we flag here as an important complement to the present tutorial.   

    \item \textbf{External fields.}  Another powerful avenue, not covered here, is coupling DDM to controlled perturbations (optical, mechanical, etc.) and fields, enabling the study of driven non-equilibrium dynamics. These approaches highlight the flexibility of DDM but fall outside the present scope. The interested reader can refer for instance to~\cite{reufer2012differential,aime2019probing,pal2020anisotropic,edera2021deformation,richards2021characterising,richards2021particle}.
\end{itemize} 

\paragraph*{Next destinations}
Finally, we sketch where DDM is headed next — possible future routes for methods, applications, and community building.
\begin{itemize}    
    \item \textbf{Volumetric and multimodal DDM.}  Confocal and light-sheet variants of DDM have been demonstrated for colloids and embryos, but remain under-utilised.  Systematic adoption for 3-D cultures and organoids, possibly with dual-colour contrast, will broaden the biological questions addressable by DDM.
    
    \item \textbf{Real-time analysis and adaptive imaging.}  GPU implementations show that on-the-fly DDM is feasible; this could lead to real-time DDM (e.g., for particle sizing or microrheology) as well as to coupling DDM to microscope control to enable feedback-based acquisition protocols.
    
    \item \textbf{Data-driven workflows.}  Machine-learning tools can regularise noisy structure functions, automate amplitude/noise estimation, and classify dynamical regimes, while DDM-derived features provide physics-informed inputs for predictive models of active matter.
    
   \item \textbf{Standardised benchmarks.}  Unlike more established scattering techniques, DDM still lacks widely accepted reference datasets and agreed protocols for assessing accuracy, dynamic range, and resolution. Establishing such benchmarks would enable meaningful inter-laboratory comparisons, support the validation of new algorithms, and strengthen pedagogical use. Because this requires community-wide adoption and shared infrastructure, we regard the creation of benchmark suites as a collective \emph{next destination} rather than an already available resource. The open design of \texttt{fastDDM} provides a natural hub to host and expand such efforts.
    
    \item \textbf{Education and outreach.}  Because DDM can be performed with entry-level microscopes and inexpensive cameras, it is ideally suited for teaching statistical physics, Fourier optics, and data science in undergraduate laboratories.
\end{itemize}

\paragraph*{Final remark}
DDM has evolved from an elegant image-difference technique into a mature framework that combines and connects microscopy and scattering.  By funneling best practices into \texttt{fastDDM} and demonstrating its breadth on publicly available data, we hope to lower the barrier for newcomers and spark innovation across soft matter, biophysics, and beyond.

In short: \emph{download the code and keep exploring}, because in DDM, as in life, dynamic information is always just...one Fourier transform away.

\begin{acknowledgments}
This research was partly funded by the Austrian Science Fund (FWF), grant number M 3250-N. We are grateful to Manuel Escobedo-S\'anchez for providing datasets from microrheology experiments, to Jasmin Di Franco for reporting bugs and testing \texttt{fastDDM}, and to the authors of Ref.~\onlinecite{bradley2023sizing} for publicly sharing the open-source microscopy videos of colloidal suspensions used in Sec.~\ref{sec:particle-sizing-edinburgh}. We also thank Nikon Austria for the loan of the water-immersion objective and Teledyne Photometrics for the loan of the Prime BSI Express CMOS camera. EL acknowledges Lorenzo Rovigatti for insightful discussions. Figures were generated with Matplotlib~\cite{matplotlib} using the SciencePlots style~\cite{SciencePlots}.

\end{acknowledgments}
\section*{Author declarations}
\subsection*{Conflict of Interest}
The authors have no conflicts to disclose.
\subsection*{Author Contributions (CRediT)}
\noindent\textbf{Enrico Lattuada}: Conceptualization, Methodology, Software, Validation, Formal Analysis, Data Curation, Visualization, Writing – Original Draft, Writing – Review \& Editing, Project Administration, Funding Acquisition.
\textbf{Fabian Krautgasser}: Software, Validation, Formal Analysis, Data Curation, Writing – Review \& Editing.
\textbf{Maxime Lavaud}: Software, Formal Analysis, Visualization, Writing – Review \& Editing.
\textbf{Fabio Giavazzi}: Methodology, Resources, Writing – Review \& Editing.
\textbf{Roberto Cerbino}: Conceptualization, Methodology, Supervision, Resources, Writing – Original Draft, Writing – Review \& Editing, Project Administration, Funding Acquisition.

\section*{Data Availability Statement}

The data supporting Sections~\ref{sec:particle-sizing-edinburgh} and~\ref{subsec:error} of this study are openly available in the Edinburgh DataShare repository at \href{https://doi.org/10.7488/ds/3851}{https://doi.org/10.7488/ds/3851}.

All other data supporting the findings of this study are openly available in Phaidra, the institutional repository of the University of Vienna, at \href{https://doi.org/10.25365/phaidra.686}{https://doi.org/10.25365/phaidra.686}.

\appendix
\section{Practical considerations for DDM setups\label{app:practical}}

The DDM methodology is compatible with any microscope–camera combination, and its key strength lies in the fact that it does not require complex setups or expensive instrumentation. This accessibility has facilitated its widespread adoption in both research and teaching contexts, including undegraduate and graduate courses and schools~\cite{ferri2011kinetics,germain2016differential,schatz2023advancing}. Nonetheless, the performance and reliability of DDM measurements still depend significantly on hardware configuration, acquisition strategy, and sample handling. This appendix collects a series of practical recommendations based on our experience with a range of imaging setups. Where appropriate, we point back to the quantitative trends and “take-home messages” discussed in Section~\ref{sec:particle-sizing}.

\vspace{-0.6\baselineskip}
\subsection*{Parameters impacting imaging conditions}
\vspace{-0.8\baselineskip}

\noindent\textbf{Microscope illumination (and alignment).}
Bright-field microscopy is the standard modality for DDM. To achieve high-quality, reproducible data, the microscope should support \emph{Köhler illumination}, which provides control over the spatial coherence of the illumination. This is essential for tuning contrast and optimizing sensitivity, particularly for weakly scattering samples. In typical commercial microscopes, one can vary the \emph{illumination NA} by changing the diameter of the condenser iris, which usually has a graduated selector (see also Section~\ref{sec:particle-sizing}). Once the condenser focal length is known, the NA can be calculated from the iris diameter—either measured directly or from an image of the back focal plane obtained with a Bertrand lens. A one-time calibration of the condenser iris (or back-focal-plane image) suffices; the calibrated setting can be reused thereafter. Although systematic studies of non-Köhler or misaligned illumination are scarce, such conditions primarily degrade field uniformity and the effective optical transfer, biasing $A(q)$ and reducing the accessible \(q\)-range.

Regarding the light source, replacing halogen lamps with LEDs improves \emph{intensity stability} and can extend the accessible lag-time range $\Delta t$ by reducing temporal noise. If spectral filtering is used, prefer a moderate bandwidth to minimize the intensity drop. We strongly discourage the use of lasers in bright-field (or other transmission DDM modes) because their high temporal and spatial coherence introduces interference patterns that are difficult to remove as they are often time-dependent.

\vspace{0.4\baselineskip}
\noindent\textbf{Camera selection and acquisition rate.}
Camera choice in DDM is a compromise between frame rate and sensitivity. Scientific CMOS (sCMOS) cameras offer an excellent balance: high speed, large dynamic range, low noise, and flexible region-of-interest selection. These features are particularly valuable when studying:
\begin{itemize}
    \item fast dynamics of dilute suspensions of nanoparticles or macromolecules,
    \item weakly scattering or low-concentration samples,
    \item time-resolved or high-throughput DDM experiments.
\end{itemize}
For slow dynamics (e.g., colloidal gels, cell monolayers), high frame rates are not required. As illustrated in this manuscript, combining acquisitions at different frame rates extends the dynamic range of lag times.

Two practical notes often asked by users: (i) \emph{monochrome vs.\ color sensors}—for quantitative DDM, monochrome sensors are preferred (no color filter array/demosaicing, higher quantum efficiency); color cameras can be used but the Bayer mosaic may imprint weak periodic artifacts in $q$. (ii) \emph{On-chip binning vs.\ full frame}—binning increases SNR and maximum frame rate (lower read noise per effective pixel) at the cost of attenuating high-$q$ content; use it when your analysis focuses on intermediate/low $q$ or when speed is critical. A practical note on cost: within a given sensor class, gains in DDM signal-to-noise with camera price tend to saturate once low read noise, high quantum efficiency, and adequate frame rate are met; beyond that threshold, investing in optics, illumination stability, or environmental control often yields larger returns.

\vspace{0.4\baselineskip}
\noindent\textbf{Exposure time and irradiance.}
Keeping exposure time constant across comparative runs helps stabilize background and noise statistics. To optimize SNR without motion blur, choose exposure so that the fastest dynamics to be measured (typically at the highest $q$ of interest) are not averaged out: a practical rule is to set the exposure shorter than the characteristic decay time at $q_{\max}$ (see also Section~\ref{sec:particle-sizing}). You can trade exposure time and illumination intensity to keep the mean image level constant; verify that $B(q)$ remains stable when doing so.

\vspace{0.4\baselineskip}
\noindent\textbf{Microscope objectives.}
Objective selection also involves trade-offs. High numerical aperture (NA) objectives improve signal contrast and spatial resolution but often have short working distances, limiting compatibility with other components of the setup, such as microscope heating stages. For objectives of 10\(\times\) or 20\(\times\) magnification, standard working distances are usually sufficient. At higher magnifications (40\(\times\), 60\(\times\), 100\(\times\)), long working distance (LWD) or extra-long working distance (ELWD) objectives may be necessary. One can begin with moderately corrected, mid-range objectives and progressively upgrade as needed.

\vspace{0.4\baselineskip}
\noindent\textbf{Contrast modes and add-ons.}
Phase-contrast (PC) microscopy is advantageous when imaging large biological objects (e.g., cells), but provides limited benefit for sub-resolution systems. In our experience, bright-field imaging with a fully closed condenser aperture can yield comparable contrast for many samples. PC rings are inexpensive accessories and allow quick switching between imaging modalities. If PC rings are available, \emph{dark-field} imaging can be achieved without specialized objectives, as a PC ring with NA larger than the objective NA blocks the direct beam and allows only scattered light to form the image~\cite{cerbino2017dark}. This basic dark-field mode is useful for detecting small particles and rotating objects (see also Section~\ref{subsubsec:non-motile bacteria}).

\vspace{0.4\baselineskip}
\noindent\textbf{Fluorescence (including confocal and light-sheet).}
For fluorescence-based DDM, suitable excitation sources (e.g., LEDs, arc lamps, or lasers) and appropriate filter sets are required. Fluorescence imaging uses \emph{epi}-illumination (illumination from the same side as observation), typically via a dichroic mirror in the objective light path. While confocal microscopy can be combined with DDM to enable optical sectioning or 3D dynamic measurements, the associated cost and complexity often exceed the needs of standard DDM applications. \emph{Confocal} DDM (ConDDM) is particularly useful for dense or optically opaque systems because it reduces multiple scattering. At small wave vectors, however, ConDDM becomes sensitive to slice-crossing \emph{number fluctuations} within the optical section (see Ref.~\onlinecite{lu2012characterizing}), which should be taken properly into account when analyzing data. An alternative fluorescence-based approach compatible with DDM is \emph{light-sheet microscopy}, which illuminates the sample with a thin optical sheet orthogonal to the detection axis. Light-sheet DDM~\cite{wulstein2016light} enables high-contrast imaging with reduced photobleaching and is particularly well suited for volumetric or long-duration studies of living systems; however, it requires specialized sample mounting and alignment and is best suited to advanced imaging platforms. Finally, \emph{photobleaching} can affect all fluorescence-based DDM measurements. Mitigate by minimizing irradiance (neutral-density filters, shorter duty cycles, shuttering between frames), choosing exposure times that avoid unnecessary dose, and using dyes with improved photostability. Always record intensity-vs.-time traces; when residual bleaching is present, analyze sub-stacks, normalize where appropriate, and verify that inferred $A(q)$ and $B(q)$ remain consistent across the run.

\vspace{0.4\baselineskip}
\noindent\textbf{Field of view and pixel resolution.}
A larger field of view at fixed pixel size grants access to smaller wavevectors $q$ and increases the number of independent $q$ values available for azimuthal averaging, thereby improving statistical reliability. Conversely, restricting the region of interest allows higher frame rates—essential for capturing fast dynamics—but comes at the cost of reduced statistics and limited access to small $q$. The optimal compromise must therefore be chosen according to the timescales of the target dynamics. In all cases, use the highest bit depth supported by the camera and avoid compressed image or video formats, which can introduce artifacts and distort the DDM signal.

\vspace{0.4\baselineskip}
\noindent\textbf{Absolute normalization.}
A distinctive advantage of bright-field DDM is that each image acts as a hologram of the sample formed with partially coherent light. As detailed in Ref.~\onlinecite{cerbino2025real}, Fourier analysis of particle holograms can yield absolute scattering intensities without the usual calibration of the instrument against a reference scatterer (e.g., toluene) needed for far-field light scattering. In practice, however, the microscope transfer function multiplies the true scattering signal, so absolute normalization based purely on hologram Fourier amplitudes is only partially effective. Removing the optical transfer function $T(q)$ typically requires calibration with a known sample or an external estimate of $T(q)$; moreover, as shown in Section~\ref{subsec:3DBF}, the sample itself may alter $T(q)$. We therefore regard absolute-normalization workflows as procedures for advanced users.

\vspace{0.4\baselineskip}
\noindent\textbf{Dark/white frames, and flat-fielding.}
A frequently asked question is whether one should adopt the “dark/white” acquisition protocols familiar from light and X-ray/neutron scattering. While no systematic studies have been published yet on this topic, we can briefly summarize a few key points based on our experience:
\begin{itemize}
  \item \textbf{Dark frames.} Modern sCMOS cameras often perform on-chip dark subtraction and defect masking; many also allow user-initiated updates of the dark reference. Otherwise, one could acquire a short \emph{master dark} (median/mean of $\gtrsim 100$ shuttered frames) and, if needed, subtract it from all images. Because DDM relies on frame differencing, dark subtraction seldom helps in bright-field DDM; it is more useful in fluorescence-based DDM when parts of the field are bright while other regions remain dark, or when fixed-pattern offsets could bias $B(\mathbf{q})$.
  \item \textbf{White (flat-field) frames.} While not required for standard DDM, flat-fielding becomes useful when the sample evolves so slowly that estimating $A(q)$ from time averages is unreliable (nonergodic or quasi-static conditions). In that case, one may estimate $A(q)$ from the \emph{Fourier power spectrum} of sample images; removing illumination inhomogeneities and fixed-pattern features by recording an empty cell—or, preferably, the same cell with \emph{solvent only}—under \emph{identical} exposure/illumination is beneficial.
\end{itemize}

\vspace{0.4\baselineskip}
\noindent\textbf{Estimate of the camera noise $B$}
While in classical far-field scattering dark/white corrections are applied in \emph{reciprocal space}, in DDM they are most naturally applied in \emph{direct space} (to the raw images) before computing structure functions. One notable exception concerns the baseline $B(\mathbf{q})$. In most applications, $B$ is treated as a fitting parameter or estimated with methods briefly described in the Supplemental material. However, one can characterize the camera by measuring $B$ from ``white'' measurements taken at several mean intensities $\overline{I}$ (with fixed exposure/gain), thereby building a calibration curve $B(\mathbf{q};\overline{I})$ specific to the camera. This approach is most effective with temperature-controlled cameras, because $B$ depends on temperature. Once $B(\mathbf{q};\overline{I})$ is calibrated, one may subtract it directly in reciprocal space from the measured image structure functions at the matching mean intensity, thus reducing the number of fit parameters by one. This is an advanced workflow; if adopted, ensure that temperature, exposure, gain, and $\overline{I}$ are matched between calibration and sample runs, and verify post-subtraction the reliability and stability of the procedure.

\vspace{0.8\baselineskip}
\subsection*{Sample-related considerations}
\vspace{-0.8\baselineskip}

\noindent\textbf{Sample preparation and optical cleanliness.}
While DDM enables experiments in the presence of static noise sources (such as dirt on the optical surfaces of the sample cell) that would not be tolerable in DLS, high-sensitivity DDM measurements (e.g., on protein solutions) remain susceptible to spurious scattering from dust or large impurities, due to the fact that the mechanical stability of the setup is never perfect and this may influence the quality of the static noise subtraction procedure on which DDM relies (see also the discussion below about Mechanical stability). In one of the advanced example, using bright-field DDM with a 20\(\times\) objective and an sCMOS camera, we were able to measure the diffusion coefficient of BSA down to 1 mg/mL, where the scattering signal was two-three orders of magnitude weaker than the background noise. This was only possible through rigorous optical cleanliness and long acquisitions. Thorough cleaning of all optical components and sample containers is essential. We recommend in addition one or more of these precautions:
\begin{itemize}
    \item Filtering all aqueous solutions (e.g., through 0.1–0.2 µm filters);
    \item Plasma-cleaning or ethanol-washing glass capillaries or slides;
    \item Using continuous-flow filtering systems for long-duration measurements of dilute protein or small nanoparticle samples.
\end{itemize}

\vspace{0.4\baselineskip}
\noindent\textbf{Sample turbidity and multiple scattering.}
At the opposite end of the signal spectrum, bright-field DDM tolerates more turbidity than DLS~\cite{giavazzi2014viscoelasticity,eitel2020hitchhiker,nixon2022probing}, but beyond a threshold multiple scattering biases both amplitudes and baselines. A practical sanity check is to inspect the intensity histogram of the \emph{difference images}: as long as the histogram is well described by a distribution symmetrically peaked at zero, multiple scattering is likely minimal. 
Alternatively, one can also consider the \textit{transmittance} of the sample, which can be estimated as the ratio between the mean intensity $\overline{I}$ of the images collected in the presence of the sample and the mean intensity $I_0$ corresponding to the case where the sample is replaced by a transparent medium with the same average refractive index. As a rule of thumb, as long as $\overline{I}/I_0\gtrsim0.5$ multiple scattering is typically not an issue \cite{nixon2022probing}.
Beyond lowering concentration, mitigation strategies include shortening the optical path length (thinner capillaries), \emph{reducing spatial coherence} (opening the condenser aperture), \emph{reducing temporal coherence} (using a broader-band source or filters with a wider bandwidth), or switching to confocal DDM~\cite{lu2012characterizing}, if feasible. In any case, even when turbidity cannot be fully mitigated, DDM provides reliable data when other approaches fail~\cite{nixon2022probing} .

\vspace{0.4\baselineskip}
\noindent\textbf{Estimation of $A$ for slowly relaxing and/or non-ergodic samples.}
For slowly relaxing and/or non-ergodic samples, the determination of $A(q)$ from fitting the image structure functions may prove impractical or even impossible. In the absence of stray light, the expectation value of the power spectrum $\left\langle |I(\mathbf{q}, t)|^2 \right\rangle_t$ of the individual images $i(\mathbf{x}, t)$ coincides with $2A(\mathbf{q})+2B(\mathbf{q})$, which makes it possible to estimate $A(\mathbf{q})$ if $B(\mathbf{q})$ is known (see also Estimate of the camera noise $B$ above). To be able to use this procedure exactly as described the cleaning conditions of all the surfaces along the train should be excellent. Under this assumption, the procedure works very well as described for instance in Ref.~\onlinecite{cho2020emergence}, where the authors studied the multiscale dynamics of colloidal gels across aggregation, geometric percolation, and the onset of nonergodicity using DDM, and exploited the single–frame Fourier power spectrum to obtain $A(q)$, and in turn the structure factor $S(q)$, without relying on full signal decorrelation. Unfortunately, this procedure cannot be used in the presence of a non-negligible stray light contribution, a case that would benefit from the "white" acquisition described above. 

\vspace{0.4\baselineskip}
\noindent\textbf{Ageing samples}  
The merging procedure described in Section~\ref{subsec:join} to combine data acquired at different rates may become delicate in time-evolving samples, in which case multi-\(\tau\) acquisition protocols inspired by DLS correlators—typically unnecessary in DDM—can be more convenient to extend the dynamic range~\cite{brown1993dynamic}. However, these protocols are scarcely compatible with saving full image sequences to disk for later processing, which makes them much less appealing for DDM than for DLS, unless one is interested in real-time analysis without storage. An interesting alternative is the efficient variable-delay acquisition scheme devised in Ref.~\onlinecite{philippe_efficient_2016}, which has been shown to work with DDM, covering $\sim 5$ decades while reducing disk usage.  

Beyond acquisition strategies, it is often crucial to disentangle intrinsic dynamical evolution from lack of stationarity. In such cases, Time Resolved Correlation (TRC) algorithms provide a complementary approach, focusing on temporal fluctuations of the correlation function at a fixed delay time~\cite{cipelletti2002time,duri2005time}. Originally developed in light scattering, TRC has been successfully adapted to imaging techniques to characterize intermittent or aging dynamics in colloidal glasses and gels. TRC-based observables give direct access to dynamical heterogeneity and rare events that are otherwise averaged out. Incorporating TRC analysis into \texttt{fastDDM} would therefore extend its capabilities towards non-stationary systems, where quantifying temporal heterogeneity is as important as the average relaxation itself.  

\vspace{0.8\baselineskip}
\subsection*{Impact of external parameters (mechanical and thermal)}
\vspace{-0.8\baselineskip}

\noindent\textbf{Mechanical stability and drift.}
Mechanical drift—especially lateral stage motion—can introduce artifacts in DDM, particularly at low $q$ or long $\Delta t$. While drift may average out over time in some cases, it can also mimic or obscure real dynamics. We recommend:
\begin{itemize}
    \item Ensuring rigid mechanical coupling between the sample stage and the microscope body,
    \item Using anti-vibration tables for high-sensitivity measurements,
    \item Performing frame-to-frame drift correction (registration) when necessary, especially in time-resolved or ageing experiments,
    \item Comparing with a drift-insensitive analysis (e.g., after image registration or using the so-called far-field DDM~\cite{philippe_efficient_2016} analysis scheme) when suspecting anomalies,
    \item Opting, whenever possible, for liquid-cooled cameras, as the mechanical vibrations generated by rotating fans in air-cooled systems can introduce artifacts during fast acquisitions, particularly when using high-magnification objectives.
\end{itemize}

\vspace{0.4\baselineskip}
\noindent\textbf{Thermal control and illumination-induced effects.}
If the dynamics are temperature-dependent, thermal control is important. However, adding a thermal stage can limit the optical access to the sample both along the optical axis and in the transverse direction and may introduce thermal gradients. Allow time for thermal equilibration and use immersion objectives or coverslip correction collars when necessary. High irradiance can heat the sample and drive \emph{convective flows}; diagnostics include slow drifts visible by eye inspection, and directional bias in DDM results (often accompanied by fringes oriented perpendicular to the direction of motion in the 2D structure functions) in the $q$-resolved signal. Mitigation strategies are to reduce irradiance, use neutral-density filters, spread the illumination, or employ light-sheet/confocal sectioning to limit dose.

\bibliography{lattuada_main}

@PREAMBLE{
 "\providecommand{\noopsort}[1]{}" 
 # "\providecommand{\singleletter}[1]{#1}%" 
}

@article{cipelletti2002time,
  title={Time-resolved correlation: a new tool for studying temporally heterogeneousdynamics},
  author={Cipelletti, Luca and Bissig, Hugo and Trappe, Veronique and Ballesta, Pierre and Mazoyer, Sylvain},
  journal={Journal of Physics: Condensed Matter},
  volume={15},
  number={1},
  pages={S257},
  year={2002},
  publisher={IOP Publishing}
}

@article{duri2005time,
  title={Time-resolved-correlation measurements of temporally heterogeneous dynamics},
  author={Duri, Agnes and Bissig, Hugo and Trappe, V{\'e}ronique and Cipelletti, Luca},
  journal={Physical Review E—Statistical, Nonlinear, and Soft Matter Physics},
  volume={72},
  number={5},
  pages={051401},
  year={2005},
  publisher={APS}
}

@article{winzor2021quantifying,
  title={Quantifying the concentration dependence of sedimentation coefficients for globular macromolecules: a continuing age-old problem},
  author={Winzor, Donald J and Dinu, Vlad and Scott, David J and Harding, Stephen E},
  journal={Biophysical Reviews},
  volume={13},
  number={2},
  pages={273--288},
  year={2021},
  publisher={Springer}
}

@article{saluja2010diffusion,
  title={Diffusion and sedimentation interaction parameters for measuring the second virial coefficient and their utility as predictors of protein aggregation},
  author={Saluja, Atul and Fesinmeyer, R Matthew and Hogan, Sabine and Brems, David N and Gokarn, Yatin R},
  journal={Biophysical journal},
  volume={99},
  number={8},
  pages={2657--2665},
  year={2010},
  publisher={Elsevier}
}

@article{meechai1999translational,
  title={Translational diffusion coefficients of bovine serum albumin in aqueous solution at high ionic strength},
  author={Meechai, Nispa and Jamieson, Alex M and Blackwell, John},
  journal={Journal of colloid and interface science},
  volume={218},
  number={1},
  pages={167--175},
  year={1999},
  publisher={Elsevier}
}

@book{hansen2013theory,
  title={Theory of simple liquids: with applications to soft matter},
  author={Hansen, Jean-Pierre and McDonald, Ian Ranald},
  year={2013},
  publisher={Academic press}
}

@article{edera2021deformation,
  title={Deformation profiles and microscopic dynamics of complex fluids during oscillatory shear experiments},
  author={Edera, Paolo and Brizioli, Matteo and Zanchetta, Giuliano and Petekidis, George and Giavazzi, Fabio and Cerbino, Roberto},
  journal={Soft Matter},
  volume={17},
  number={37},
  pages={8553--8566},
  year={2021},
  publisher={Royal Society of Chemistry}
}

@article{richards2021particle,
  title={Particle sizing for flowing colloidal suspensions using flow-differential dynamic microscopy},
  author={Richards, James A and Martinez, Vincent A and Arlt, Jochen},
  journal={Soft Matter},
  volume={17},
  number={14},
  pages={3945--3953},
  year={2021},
  publisher={Royal Society of Chemistry}
}

@article{richards2021characterising,
  title={Characterising shear-induced dynamics in flowing complex fluids using differential dynamic microscopy},
  author={Richards, James A and Martinez, Vincent A and Arlt, Jochen},
  journal={Soft Matter},
  volume={17},
  number={39},
  pages={8838--8849},
  year={2021},
  publisher={Royal Society of Chemistry}
}

@article{poy2023hidden,
  title={Hidden traces of chirality in the fluctuations of a fully unwound cholesteric},
  author={Poy, Guilhem},
  journal={Soft Matter},
  volume={19},
  number={6},
  pages={1115--1130},
  year={2023},
  publisher={Royal Society of Chemistry}
}

@article{pastore2022multiscale,
  title={Multiscale heterogeneous dynamics in two-dimensional glassy colloids},
  author={Pastore, Raffaele and Giavazzi, Fabio and Greco, Francesco and Cerbino, Roberto},
  journal={The Journal of Chemical Physics},
  volume={156},
  number={16},
  year={2022},
  publisher={AIP Publishing}
}

@article{sebastian2024distinctive,
  title={Distinctive features of pretransitional behaviour between nematic phases as revealed by DDM},
  author={Sebasti{\'a}n, Nerea and Cmok, Luka and Petelin, Andrej and Mandle, Richard J and Mertelj, Alenka},
  journal={Liquid Crystals},
  volume={51},
  number={6},
  pages={1047--1063},
  year={2024},
  publisher={Taylor \& Francis}
}

@article{mailer2015particle,
  title={Particle sizing by dynamic light scattering: non-linear cumulant analysis},
  author={Mailer, Alastair G and Clegg, Paul S and Pusey, Peter N},
  journal={Journal of Physics: Condensed Matter},
  volume={27},
  number={14},
  pages={145102},
  year={2015},
  publisher={IOP Publishing}
}

@article{philippe_efficient_2016,
    title = {An efficient scheme for sampling fast dynamics at a low average data acquisition rate},
    volume = {28},
    url = {https://www.scopus.com/inward/record.uri?eid=2-s2.0-84955485145&doi=10.1088%2f0953-8984%2f28%2f7%2f075201&partnerID=40&md5=e7fca798750c964ba16d491f456f2fec},
    doi = {10.1088/0953-8984/28/7/075201},
    number = {7},
    journal = {Journal of Physics Condensed Matter},
    author = {Philippe, A. and Aime, S. and Roger, V. and Jelinek, R. and Prévot, G. and Berthier, L. and Cipelletti, L.},
    year = {2016},
    note = {8 citations (Crossref) [2024-01-03]},
}

@article{brown1993dynamic,
  title={Dynamic light scattering},
  author={Brown, Wyn},
  journal={(No Title)},
  year={1993},
  publisher={Oxford University PressOxford}
}

@incollection{cerbino2025real,
  title={From Real to Reciprocal Space: Scattering Information From Real Space Images},
  author={Cerbino, Roberto},
  booktitle={Neutrons, X-rays, and Light},
  pages={615--647},
  year={2025},
  publisher={Elsevier}
}

@article{schatz2023advancing,
  title={Advancing access to cutting-edge tabletop science},
  author={Schatz, Michael F and Cicuta, Pietro and Gordon, Vernita D and Pilizota, Teuta and Rodenborn, Bruce and Shattuck, Mark D and Swinney, Harry L},
  journal={Annual Review of Fluid Mechanics},
  volume={55},
  number={1},
  pages={213--235},
  year={2023},
  publisher={Annual Reviews}
}

@article{volk2018density,
  title={Density model for aqueous glycerol solutions},
  author={Volk, Andreas and K{\"a}hler, Christian J},
  journal={Experiments in Fluids},
  volume={59},
  number={5},
  pages={75},
  year={2018},
  publisher={Springer}
}

@article{giavazzi2021probing,
  title={Probing roto-translational diffusion of small anisotropic colloidal particles with a bright-field microscope},
  author={Giavazzi, Fabio and Pal, Antara and Cerbino, Roberto},
  journal={The European Physical Journal E},
  volume={44},
  number={4},
  pages={61},
  year={2021},
  publisher={Springer}
}

@article{kamal2024dynamics,
  title={Dynamics of anisotropic colloidal systems: What to choose, DLS, DDM or XPCS?},
  author={Kamal, Md Arif and Brizioli, Matteo and Zinn, Thomas and Narayanan, Theyencheri and Cerbino, Roberto and Giavazzi, Fabio and Pal, Antara},
  journal={Journal of Colloid and Interface Science},
  volume={660},
  pages={314--320},
  year={2024},
  publisher={Elsevier}
}

@article{reufer2012differential,
  title={Differential dynamic microscopy for anisotropic colloidal dynamics},
  author={Reufer, Mathias and Martinez, Vincent A and Schurtenberger, Peter and Poon, Wilson CK},
  journal={Langmuir},
  volume={28},
  number={10},
  pages={4618--4624},
  year={2012},
  publisher={ACS Publications}
}

@article{martinez2012differential,
  title={Differential dynamic microscopy: a high-throughput method for characterizing the motility of microorganisms},
  author={Martinez, Vincent A and Besseling, Rut and Croze, Ottavio A and Tailleur, Julien and Reufer, Mathias and Schwarz-Linek, Jana and Wilson, Laurence G and Bees, Martin A and Poon, Wilson CK},
  journal={Biophysical journal},
  volume={103},
  number={8},
  pages={1637--1647},
  year={2012},
  publisher={Elsevier}
}

@article{giavazzi2014viscoelasticity,
  title={Viscoelasticity of nematic liquid crystals at a glance},
  author={Giavazzi, Fabio and Crotti, Stefano and Speciale, Antonio and Serra, Francesca and Zanchetta, Giuliano and Trappe, Veronique and Buscaglia, Marco and Bellini, Tommaso and Cerbino, Roberto},
  journal={Soft Matter},
  volume={10},
  number={22},
  pages={3938--3949},
  year={2014},
  publisher={Royal Society of Chemistry}
}

@article{eitel2020hitchhiker,
  title={A Hitchhiker's guide to particle sizing techniques},
  author={Eitel, Kathrin and Bryant, Gary and Sch{\"o}pe, Hans Joachim},
  journal={Langmuir},
  volume={36},
  number={35},
  pages={10307--10320},
  year={2020},
  publisher={ACS Publications}
}

@article{nixon2022probing,
  title={Probing the dynamics of turbid colloidal suspensions using differential dynamic microscopy},
  author={Nixon-Luke, Reece and Arlt, Jochen and Poon, Wilson CK and Bryant, Gary and Martinez, Vincent A},
  journal={Soft Matter},
  volume={18},
  number={9},
  pages={1858--1867},
  year={2022},
  publisher={Royal Society of Chemistry}
}

@article{aime2019probing,
  title={Probing shear-induced rearrangements in Fourier space. II. Differential dynamic microscopy},
  author={Aime, Stefano and Cipelletti, Luca},
  journal={Soft Matter},
  volume={15},
  number={2},
  pages={213--226},
  year={2019},
  publisher={Royal Society of Chemistry},
  doi={https://doi.org/10.1039/C8SM01564C},
}

@article{croccolo2006effect,
  title={Effect of Gravity on the Dynamics of Nonequilibrium Fluctuations in a Free-Diffusion Experiment},
  author={Croccolo, Fabrizio and Brogioli, Doriano and Vailati, Alberto and Giglio, Marzio and Cannell, David S},
  journal={Ann. N. Y. Acad. Sci.},
  volume={1077},
  number={1},
  pages={365--379},
  year={2006},
  publisher={Wiley Online Library},
  doi={https://doi.org/10.1196/annals.1362.030},
}

@article{giavazzi2016structure,
  title={Structure and dynamics of concentration fluctuations in a non-equilibrium dense colloidal suspension},
  author={Giavazzi, Fabio and Savorana, Giovanni and Vailati, Alberto and Cerbino, Roberto},
  journal={Soft Matter},
  volume={12},
  number={31},
  pages={6588--6600},
  year={2016},
  publisher={Royal Society of Chemistry},
  doi={https://doi.org/10.1039/C6SM00935B},
}

@article{cho2020emergence,
  title={Emergence of multiscale dynamics in colloidal gels},
  author={Cho, Jae Hyung and Cerbino, Roberto and Bischofberger, Irmgard},
  journal={Phys. Rev. Lett.},
  volume={124},
  number={8},
  pages={088005},
  year={2020},
  publisher={APS},
  doi={https://doi.org/10.1103/PhysRevLett.124.088005},
}

@article{gao2015microdynamics,
  title={Microdynamics and arrest of coarsening during spinodal decomposition in thermoreversible colloidal gels},
  author={Gao, Yongxiang and Kim, Juntae and Helgeson, Matthew E},
  journal={Soft Matter},
  volume={11},
  number={32},
  pages={6360--6370},
  year={2015},
  publisher={The Royal Society of Chemistry},
  doi={https://doi.org/10.1039/C5SM00851D},
}

@article{martineau2022engineering,
  title={Engineering gelation kinetics in living silk hydrogels by differential dynamic microscopy microrheology and machine learning},
  author={Martineau, Rhett L and Bayles, Alexandra V and Hung, Chia-Suei and Reyes, Kristofer G and Helgeson, Matthew E and Gupta, Maneesh K},
  journal={Adv. Biol.},
  volume={6},
  number={1},
  pages={2101070},
  year={2022},
  publisher={Wiley Online Library},
  doi={https://doi.org/10.1002/adbi.202101070},
}

@article{ferri2011kinetics,
  title={Kinetics of colloidal fractal aggregation by differential dynamic microscopy},
  author={Ferri, F and D’Angelo, A and Lee, M and Lotti, A and Pigazzini, MC and Singh, Kanwarpal and Cerbino, R},
  journal={Eur. Phys. J. Spec. Top.},
  volume={199},
  pages={139--148},
  year={2011},
  publisher={Springer},
  doi={https://doi.org/10.1140/epjst/e2011-01509-9},
}

@book{vandeHulst1981,
  title={Light Scattering by Small Particles},
  author={Van de Hulst, H. C.},
  year={1981},
  publisher={Dover Publications},
  address={New York},
  isbn={978-0486642284}
}

@article{barrat2023soft,
  title={Soft matter roadmap},
  author={Barrat, Jean-Louis and Del Gado, Emanuela and Egelhaaf, Stefan U and Mao, Xiaoming and Dijkstra, Marjolein and Pine, David J and Kumar, Sanat K and Bishop, Kyle and Gang, Oleg and Obermeyer, Allie and others},
  journal={J. Phys.: Mater.},
  volume={7},
  number={1},
  pages={012501},
  year={2023},
  publisher={IOP Publishing},
  doi={https://doi.org/10.1088/2515-7639/ad06cc},
}

@ARTICLE{norouzisadeh2021modern,
  title={The modern structurator: increased performance for calculating the structure function},
  author={Norouzisadeh, Mojtaba and Chraga, Mohammed and Cerchiari, Giovanni and Croccolo, Fabrizio},
  journal={Eur. Phys. J. E},
  volume={44},
  number={12},
  pages={146},
  year={2021},
  publisher={Springer},
  doi={https://doi.org/10.1140/epje/s10189-021-00146-2},
}

@article{germain2016differential,
  title={Differential dynamic microscopy to characterize Brownian motion and bacteria motility},
  author={Germain, David and Leocmach, Mathieu and Gibaud, Thomas},
  journal={Am. J. Phys.},
  volume={84},
  number={3},
  pages={202--210},
  year={2016},
  publisher={American Association of Physics Teachers},
  doi={https://doi.org/10.1119/1.4939516},
}

@article{shokeen2017comparison,
  title={Comparison of nanoparticle diffusion using fluorescence correlation spectroscopy and differential dynamic microscopy within concentrated polymer solutions},
  author={Shokeen, Namita and Issa, Christopher and Mukhopadhyay, Ashis},
  journal={Appl. Phys. Lett.},
  volume={111},
  number={26},
  year={2017},
  publisher={AIP Publishing},
  doi={https://doi.org/10.1063/1.5016062},
}

@article{hitimana2022diffusive,
  title={Diffusive dynamics of charged nanoparticles in convex lens-induced confinement},
  author={Hitimana, Emmanuel and Roopnarine, Brittany K and Morozova, Svetlana},
  journal={Soft Matter},
  volume={18},
  number={4},
  pages={832--840},
  year={2022},
  publisher={Royal Society of Chemistry},
  doi={https://doi.org/10.1039/D1SM01554K},
}

@article{cerbino2022differential,
  title={Differential dynamic microscopy for the characterization of polymer systems},
  author={Cerbino, Roberto and Giavazzi, Fabio and Helgeson, Matthew E},
  journal={J. Polym. Sci.},
  volume={60},
  number={7},
  pages={1079--1089},
  year={2022},
  publisher={Wiley Online Library},
  doi={https://doi.org/10.1002/pol.20210217},
}

@article{al2022differential,
  title={Differential dynamic microscopy for the characterisation of motility in biological systems},
  author={Al-Shahrani, Monerh and Bryant, Gary},
  journal={Phys. Chem. Chem. Phys.},
  volume={24},
  number={35},
  pages={20616--20623},
  year={2022},
  publisher={Royal Society of Chemistry},
  doi={https://doi.org/10.1039/D2CP02034C},
}

@article{safari2015differential,
  title={Differential dynamic microscopy of weakly scattering and polydisperse protein-rich clusters},
  author={Safari, Mohammad S and Vorontsova, Maria A and Poling-Skutvik, Ryan and Vekilov, Peter G and Conrad, Jacinta C},
  journal={Phys. Rev. E},
  volume={92},
  number={4},
  pages={042712},
  year={2015},
  publisher={APS},
  doi={https://doi.org/10.1103/PhysRevE.92.042712},
}

@article{he2012diffusive,
  title={Diffusive dynamics of nanoparticles in aqueous dispersions},
  author={He, Kai and Spannuth, Melissa and Conrad, Jacinta C and Krishnamoorti, Ramanan},
  journal={Soft Matter},
  volume={8},
  number={47},
  pages={11933--11938},
  year={2012},
  publisher={Royal Society of Chemistry},
  doi={https://doi.org/10.1039/C2SM26392K},
}

@article{wulstein2016light,
  title={Light-sheet microscopy with digital Fourier analysis measures transport properties over large field-of-view},
  author={Wulstein, Devynn M and Regan, Kathryn E and Robertson-Anderson, Rae M and McGorty, Ryan},
  journal={Opt. Express},
  volume={24},
  number={18},
  pages={20881--20894},
  year={2016},
  publisher={Optica Publishing Group},
  doi={https://doi.org/10.1364/OE.24.020881},
}

@article{verwei2022quantifying,
  title={Quantifying cytoskeleton dynamics using differential dynamic microscopy},
  author={Verwei, Hannah N and Lee, Gloria and Leech, Gregor and Petitjean, Irene Ist{\'u}riz and Koenderink, Gijsje H and Robertson-Anderson, Rae M and McGorty, Ryan James},
  journal={Journal of visualized experiments: JoVE},
  number={184},
  year={2022},
  publisher={NIH Public Access},
  doi={https://dx.doi.org/10.3791/63931},
}

@article{dzakpasu2004dynamic_I,
  title={Dynamic light scattering microscopy. A novel optical technique to image submicroscopic motions. I: Theory},
  author={Dzakpasu, Rhonda and Axelrod, Daniel},
  journal={Biophys. J.},
  volume={87},
  number={2},
  pages={1279--1287},
  year={2004},
  publisher={Elsevier},
  doi={https://doi.org/10.1529/biophysj.103.033837},
}

@article{dzakpasu2004dynamic_II,
  title={Dynamic light scattering microscopy. A novel optical technique to image submicroscopic motions. II: Experimental applications},
  author={Dzakpasu, Rhonda and Axelrod, Daniel},
  journal={Biophys. J.},
  volume={87},
  number={2},
  pages={1288--1297},
  year={2004},
  publisher={Elsevier},
  doi={https://doi.org/10.1529/biophysj.104.041400},
}

@BOOK{goodman2017introduction,
  title={Introduction to Fourier optics},
  author={Goodman, Joseph W},
  year={2017},
  publisher={Macmillan Learning},
  ISBN={9781319119164},
}

@article{scheffold2007new,
  title={New trends in light scattering},
  author={Scheffold, Frank and Cerbino, Roberto},
  journal={Curr. Opin. Colloid Interface Sci.},
  volume={12},
  number={1},
  pages={50--57},
  year={2007},
  publisher={Elsevier},
  doi={https://doi.org/10.1016/j.cocis.2007.03.005},
}

@article{cerbino2017perspective,
  title={Perspective: Differential dynamic microscopy extracts multi-scale activity in complex fluids and biological systems},
  author={Cerbino, Roberto and Cicuta, Pietro},
  journal={J. Chem. Phys.},
  volume={147},
  number={11},
  year={2017},
  publisher={AIP Publishing},
  doi={https://doi.org/10.1063/1.5001027},
}

@MISC{edinburgh,
    author = {J. J. Bradley and V. A. Martinez and J. Arlt and J. R. Royer and W. C. K. Poon},
    title = {Sizing Multimodal Suspensions with Differential Dynamic Microscopy [dataset]},
    year = {2023},
    howpublished = "\url{https://datashare.ed.ac.uk/handle/10283/4858}",
    note = "[Online; accessed 15-December-2023]",
    doi={https://doi.org/10.7488/ds/3851},
}

@article{bradley2023sizing,
  title={Sizing multimodal suspensions with differential dynamic microscopy},
  author={Bradley, Joe J and Martinez, Vincent A and Arlt, Jochen and Royer, John R and Poon, Wilson CK},
  journal={Soft Matter},
  volume={19},
  number={42},
  pages={8179--8192},
  year={2023},
  publisher={Royal Society of Chemistry},
  doi={https://doi.org/10.1039/D3SM00593C},
}

@ARTICLE{giavazzi2017image,
  title={Image windowing mitigates edge effects in Differential Dynamic Microscopy},
  author={Giavazzi, Fabio and Edera, Paolo and Lu, Peter J and Cerbino, Roberto},
  journal={Eur. Phys. J. E},
  volume={40},
  pages={1--9},
  year={2017},
  doi={https://doi.org/10.1140/epje/i2017-11587-3},
}

@ARTICLE{cerbino2008differential,
  title={Differential dynamic microscopy: probing wave vector dependent dynamics with a microscope},
  author={Cerbino, Roberto and Trappe, Veronique},
  journal={Phys. Rev. Lett.},
  volume={100},
  number={18},
  pages={188102},
  year={2008},
  doi={https://doi.org/10.1103/PhysRevLett.100.188102},
}

@ARTICLE{giavazzi2009scattering,
  title={Scattering information obtained by optical microscopy: differential dynamic microscopy and beyond},
  author={Giavazzi, Fabio and Brogioli, Doriano and Trappe, Veronique and Bellini, Tommaso and Cerbino, Roberto},
  journal={Phys. Rev. E},
  volume={80},
  number={3},
  pages={031403},
  year={2009},
  doi={https://doi.org/10.1103/PhysRevE.80.031403},
}

@ARTICLE{gu2021uncertainty,
  title={Uncertainty quantification and estimation in differential dynamic microscopy},
  author={Gu, Mengyang and Luo, Yimin and He, Yue and Helgeson, Matthew E and Valentine, Megan T},
  journal={Phys. Rev. E},
  volume={104},
  number={3},
  pages={034610},
  year={2021},
  doi={https://doi.org/10.1103/PhysRevE.104.034610},
}

@ARTICLE{cipelletti2003universal,
  title={Universal non-diffusive slow dynamics in aging soft matter},
  author={Cipelletti, Luca and Ramos, Laurence and Manley, Suliana and Pitard, Estelle and Weitz, David A and Pashkovski, Eugene E and Johansson, Marie},
  journal={Faraday Discuss.},
  volume={123},
  pages={237--251},
  year={2003},
  doi={https://doi.org/10.1039/B204495A},
}

@ARTICLE{lu2012characterizing,
  title={Characterizing concentrated, multiply scattering, and actively driven fluorescent systems with confocal differential dynamic microscopy},
  author={Lu, Peter J and Giavazzi, Fabio and Angelini, Thomas E and Zaccarelli, Emanuela and Jargstorff, Frank and Schofield, Andrew B and Wilking, James N and Romanowsky, Mark B and Weitz, David A and Cerbino, Roberto and others},
  journal={Phys. Rev. Lett.},
  volume={108},
  number={21},
  pages={218103},
  year={2012},
  doi={https://doi.org/10.1103/PhysRevLett.108.218103},
}

@ARTICLE{guidolin2023protein,
  title={Protein Sizing with Differential Dynamic Microscopy},
  author={Guidolin, Chiara and Heim, Christopher and Adams, Nathan B P and Baaske, Philipp and Rondelli, Valeria and Cerbino, Roberto and Giavazzi, Fabio},
  journal={Macromolecules},
  volume={56},
  number={20},
  pages={8290--8297},
  year={2023},
  publisher={ACS Publications},
  doi={https://doi.org/10.1021/acs.macromol.3c00782},
}

@ARTICLE{harding1985concentration,
  title={The concentration-dependence of macromolecular parameters},
  author={Harding, Stephen E and Johnson, Paley},
  journal={Biochem. J.},
  volume={231},
  number={3},
  pages={543--547},
  year={1985},
  publisher={Portland Press Ltd.},
  doi={https://doi.org/10.1042/bj2310543},
}

@ARTICLE{connolly2012weak,
  title={Weak interactions govern the viscosity of concentrated antibody solutions: high-throughput analysis using the diffusion interaction parameter},
  author={Connolly, Brian D and Petry, Chris and Yadav, Sandeep and Demeule, Barth{\'e}lemy and Ciaccio, Natalie and Moore, Jamie MR and Shire, Steven J and Gokarn, Yatin R},
  journal={Biophys. J.},
  volume={103},
  number={1},
  pages={69--78},
  year={2012},
  publisher={Elsevier},
  doi={https://doi.org/10.1016/j.bpj.2012.04.047},
}

@ARTICLE{wei2023improved,
  title={Improved Diffusion Interaction Parameter Measurement to Predict the Viscosity of Concentrated mAb Solutions},
  author={Wei, Yangjie and Qi, Wei and Maglalang, Erick and Pelegri-O’Day, Emma M and Luong, Michelle and Razinkov, Vladimir and Sloey, Christopher},
  journal={Mol. Pharmaceutics},
  volume={20},
  number={12},
  pages={6420--6428},
  year={2023},
  publisher={ACS Publications},
  doi={https://doi.org/10.1021/acs.molpharmaceut.3c00797},
}

@ARTICLE{zimm1948scattering,
  title={The scattering of light and the radial distribution function of high polymer solutions},
  author={Zimm, Bruno H},
  journal={J. Chem. Phys.},
  volume={16},
  number={12},
  pages={1093--1099},
  year={1948},
  publisher={American Institute of Physics},
  doi={https://doi.org/10.1063/1.1746738},
}

@ARTICLE{cerbino2017dark,
  title={Dark field differential dynamic microscopy enables accurate characterization of the roto-translational dynamics of bacteria and colloidal clusters},
  author={Cerbino, Roberto and Piotti, Davide and Buscaglia, Marco and Giavazzi, Fabio},
  journal={J. Phys.: Condens. Matter},
  volume={30},
  number={2},
  pages={025901},
  year={2017},
  doi={https://doi.org/10.1088/1361-648X/aa9bc5},
}

@ARTICLE{pertoft2000fractionation,
  title={Fractionation of cells and subcellular particles with Percoll},
  author={Pertoft, H{\aa}kan},
  journal={J. Biochem. Biophys. Methods},
  volume={44},
  number={1-2},
  pages={1--30},
  year={2000},
  doi={https://doi.org/10.1016/S0165-022X(00)00066-X},
}

@MISC{diffmicro,
    author={Cerchiari, Giovanni and Norouzisadeh, Mojtaba and Chraga, Mohammed},
    title={Diffmicro [Version 3.1]},
    year={2021},
    howpublished="\url{https://doi.org/10.5281/zenodo.5720223}",
}

@ARTICLE{giavazzi2014digital,
  title={Digital Fourier microscopy for soft matter dynamics},
  author={Giavazzi, Fabio and Cerbino, Roberto},
  journal={J. Opt.},
  volume={16},
  number={8},
  pages={083001},
  year={2014},
  doi={https://doi.org/10.1088/2040-8978/16/8/083001},
}

@BOOK{berne2000dynamic,
  title={Dynamic light scattering: with applications to chemistry, biology, and physics},
  author={Berne, Bruce J and Pecora, Robert},
  year={2000},
  publisher={Dover Publications},
  ISBN={0486411559},
}

@ARTICLE{bayles2016dark,
  title={Dark-field differential dynamic microscopy},
  author={Bayles, Alexandra V and Squires, Todd M and Helgeson, Matthew E},
  journal={Soft Matter},
  volume={12},
  number={8},
  pages={2440--2452},
  year={2016},
  doi={https://doi.org/10.1039/C5SM02576A},
}

@ARTICLE{brizioli2022reciprocal,
  title={Reciprocal space study of brownian yet non-Gaussian diffusion of small tracers in a hard-sphere glass},
  author={Brizioli, Matteo and Sentjabrskaja, Tatjana and Egelhaaf, Stefan U and Laurati, Marco and Cerbino, Roberto and Giavazzi, Fabio},
  journal={Front. Phys.},
  volume={10},
  pages={893777},
  year={2022},
  doi={https://doi.org/10.3389/fphy.2022.893777},
}

@ARTICLE{pal2020anisotropic,
  title={Anisotropic dynamics and kinetic arrest of dense colloidal ellipsoids in the presence of an external field studied by differential dynamic microscopy},
  author={Pal, Antara and Martinez, Vincent A and Ito, Thiago H and Arlt, Jochen and Crassous, J{\'e}r{\^o}me J and Poon, Wilson CK and Schurtenberger, Peter},
  journal={Sci. Adv.},
  volume={6},
  number={3},
  pages={eaaw9733},
  year={2020},
  publisher={American Association for the Advancement of Science},
  doi={https://doi.org/10.1126/sciadv.aaw9733},
}

@ARTICLE{giavazzi2018tracking,
  title={Tracking-free determination of single-cell displacements and division rates in confluent monolayers},
  author={Giavazzi, Fabio and Malinverno, Chiara and Scita, Giorgio and Cerbino, Roberto},
  journal={Front. Phys.},
  volume={6},
  pages={120},
  year={2018},
  doi={https://doi.org/10.3389/fphy.2018.00120},
}

@ARTICLE{kurzthaler2018probing,
  title={Probing the spatiotemporal dynamics of catalytic Janus particles with single-particle tracking and differential dynamic microscopy},
  author={Kurzthaler, Christina and Devailly, Cl{\'e}mence and Arlt, Jochen and Franosch, Thomas and Poon, Wilson CK and Martinez, Vincent A and Brown, Aidan T},
  journal={Phys. Rev. Lett.},
  volume={121},
  number={7},
  pages={078001},
  year={2018},
  doi={https://doi.org/10.1103/PhysRevLett.121.078001},
}

@ARTICLE{bayles2017probe,
  title={Probe microrheology without particle tracking by differential dynamic microscopy},
  author={Bayles, Alexandra V and Squires, Todd M and Helgeson, Matthew E},
  journal={Rheol. Acta},
  volume={56},
  pages={863--869},
  year={2017},
  doi={https://doi.org/10.1007/s00397-017-1047-7},
}

@ARTICLE{escobedo2018microliter,
  title={Microliter viscometry using a bright-field microscope: $\eta$-DDM},
  author={Escobedo-S{\'a}nchez, MA and Segovia-Guti{\'e}rrez, JP and Zuccolotto-Bernez, AB and Hansen, J and Marciniak, CC and Sachowsky, K and Platten, F and Egelhaaf, SU},
  journal={Soft Matter},
  volume={14},
  number={34},
  pages={7016--7025},
  year={2018},
  doi={https://doi.org/10.1039/C8SM00784E},
}

@ARTICLE{edera2017differential,
  title={Differential dynamic microscopy microrheology of soft materials: A tracking-free determination of the frequency-dependent loss and storage moduli},
  author={Edera, Paolo and Bergamini, Davide and Trappe, V{\'e}ronique and Giavazzi, Fabio and Cerbino, Roberto},
  journal={Phys. Rev. Mater.},
  volume={1},
  number={7},
  pages={073804},
  year={2017},
  doi={https://doi.org/10.1103/PhysRevMaterials.1.073804},
}

@MISC{SciencePlots,
    author={Garrett, John D.},
    title={SciencePlots [Version 2.1.1]},
    year={2021},
    howpublished="\url{http://doi.org/10.5281/zenodo.4106649}",
}

@ARTICLE{matplotlib,
  Author    = {Hunter, J. D.},
  Title     = {Matplotlib: A 2D graphics environment},
  Journal   = {Comput. Sci. Eng.},
  Volume    = {9},
  Number    = {3},
  Pages     = {90--95},
  publisher = {IEEE COMPUTER SOC},
  doi       = {10.1109/MCSE.2007.55},
  year      = 2007
}

@BOOK{priemer1991introductory,
  title={Introductory signal processing},
  author={Priemer, Roland},
  volume={6},
  year={1990},
  publisher={World scientific},
  doi={https://doi.org/10.1142/0864},
}

@ARTICLE{harris1978use,
  title={On the use of windows for harmonic analysis with the discrete Fourier transform},
  author={Harris, Fredric J},
  journal={Proc. IEEE},
  volume={66},
  number={1},
  pages={51--83},
  year={1978},
  publisher={IEEE},
  doi={https://doi.org/10.1109/PROC.1978.10837},
}

@ARTICLE{larsen2021probing,
  title={Probing interactions of therapeutic antibodies with serum via second virial coefficient measurements},
  author={Larsen, Hayli A and Atkins, William M and Nath, Abhinav},
  journal={Biophys. J.},
  volume={120},
  number={18},
  pages={4067--4078},
  year={2021},
  doi={https://doi.org/10.1016/j.bpj.2021.08.007},
}

@ARTICLE{drechsler2017active,
  title={Active diffusion and advection in Drosophila oocytes result from the interplay of actin and microtubules},
  author={Drechsler, Maik and Giavazzi, Fabio and Cerbino, Roberto and Palacios, Isabel M},
  journal={Nat. Commun.},
  volume={8},
  number={1},
  pages={1520},
  year={2017},
  publisher={Nature Publishing Group UK London},
  doi={https://doi.org/10.1038/s41467-017-01414-6},
}

@BOOK{chatfield2013analysis,
  title={The analysis of time series: theory and practice},
  author={Chatfield, Christopher},
  year={2013},
  publisher={Springer}
}

@BOOK{goodman2015statistical,
  title={Statistical optics},
  author={Goodman, Joseph W},
  year={2015},
  publisher={John Wiley \& Sons}
}

@MISC{ddm-toolkit,
  author={Barthe, Lancelot and Werts, Martinus},
  title={{ddm-toolkit}},
  howpublished="\url{https://github.com/mhvwerts/ddm-toolkit}",
  year={2020},
  note={Accessed: 2024-04-30},
}

@MISC{muntz_openddm,
  author={Muntz, Iain and Conboy, James and Ist\'{u}riz, Irene and Kok, Maurits},
  title={{openddm}},
  howpublished="\url{https://github.com/koenderinklab/openddm}",
  year={2022},
  note={Accessed: 2024-04-30},
}

@MISC{ddmsoft,
  author={Dux, Fr\'{e}d\'{e}ric},
  title={{ddmsoft}},
  howpublished="\url{https://github.com/duxfrederic/ddmsoft}",
  year={2019},
  note={Accessed: 2024-04-30},
}

@MISC{cddm,
  author={Petelin, Andrej},
  title={{cddm [Version 0.3.0]}},
  howpublished="\url{https://zenodo.org/records/4603716}",
  year={2021},
}

@article{arko2019cross,
  title={Cross-differential dynamic microscopy},
  author={Arko, Matej and Petelin, Andrej},
  journal={Soft Matter},
  volume={15},
  number={13},
  pages={2791--2797},
  year={2019},
  publisher={Royal Society of Chemistry},
  doi={https://doi.org/10.1039/C9SM00121B},
}

@MISC{pyddm,
  author={McGorty, Ryan},
  title={{PyDDM}},
  howpublished="\url{https://github.com/rmcgorty/PyDDM}",
  year={2022},
  note={Accessed: 2024-04-30},
}

@MISC{lmfit,
    author={Newville, Matthew and Stensitzki, Till and Allen, Daniel B. and Ingargiola, Antonino},
    title={lmfit [Version 1.3.1]},
    year={2024},
    howpublished="\url{https://doi.org/10.5281/zenodo.11813}",
}

@MISC{fastddm,
  author={Lattuada, Enrico and Krautgasser, Fabian and Lavaud, Maxime and Cerbino, Roberto},
  title={{fastDDM [Version 0.3.15]}},
  howpublished="\url{https://github.com/somexlab/fastddm}",
  year={2022},
}

@MISC{fastddm-tutorials,
  author={Lattuada, Enrico and Krautgasser, Fabian and Lavaud, Maxime and Cerbino, Roberto},
  title={{fastDDM tutorials}},
  howpublished="\url{https://github.com/somexlab/fastddm-tutorials}",
  year={2024},
}

@ARTICLE{ladner1980parallel,
  title={Parallel prefix computation},
  author={Ladner, Richard E and Fischer, Michael J},
  journal={J. Assoc. Comput. Mach.},
  volume={27},
  number={4},
  pages={831--838},
  year={1980},
  publisher={ACM New York, NY, USA},
}

@ARTICLE{zhang2023differential,
  title={Differential Dynamic Microscopy: Diffusion Measurements Where You Want Them},
  author={Zhang, Xujun and Fu, Jinxin and Zhang, Zhaoxian and Jangda, Mateen and Rosu, Cornelia and Parkinson, Graham D. B. and Russo, Paul S.},
  journal={Macromolecules},
  volume={57},
  number={1},
  pages={3--20},
  year={2024},
  doi={https://doi.org/10.1021/acs.macromol.3c01166},
}

@article{mason2000estimating,
  title={Estimating the viscoelastic moduli of complex fluids using the generalized Stokes-Einstein equation},
  author={Mason, Thomas G},
  journal={Rheol. Acta},
  volume={39},
  pages={371--378},
  year={2000},
  publisher={Springer},
  doi={https://doi.org/10.1007/s003970000094},
}

@book{bray2000cell,
  title={Cell movements: from molecules to motility},
  author={Bray, Dennis},
  year={2000},
  publisher={Garland Science}
}

@article{nossal1971use,
  title={Use of laser scattering for quantitative determinations of bacterial motility},
  author={Nossal, R and Chen, S-H and Lai, C-C},
  journal={Opt. Commun.},
  volume={4},
  number={1},
  pages={35--39},
  year={1971},
  publisher={Elsevier},
  doi={https://doi.org/10.1016/0030-4018(71)90122-2},
}

@article{boon1974light,
  title={Light-scattering spectrum due to wiggling motions of bacteria},
  author={Boon, Jean Pierre and Nossal, Ralph and Chen, Sow-Hsin},
  journal={Biophys. J.},
  volume={14},
  number={11},
  pages={847--864},
  year={1974},
  publisher={Elsevier},
  doi={https://doi.org/10.1016/s0006-3495(74)85954-0},
}

@article{giavazzi2016simultaneous,
  title={Simultaneous characterization of rotational and translational diffusion of optically anisotropic particles by optical microscopy},
  author={Giavazzi, Fabio and Haro-P{\'e}rez, Catalina and Cerbino, Roberto},
  journal={J. Phys.: Condens. Matter},
  volume={28},
  number={19},
  pages={195201},
  year={2016},
  publisher={IOP Publishing},
  doi={https://doi.org/10.1088/0953-8984/28/19/195201},
}

@article{mason1995optical,
  title={Optical measurements of frequency-dependent linear viscoelastic moduli of complex fluids},
  author={Mason, Thomas G and Weitz, David A},
  journal={Phys. Rev. Lett.},
  volume={74},
  number={7},
  pages={1250},
  year={1995},
  publisher={APS},
  doi={https://doi.org/10.1103/PhysRevLett.74.1250},
}

@article{marple1987digital,
  title={Digital spectral analysis with applications},
  author={Marple Jr, S Lawrence},
  journal={Englewood Cliffs},
  year={1987}
}

@article{vyas2016passive,
  title={Passive microrheology in the effective time domain: analyzing time dependent colloidal dispersions},
  author={Vyas, Bhavna M and Orpe, Ashish V and Kaushal, Manish and Joshi, Yogesh M},
  journal={Soft Matter},
  volume={12},
  number={39},
  pages={8167--8176},
  year={2016},
  publisher={Royal Society of Chemistry},
  doi={https://doi.org/10.1039/C6SM00829A},
}

@article{evans2009direct,
  title={Direct conversion of rheological compliance measurements into storage and loss moduli},
  author={Evans, RML and Tassieri, Manlio and Auhl, Dietmar and Waigh, Thomas A},
  journal={Phys. Rev. E},
  volume={80},
  number={1},
  pages={012501},
  year={2009},
  publisher={APS},
  doi={https://doi.org/10.1103/PhysRevE.80.012501},
}

@article{lennon2023data,
  title={A data-driven method for automated data superposition with applications in soft matter science},
  author={Lennon, Kyle R and McKinley, Gareth H and Swan, James W},
  journal={Data-Centric Engineering},
  volume={4},
  pages={e13},
  year={2023},
  publisher={Cambridge University Press},
  doi={https://doi.org/10.1017/dce.2023.3},
}

@inproceedings{schmidt2018,
  author    = {Uwe Schmidt and Martin Weigert and Coleman Broaddus and Gene Myers},
  title     = {Cell Detection with Star-Convex Polygons},
  booktitle = {Medical Image Computing and Computer Assisted Intervention - {MICCAI} 
  2018 - 21st International Conference, Granada, Spain, September 16-20, 2018, Proceedings, Part {II}},
  pages     = {265--273},
  year      = {2018},
  doi       = {10.1007/978-3-030-00934-2_30}
}

@article{freud2020,
    title = {freud: A Software Suite for High Throughput
             Analysis of Particle Simulation Data},
    author = {Vyas Ramasubramani and
              Bradley D. Dice and
              Eric S. Harper and
              Matthew P. Spellings and
              Joshua A. Anderson and
              Sharon C. Glotzer},
    journal = {Comput. Phys. Commun.},
    volume = {254},
    pages = {107275},
    year = {2020},
    issn = {0010-4655},
    doi = {https://doi.org/10.1016/j.cpc.2020.107275},
    url = {http://www.sciencedirect.com/science/article/pii/S0010465520300916},
}

@article{angelini2011glass,
  title={Glass-like dynamics of collective cell migration},
  author={Angelini, Thomas E and Hannezo, Edouard and Trepat, Xavier and Marquez, Manuel and Fredberg, Jeffrey J and Weitz, David A},
  journal={Proc. Natl. Acad. Sci. U. S. A.},
  volume={108},
  number={12},
  pages={4714--4719},
  year={2011},
  publisher={National Acad Sciences},
  doi={https://doi.org/10.1073/pnas.1010059108},
}

@article{fajner2021hecw,
  title={Hecw controls oogenesis and neuronal homeostasis by promoting the liquid state of ribonucleoprotein particles},
  author={Fajner, Valentina and Giavazzi, Fabio and Sala, Simona and Oldani, Amanda and Martini, Emanuele and Napoletano, Francesco and Parazzoli, Dario and Cesare, Giuliana and Cerbino, Roberto and Maspero, Elena and others},
  journal={Nat. Commun.},
  volume={12},
  number={1},
  pages={5488},
  year={2021},
  publisher={Nature Publishing Group UK London},
  doi={https://doi.org/10.1038/s41467-021-25809-8},
}

@article{park2015unjamming,
  title={Unjamming and cell shape in the asthmatic airway epithelium},
  author={Park, Jin-Ah and Kim, Jae Hun and Bi, Dapeng and Mitchel, Jennifer A and Qazvini, Nader Taheri and Tantisira, Kelan and Park, Chan Young and McGill, Maureen and Kim, Sae-Hoon and Gweon, Bomi and others},
  journal={Nat. Mater.},
  volume={14},
  number={10},
  pages={1040--1048},
  year={2015},
  publisher={Nature Publishing Group UK London},
  doi={https://doi.org/10.1038/nmat4357},
}

@article{devany2021cell,
  title={Cell cycle-dependent active stress drives epithelia remodeling},
  author={Devany, John and Sussman, Daniel M and Yamamoto, Takaki and Manning, M Lisa and Gardel, Margaret L},
  journal={Proc. Natl. Acad. Sci. U. S. A.},
  volume={118},
  number={10},
  pages={e1917853118},
  year={2021},
  publisher={National Acad Sciences},
  doi={https://doi.org/10.1073/pnas.1917853118},
}

@article{atia2018geometric,
  title={Geometric constraints during epithelial jamming},
  author={Atia, Lior and Bi, Dapeng and Sharma, Yasha and Mitchel, Jennifer A and Gweon, Bomi and A. Koehler, Stephan and DeCamp, Stephen J and Lan, Bo and Kim, Jae Hun and Hirsch, Rebecca and others},
  journal={Nat. Phys.},
  volume={14},
  number={6},
  pages={613--620},
  year={2018},
  publisher={Nature Publishing Group UK London},
  doi={https://doi.org/10.1038/s41567-018-0089-9},
}

@article{potenza_how_2010,
    title = {How to Measure the Optical Thickness of Scattering Particles from the Phase Delay of Scattered Waves: Application to Turbid Samples},
    volume = {105},
    url = {https://link.aps.org/doi/10.1103/PhysRevLett.105.193901},
    doi = {10.1103/PhysRevLett.105.193901},
    number = {19},
    urldate = {2025-05-24},
    journal = {Physical Review Letters},
    author = {Potenza, M. A. C. and Sabareesh, K. P. V. and Carpineti, M. and Alaimo, M. D. and Giglio, M.},
    month = nov,
    year = {2010},
    note = {Publisher: American Physical Society},
    pages = {193901},
}

@article{duri_resolving_2009,
    title = {Resolving {Long}-{Range} {Spatial} {Correlations} in {Jammed} {Colloidal} {Systems} {Using} {Photon} {Correlation} {Imaging}},
    volume = {102},
    url = {https://link.aps.org/doi/10.1103/PhysRevLett.102.085702},
    doi = {10.1103/PhysRevLett.102.085702},
    number = {8},
    urldate = {2025-05-25},
    journal = {Physical Review Letters},
    author = {Duri, A. and Sessoms, D. A. and Trappe, V. and Cipelletti, L.},
    month = feb,
    year = {2009},
    note = {Publisher: American Physical Society},
    pages = {085702},
}

@book{lindner_neutrons_2024,
    title = {Neutrons, {X}-rays, and {Light}: {Scattering} {Methods} {Applied} to {Soft} {Condensed} {Matter}},
    isbn = {978-0-443-29117-3},
    shorttitle = {Neutrons, {X}-rays, and {Light}},
    publisher = {Elsevier},
    author = {Lindner, Peter and Oberdisse, Julian},
    month = dec,
    year = {2024},
    note = {Google-Books-ID: JGkGEQAAQBAJ},
}

@misc{newville_lmfit_2025,
	title = {{LMFIT}: {Non}-{Linear} {Least}-{Squares} {Minimization} and {Curve}-{Fitting} for {Python}},
	url = {https://doi.org/10.5281/zenodo.16175987},
	publisher = {Zenodo},
	author = {Newville, Matthew and Otten, Renee and Nelson, Andrew and Stensitzki, Till and Ingargiola, Antonino and Allan, Daniel and Fox, Austin and Carter, Faustin and Rawlik, Michal},
	month = jul,
	year = {2025},
	doi = {10.5281/zenodo.16175987},
}
\end{document}


\maketitle
\section{Estimation of the noise baseline $B(q)$}
\begin{figure}[h]
    \centering
\includegraphics[width=\columnwidth]{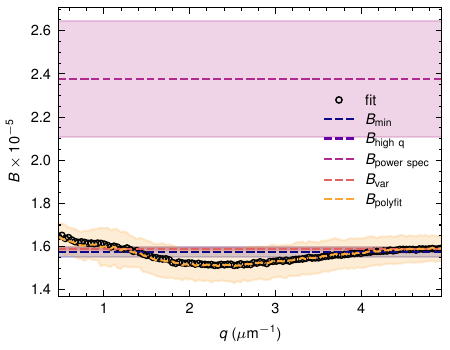}
    \caption{Dashed lines represent estimates of the noise $B$ as a function of the wave vector $q$. Black symbols show the result of the fit to the structure function of 250 nm colloids, as described in Section V.A.}
    \label{fig:part-siz-noise}
\end{figure}

In Fig.~\ref{fig:part-siz-noise} we show the results of using different methods to estimate the noise baseline $B(q)$. In addition to the polynomial $B_\mathrm{polyfit}$ fit described in section V. A. of the main text, other alternatives are:

\begin{align*}
    &B_{\mathrm{min}} = \min{\left\{ d(q, \Delta t = \Delta t_{\mathrm{min}}) \right\}} ~, \\
    &B_{\mathrm{high \, q}} = \langle d(q, \Delta t) \rangle_{q > q^*}~,\\
    &B_{\mathrm{power \, spec}} = \langle \lvert \tilde{i}(q, t) \rvert^2 \rangle_{t, q > q^*}~,\\
    &B_{\mathrm{var}} = \langle \lvert \tilde{i}(q, t) - \tilde{i}_0 (q) \rvert^2 \rangle_{t, q > q^*}~.
\end{align*}

\section{Intrinsic variance of the structure function}

To the best of our knowledge, no closed–form expression for the variance of the structure function 
$D(q,\Delta t)$ has previously appeared in the DDM literature. 
The derivation presented here is therefore original to this work and constitutes one of the methodological contributions of the paper. This expression enables a principled estimate of the statistical uncertainty of $D(q,\Delta t)$ directly from 
the raw image series, which is essential for robust weighting in subsequent fits.

We start from the definition of the structure function $D(\mathbf{q},\Delta t)$, as an estimator obtained by using $N_{im}$ consecutive images (see Eq. 44 in the main text). We then focus on the variance of $D$ defined as $V_0(q,\Delta t)=\langle D(\mathbf{q},\Delta t)^2 \rangle-\langle D(\mathbf{q},\Delta t) \rangle^2$.

Under the following hypotheses:
\begin{itemize}
\item Each image contains a large number $N_\text{p}$ of identical particles, each one of them performing an uncorrelated Brownian motion with the same diffusivity.
\item Noise is delta-correlated (both in space and time) and represents an additive contribution. Each pixel fluctuates with the same variance.  
\item For a given $q$-value, the characteristic decorrelation rate $\Gamma(q)$ is lower than the sampling rate $\gamma_0=1/\Delta t_0$. Under this hypothesis, all functions $F$ of interest depend "smoothly" on the image index $m$. This enables evaluating sums performed over the index $m$ as time integrals 
\begin{equation}
\frac{1}{m_2-m_1}\sum^{m_2}_{m=m_1} F(m)\sim \frac{1}{t_2-t_1}\int^{t_2}_{t_1}dt F(\gamma_0 t),
\end{equation}
where $t_1=m_1\Delta t_0$ and $t_2=m_2\Delta t_0$.
\end{itemize}

we obtain, to the leading order in the number $N_\text{p}$ of particles, the following expression
\begin{align*}
V_0=\sigma^2_{\text{sig}}+\sigma^2_{\text{noi}}+\sigma^2_{\text{mix}},
\end{align*}
where 
\begin{align*}
    \sigma^2_{\text{sig}} &= \frac{A^2(q)}{4} \frac{a+b+c+d}{\mathcal{T}^2} ~,\\
    \sigma^2_{\text{noi}} &= \frac{B^2(q)}{2} \frac{3 \tilde{T} - \Delta t \gamma_0}{\tilde{T}^2}~, \\
    \sigma^2_{\text{mix}} &= A(q)B(q) \frac{2\tilde{T}\mathcal{D} + (\tilde{T}-\Delta t \gamma_0) \mathcal{D}^2}{\tilde{T}^2} ~,\\
\end{align*}

and where the various variables appearing are defined as

\begin{align*}
    \tilde{T} (\Delta t) &= N_{im} - \Delta t \gamma_0 ~, \\
    \mathcal{T} (q, \Delta t) &= \frac{\Gamma(q)}{\gamma_0} \tilde{T} ~,\\
    Y(q, \Delta t) &= \Delta t \Gamma(q) ~,\\
    \mathcal{D} (q, \Delta t) &= 1 - \exp(-Y) = 1 - f(q, \Delta t) ~,\\
    a &= \frac{1}{2} \{2 \mathcal{T} - [1 - \exp(-2\mathcal{T})]\} [2 - \exp(-Y)]^2 ~,\\
    b &= \frac{1}{2} - \left( \frac{1}{2} + Y \right) \exp(-2Y) + [2 - \exp(-2Y)] (\mathcal{T}-Y) - \frac{1}{2} \left[ 1 - \exp(-2(\mathcal{T} - Y)) \right] ~,\\
    c &= -4 (\mathcal{T} + 2 \mathcal{T} Y - Y - Y^2) \exp(-Y) + 2 \left[ \exp(-Y) - \exp(-2\mathcal{T} + Y) \right] ~,\\
    d &= 2 (\mathcal{T} + 2 \mathcal{T} Y - Y - Y^2) \exp(-2Y) - \left[ \exp(-2Y) - \exp(-2\mathcal{T}) \right]~ . \\
\end{align*}

This expression is expected to cease to work whenever one or more of the above-mentioned hypotheses is not fulfilled, in particular when the particles are not identical, they do interact, or they do not perform Brownian motion.

\section{FastDDM performance}

\begin{figure}[h]
    \centering
    \includegraphics[width=0.7\columnwidth]{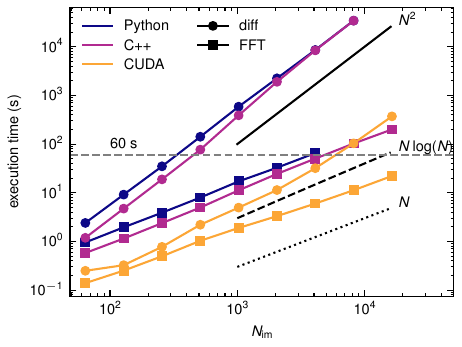}
    \caption{Compute time (symbols) needed to calculate the structure function from a sequence of $N_{im}$ 16-bit images, $512\times512$ pixels each. The lines represent the expected scaling of the different algorithms: the ``difference'' scheme should scale as $N_{im}^2$, while the FFT-based algorithm is $N_{im} \log{N_{im}}$. The dashed horizontal line indicates an execution time of 1 minute.}
    \label{fig:performance}
\end{figure}

Fig.~\ref{fig:performance} shows the compute time needed for each \emph{core} and \emph{mode} of fastDDM to calculate the structure function from a sequence of $N_{im}$ 16-bit synthetic images ($512\times512$ pixels each).
Performance evaluation was conducted using a workstation with the following specifications:

\begin{itemize}
    \item CPU: AMD\textregistered~Ryzen 7 5800x
    \item RAM: 128 GB DDR4
    \item GPU: NVIDIA GeForce RTX 3080 Ti with 12 GB of dedicated VRAM
    \item OS: Ubuntu 20.04.6 LTS
\end{itemize}

In our initial analysis, we focused on image sequences where $N_{im}$ is a power of two.
The following script was used to measure the execution time.

\begin{lstlisting}[language=Python, caption=Execution time script]
import fastddm as fd

import numpy as np
import pandas as pd
from itertools import product
from time import time

# create dummy image sequence (size is largest required)
images = np.random.randint(65536, size=(16384, 512, 512), dtype=np.uint16)

num_images = [2**i for i in range(6, 15)]
cores = ['py', 'cpp', 'cuda']
modes = ['diff', 'fft']

# prepare lists for dataframe
N, c, m, t = [], [], [], []

for num, core, mode in product(num_images, cores, modes):
    # skip execution of 'diff' for 'py' and 'cpp' core for largest N
    if num == 16384 and mode == 'diff' and core != 'cuda':
        continue
    # skip execution for 'py' also when N is 8000 or larger
    if num >= 8000 and core == 'py' and mode == 'fft':
        continue
    # get initial time
    t0 = time()
    # compute structure function
    dqt = fddm.ddm(img_seq=images[:num],
                   lags=range(1, num),
                   core=core,
                   mode=mode)
    # get final time
    t_exec = time() - t0

    # add values to lists
    N.append(num)
    c.append(core)
    m.append(mode)
    t.append(t_exec)

# create pandas dataframe and save as csv
df = pd.DataFrame({'num_images': N,
                   'core': c,
                   'mode': m,
                   'exec_time': t})
df.to_csv('exec_times_powerof2.csv', index=False)
\end{lstlisting}

\begin{figure}[h]
    \centering
    \includegraphics[width=0.7\columnwidth]{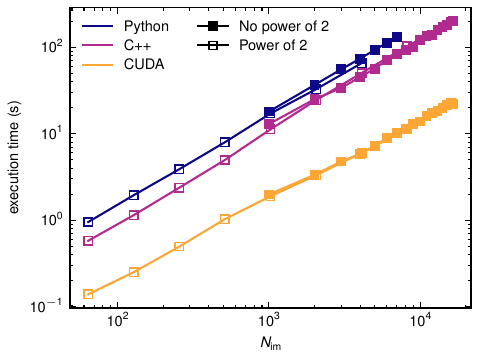}
    \caption{Comparative analysis of the FFT-based algorithm performance on image sequences with $N_\mathrm{im}$ as a power of 2 (open symbols) versus non-power of 2 (closed symbols).}
    \label{fig:no-power-of-2}
\end{figure}

The power-of-two condition is particularly critical when calculating the structure function via the FFT algorithm, as it traditionally ensures computational efficiency and algorithmic simplicity.
Nevertheless, advancements in FFT libraries have expanded their capability to process sequence lengths that are factorizable into a limited set of prime numbers, thereby offering greater flexibility without significantly compromising performance.
Fig.~\ref{fig:no-power-of-2} presents a comparative analysis of the execution times for the FFT-based algorithm across a series of data points where $N_\mathrm{im}$ is multiple of 1000 in the range $N_\mathrm{im} \in [1000, 16000]$.
The data consistently indicate no significant degradation in performance.

\clearpage
\newpage
\section{Intermediate scattering functions}

\subsection{A (mostly harmless) introduction to particle sizing}

\begin{figure}[h]
    \centering
    \includegraphics{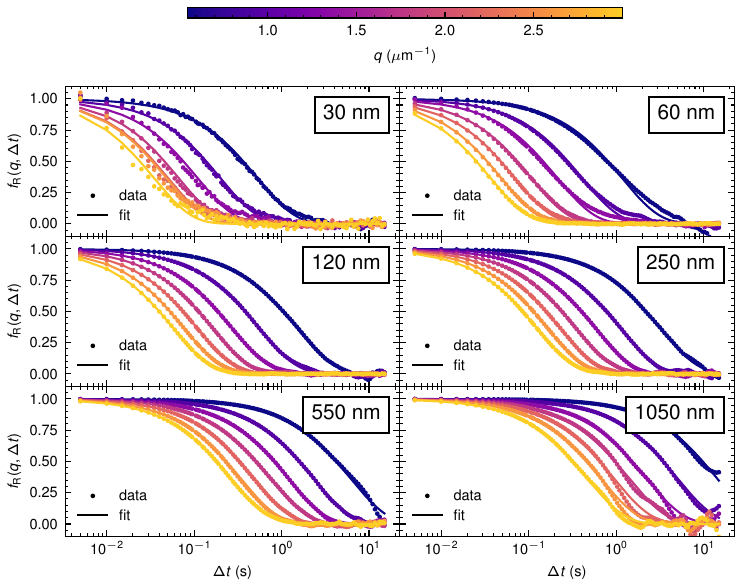}
    \caption{Intermediate scattering function $f(q, \Delta t)$ (symbols) and best fit (lines) obtained from the data presented in Fig.~5 for the different particle sizes, as indicated by the labels.}
    \label{fig:fqt-edinburgh}
\end{figure}

\clearpage
\newpage
\subsection{Estimation of the Mean Squared Displacement}

\begin{figure}[h]
    \centering
    \includegraphics{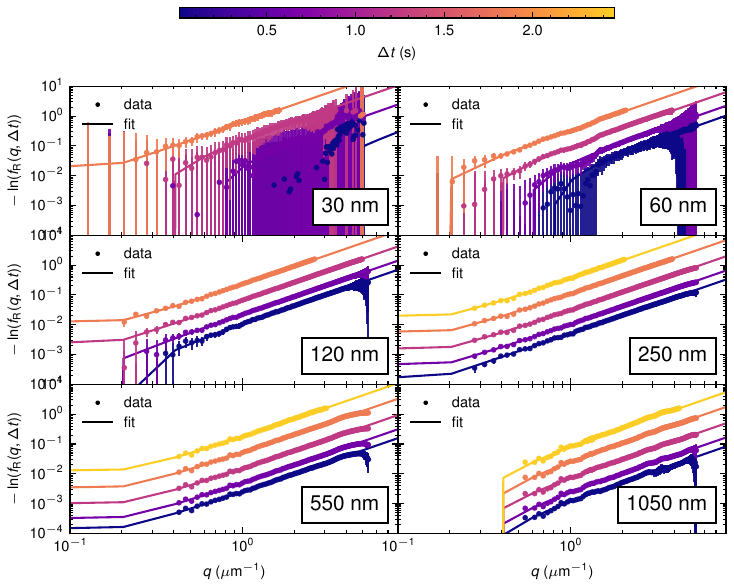}
    \caption{Logarithm of Intermediate scattering function $f(q, \Delta t)$ (symbols) and best fit (lines) obtained from the data presented in Fig.~5 for the different particle sizes, as indicated by the labels.}
    \label{fig:fqt-edinburgh}
\end{figure}

\clearpage
\newpage

\subsection{Effect of image windowing}

\begin{figure}[h]
    \centering
    \includegraphics{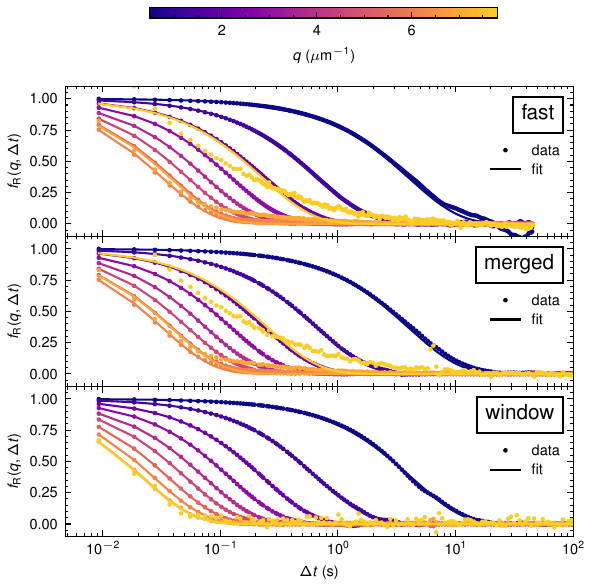}
    \caption{Intermediate scattering function $f(q, \Delta t)$ (symbols) and best fit (lines) obtained from the data presented in Fig.~8 for the different cases, as indicated by the labels.}
    \label{fig:fqt-fast-melt-windowing}
\end{figure}

\clearpage
\newpage
\subsection{Effect of the objective lens magnification}

\begin{figure}[h]
    \centering
    \includegraphics{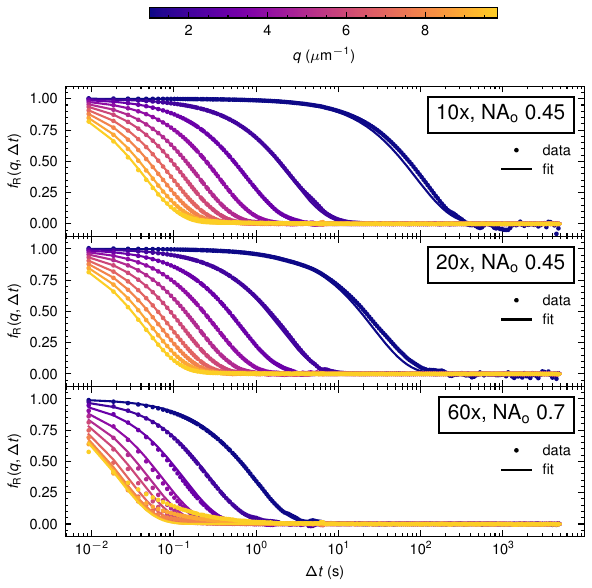}
    \caption{Intermediate scattering function $f(q, \Delta t)$ (symbols) and best fit (lines) obtained from the data presented in Fig.~9 for the different objectives used, as indicated by the labels.}
    \label{fig:fqt-objective-mag}
\end{figure}

\begin{figure}[h]
    \centering
    \includegraphics{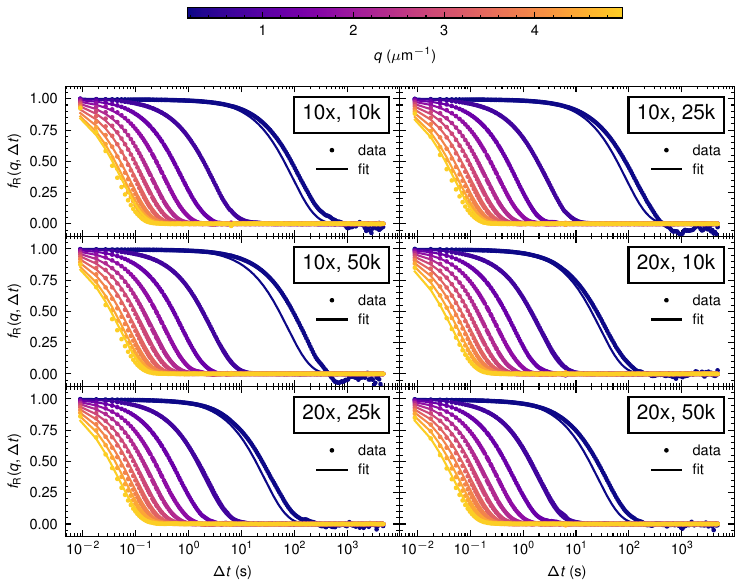}
    \caption{Intermediate scattering function $f(q, \Delta t)$ (symbols) and best fit (lines) obtained from the data presented in Fig.~10 for the different objectives used and average intensities on the camera sensor, as indicated by the labels.}
    \label{fig:fqt-objective-mag-int}
\end{figure}

\clearpage
\newpage
\subsection{Effect of the objective numerical aperture}

\begin{figure}[h]
    \centering
    \includegraphics{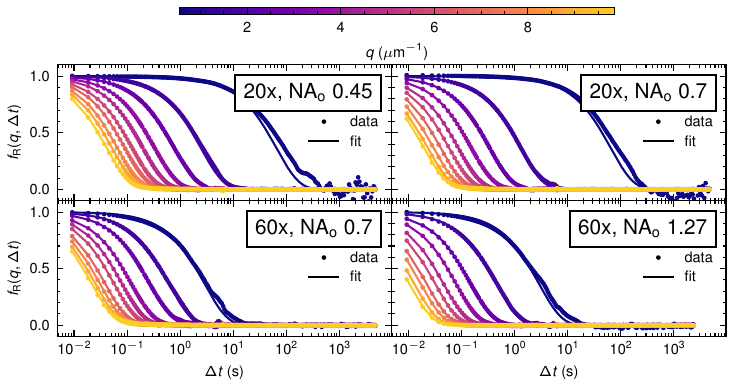}
    \caption{Intermediate scattering function $f(q, \Delta t)$ (symbols) and best fit (lines) obtained from the data presented in Fig.~11 for the different objectives used, as indicated by the labels.}
    \label{fig:fqt-objective-na}
\end{figure}

\clearpage
\newpage
\subsection{Effect of the condenser numerical aperture}

\begin{figure}[h]
    \centering
    \includegraphics{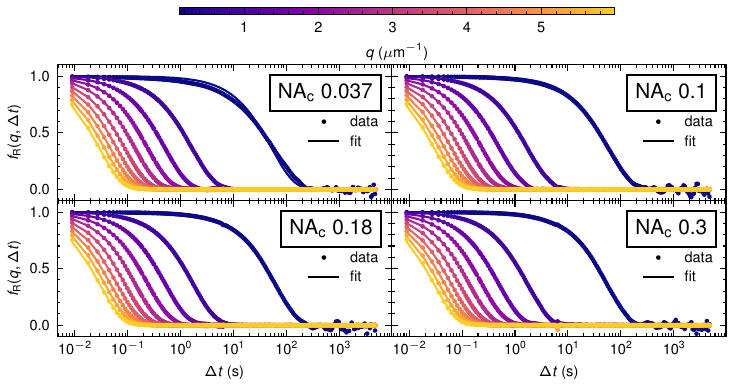}
    \caption{Intermediate scattering function $f(q, \Delta t)$ (symbols) and best fit (lines) obtained from the data presented in Fig.~12 for the different condenser numerical apertures used, as indicated by the labels.}
    \label{fig:fqt-condenser-na}
\end{figure}


\clearpage
\newpage